\input harvmac.tex

% AMS fonts required!

\input amssym.def
\input amssym.tex

\input epsf.tex

\noblackbox

\def\urlfont{\hyphenpenalty=10000 \hyphenchar\tentt='057 \tt }

\def\boldone{\relax{\rm 1\kern-.35em 1}}

\def\ZZ{{\Bbb Z}}
\def\Spin{\mathop{\rm Spin}\nolimits}
\def\dvol{\mathop{d\rm vol}\nolimits}

\def\ads{\hbox{$AdS$}}
\def\diag{\mathop{\rm diag}\nolimits}
\def\Hol{\mathop{\rm Hol}\nolimits}
\def\timesw{\mathbin{\times_w}}
\def\go{\frak{o}}
\def\gu{\frak{u}}
\def\gso{\frak{so}}
\def\gsu{\frak{su}}
\def\gsp{\frak{sp}}
\def\AdS{$AdS$}
\def\CFT{$CFT$}

\def\CN{{\cal N}}
\def\CM{{\cal M}}
\def\BC{{\Bbb C}}
\def\BP{{\Bbb P}}
\def\BZ{{\Bbb Z}}
\def\BW{{\Bbb W}}
\def\BF{{\Bbb F}}
\def\bk{{\bf N}}
\def\bkb{{\bf\bar N}}

\def\catquot{\mathchoice
{\mathrel{\mskip-4.5mu/\!/\mskip-4.5mu}}
{\mathrel{\mskip-4.5mu/\!/\mskip-4.5mu}}
{\mathrel{\mskip-3mu/\mskip-4.5mu/\mskip-3mu}}
{\mathrel{\mskip-3mu/\mskip-4.5mu/\mskip-3mu}}
}

\newbox\tmpbox\setbox\tmpbox\hbox{\abstractfont DUKE-CGTP-98-08}
\Title{\vbox{\baselineskip12pt\hbox to\wd\tmpbox{\hss
hep-th/9810201}\hbox{DUKE-CGTP-98-08}}}
{\vbox{
\centerline{Non-Spherical Horizons, I} }}
\centerline{David R. Morrison and M. Ronen Plesser\footnote{$^*$}{On leave from
Department of Particle Physics, Weizmann Institute of Science, Rehovot,
Israel}}
\bigskip
\bigskip
\centerline{Center for Geometry and Theoretical Physics}
\centerline{Box 90318, Duke University}
\centerline{Durham, NC 27708-0318}
\bigskip
\bigskip
\bigskip
\bigskip
\noindent
{\bf Abstract:}
We formulate an extension of Maldacena's \AdS/CFT conjectures to the case of
branes located at singular points in the ambient transverse space.  For
singularities which occur at finite distance in the moduli space of M or $F$
theory models with spacetime-filling branes, the conjectures identify the
worldvolume theory on the $p$-branes with a compactification of M or IIB
theory on $\AdS_{p+2} \times H^{D-p-2}$.  We show how the singularity
determines the horizon $H$, and demonstrate the relationship between global
symmetries on the worldvolume and gauge symmetries in the \AdS\ model.  As a
first application, we study some singularities relevant to the D3-branes
required in four-dimensional $F$-theory.  For these we are able to
explicitly derive the low-energy field theory on the worldvolume and
compare its properties to predictions from the dual \AdS\ model. In
particular, we examine the baryon spectra of the models and the fate of the
Abelian factors in the gauge group.

\Date{October 1998}

\lref\AFHS{
B.~S. Acharya, J.~M. Figueroa-O'Farrill, C.~M. Hull, and B.~Spence, ``Branes at
  conical singularities and holography,'' {\urlfont hep-th/9808014}.}

\lref\AFM{
O.~Aharony, A.~Fayyazuddin, and J.~Maldacena, ``The large {N} limit of {${\cal
  N}=2, 1$} field theories from threebranes in {F}-theory,'' J. High Energy
  Phys. {\bf 07} (1998) 013, {\urlfont hep-th/9806159}.}

\lref\AKY{
C.~Ahn, H.~Kim, and H.~S. Yang, ``{$SO(2N)$} (0,2) {SCFT} and {M} theory on
  {$AdS_7 \times {\bf RP}^4$},'' {\urlfont hep-th/9808182}.}

\lref\AOY{
O.~Aharony, Y.~Oz, and Z.~Yin, ``{M} theory on {$AdS_p\times S^{11-p}$} and
  superconformal field theories,'' Phys. Lett. B {\bf 430} (1998) 87--93,
  {\urlfont hep-th/9803051}.}

\lref\BBMeight{
K.~Becker and M.~Becker, ``{\cal M}-theory on eight-manifolds,'' Nucl. Phys. B
  {\bf 477} (1996) 155--167, {\urlfont hep-th/9605053}.}

\lref\BGM{
C.~P. Boyer, K.~Galicki, and B.~M. Mann, ``The geometry and topology of
  3-{S}asakian manifolds,'' J. Reine Angew. Math. {\bf 455} (1994) 163--220.}

\lref\CRW{
L.~Castellani, L.~J. Romans, and N.~P. Warner, ``A classification of
  compactifying solutions for {D=11} supergravity,'' Nucl. Phys. B {\bf 241}
  (1984) 429--462.}

\lref\CdO{
P.~Candelas and X.~C. de~la Ossa, ``Comments on conifolds,'' Nucl. Phys. B {\bf
  342} (1990) 246--268.}

\lref\DKV{
M.~R. Douglas, S.~Katz, and C.~Vafa, ``Small instantons, del {P}ezzo surfaces
  and type {I$'$} theory,'' Nucl. Phys. B {\bf 497} (1997) 155--172, {\urlfont
  hep-th/9609071}.}

\lref\DLPS{
M.~J. Duff, H.~L\"u, C.~N. Pope, and E.~Sezgin, ``Supermembranes with fewer
  supersymmetries,'' Phys. Lett. B {\bf 371} (1996) 206--214, {\urlfont
  hep-th/9511162}.}

\lref\DLP{
M.~J. Duff, H.~L\"{u}, and C.~N. Pope, ``{${\rm AdS}_5 \times S^5$}
  untwisted,'' Nucl. Phys. B {\bf 532} (1998) 181--209,
  {\urlfont hep-th/9803061}.}

\lref\DNKK{
M.~J. Duff, B.~E.~W. Nilsson, and C.~N. Pope, ``{K}aluza--{K}lein
  supergravity,'' Phys. Rep. {\bf 130} (1986) 1--142.}

\lref\DuffMinasianWitten{
M.~J. Duff, R.~Minasian, and E.~Witten, ``Evidence for heterotic/heterotic
  duality,'' Nucl. Phys. B {\bf 465} (1996) 413--438, {\urlfont
  hep-th/9601036}.}

\lref\DuffStelle{
M.~J. Duff and K.~S. Stelle, ``Multimembrane solutions of {$D=11$}
  supergravity,'' Phys. Lett. B {\bf 253} (1991) 113--118.}

\lref\FKPZ{
S.~Ferrara, A.~Kehagias, H.~Partouche, and A.~Zaffaroni, ``Membranes and
  fivebranes with lower supersymmetry and their {AdS} supergravity duals,''
  Phys. Lett. B {\bf 431} (1998) 42--48, {\urlfont hep-th/9803109}.}

\lref\FKone{
Th. Friedrich and I.~Kath, ``{E}instein manifolds of dimension five with small
  first eigenvalue of the {D}irac operator,'' J. Differential Geom. {\bf 29}
  (1989) 263--279.}

\lref\FKtwo{
Th. Friedrich and I.~Kath, ``Real {K}illing spinors and holonomy,'' Comm. Math.
  Phys. {\bf 133} (1990) 543--561.}

\lref\FZ{
S.~Ferrara and A.~Zaffaroni, ``{N=1,2} {4D} superconformal field theories and
  supergravity in {$AdS_5$},'' Phys. Lett. B {\bf 431} (1998) 49--56,
  {\urlfont hep-th/9803060}.}

\lref\FaySpa{
A.~Fayyazuddin and M.~Spali\'nski, ``Large {N} superconformal gauge theories
  and supergravity orientifolds,'' Nucl. Phys. B {\bf 535} (1998) 219--232,
  {\urlfont hep-th/9805096}.}

\lref\GH{
P.~Griffiths and J.~Harris, ``Principles of algebraic geometry,'' Wiley, 1978.}

\lref\GibbonsRych{
G.~W. Gibbons and P.~Rychenkova, ``Cones, tri-{S}asakian structures and
  superconformal invariance,'' {\urlfont hep-th/9809158}.}

\lref\GKPo{
S.~S. Gubser, I.~R. Klebanov, and A.~M. Polyakov, ``Gauge theory correlators
  from non-critical string theory,'' Phys. Lett. B {\bf 428} (1998) 105--114,
  {\urlfont hep-th/9802109}.}

\lref\GMS{
B.~R. Greene, D.~R. Morrison, and A.~Strominger, ``Black hole condensation and
  the unification of string vacua,'' Nucl. Phys. B {\bf 451} (1995) 109--120,
  {\urlfont hep-th/9504145}.}

\lref\GanorHanany{
O.~J. Ganor and A.~Hanany, ``Small {$E_8$} instantons and tensionless non
  critical strings,'' Nucl. Phys. B {\bf 474} (1996) 122--140, {\urlfont
  hep-th/9602120}.}

\lref\GmodH{
L.~Castellani, A.~Ceresole, R.~D'Auria, S.~Ferrara, P.~Fr\'e, and M.~Trigiante,
  ``{$G/H$} {M}-branes and {$AdS_{p+2}$} geometries,'' Nucl. Phys. B {\bf 527}
  (1998) 142--170, {\urlfont hep-th/9803039}.}

\lref\Guven{
R.~G\"uven, ``Black p-brane solutions of {$D=11$} supergravity,'' Phys. Lett. B
  {\bf 276} (1992) 49--55.}

\lref\HSU{
A.~Hanany, M.~J. Strassler, and A.~M. Uranga, ``Finite theories and marginal
  operators on the brane,'' J. High Energy Phys. {\bf 06} (1998) 011,
  {\urlfont hep-th/9803086}.}

\lref\HoravaWitten{
P.~Ho\v{r}ava and E.~Witten, ``Heterotic and type {I} string dynamics from
  eleven dimensions,'' Nucl. Phys. B {\bf 460} (1996) 506--524, {\urlfont
  hep-th/9510209}.}

\lref\HorowitzStrominger{
G.~Horowitz and A.~Strominger, ``Black strings and p-branes,'' Nucl. Phys. B
  {\bf 360} (1991) 197--209.}

\lref\IRU{
L.~I. Ib\'a{\~n}ez, R.~Rabad\'an, and A.~M. Uranga, ``Anomalous {$U(1)$'s} in
  type {I} and type {IIB} {$D=4$}, {$N=1$} string vacua,'' {\urlfont
  hep-th/9808139}.}

\lref\KMP{
S.~Katz, D.~R. Morrison, and M.~R. Plesser, ``Enhanced gauge symmetry in type
  {II} string theory,'' Nucl. Phys. B {\bf 477} (1996) 105--140, {\urlfont
  hep-th/9601108}.}

\lref\MS{
D.~R. Morrison and N.~Seiberg, ``Extremal transitions and five-dimensional
  supersymmetric field theories,'' Nucl. Phys. B {\bf 483} (1997) 229--247,
  {\urlfont hep-th/9609070}.}

\lref\MVtwo{
D.~R. Morrison and C.~Vafa, ``Compactifications of {F}-theory on
  {C}alabi--{Y}au threefolds, {II},'' Nucl. Phys. B {\bf 476} (1996) 437--469,
  {\urlfont hep-th/9603161}.}

\lref\PRmirror{
E.~Perevalov and G.~Rajesh, ``Mirror symmetry via deformation of bundles on
  {K3} surfaces,'' Phys. Rev. Lett. {\bf 79} (1997) 2931--2934, {\urlfont
  hep-th/9706005}.}

\lref\SVW{
S.~Sethi, C.~Vafa, and E.~Witten, ``Constraints on low-dimensional string
  compactifications,'' Nucl. Phys. B {\bf 480} (1996) 213--224, {\urlfont
  hep-th/9606122}.}

\lref\SeibergWittensix{
N.~Seiberg and E.~Witten, ``Comments on string dynamics in six dimensions,''
  Nucl. Phys. B {\bf 471} (1996) 121--134, {\urlfont hep-th/9603003}.}

\lref\SusskindWitten{
L.~Susskind and E.~Witten, ``The holographic bound in anti-de {S}itter space,''
  {\urlfont hep-th/9805114}.}

\lref\Whol{
E.~Witten, ``Anti-de {S}itter space and holography,'' Adv. Theor. Math. Phys.
  {\bf 2} (1998) 253--291, {\urlfont hep-th/9802150}.}

\lref\Wilson{
P.~M.~H. Wilson, ``The {K\"a}hler cone on {C}alabi--{Y}au threefolds,'' Invent.
  Math. {\bf 107} (1992) 561--583, Erratum, ibid. {\bf 114} (1993) 231--233.}

\lref\WitMF{
E.~Witten, ``Phase transitions in {$M$}-theory and {$F$}-theory,'' Nucl. Phys.
  B {\bf 471} (1996) 195--216, {\urlfont hep-th/9603150}.}

\lref\ahnone{
C.~Ahn, K.~Oh, and R.~Tatar, ``Orbifolds of {$AdS_7\times S^4$} and six
  dimensional (0,1) {SCFT},'' Phys. Lett. B {\bf 442} (1998) 109--116,
  {\urlfont hep-th/9804093}.}

\lref\ahntwo{
C.~Ahn, K.~Oh, and R.~Tatar, ``Branes, orbifolds and the three dimensional
  {${\cal N}=2$} {SCFT} in the large {$N$} limit,'' J. High Energy
  Phys. {\bf 11} (1998) 024, {\urlfont hep-th/9806041}.}

\lref\altmann{
K.~Altmann, ``The versal deformation of an isolated toric {G}orenstein
  singularity,'' Invent. Math. {\bf 128} (1997) 443--479, {\urlfont
  alg-geom/9403004}.}

\lref\ami{
A.~Hanany and A.~M. Uranga, ``Brane boxes and branes on singularities,'' J.
  High Energy Phys. {\bf 05} (1998) 013, {\urlfont hep-th/9805139}.}

\lref\aotfth{
C.~Ahn, K.~Oh, and R.~Tatar, ``The large {$N$} limit of {${\cal N}=1$} field
  theories from {F} theory,'' {\urlfont hep-th/9808143}.}

\lref\baer{
C.~B\"ar, ``Real {K}illing spinors and holonomy,'' Comm. Math. Phys. {\bf 154}
  (1993) 509--521.}

\lref\berardbergery{
L.~B\'erard~Bergery, ``Sur de nouvelles vari\'et\'es riemanniennes
  d'{E}instein,'' Institut Elie Cartan {\bf 6} (1982) 1--60.}

\lref\berger{
M.~Berger, ``Sur les groups d'holonomie des vari\'et\'es \`a connexion affine
  et des vari\'et\'es riemanniennes,'' Bull. Soc. Math. France {\bf 83} (1955)
  279--330.}

\lref\berkooz{
M.~Berkooz, ``A supergravity dual of a (1,0) field theory in six dimensions,''
  Phys. Lett. B {\bf 437} (1998) 315--317, {\urlfont hep-th/9802195}.}

\lref\bj{
M.~Bershadsky and A.~Johansen, ``Large {N} limit of orbifold field theories,''
  Nucl. Phys. B {\bf 536} (1998) 141--148, {\urlfont hep-th/9803249}.}

\lref\bkv{
M.~Bershadsky, Z.~Kakushadze, and C.~Vafa, ``String expansion as large {$N$}
  expansion of gauge theories,'' Nucl. Phys. B {\bf 523} (1998) 59--72,
  {\urlfont hep-th/9803076}.}

\lref\botttu{
R.~Bott and L.~W. Tu, ``Differential forms in algebraic topology,'' Springer
  Verlag, New York, 1982.}

\lref\brg{
B.~R. Greene, ``{D}-brane topology changing transitions,'' Nucl. Phys. B {\bf
  525} (1998) 284--296, {\urlfont hep-th/9711124}.}

\lref\brieskorn{
E.~Brieskorn, ``{B}eispiele zur {D}ifferentialtopologie von
  {S}ingularit\"aten,'' Invent. Math. {\bf 2} (1966) 1--14.}

\lref\bryantsalamon{
R.~L. Bryant and S.~M. Salamon, ``On the construction of some complete metrics
  with exceptional holonomy,'' Duke Math. J. {\bf 58} (1989) 829--850.}

\lref\bryant{
R.~Bryant, ``Metrics with exceptional holonomy,'' Ann. of Math. (2) {\bf 126}
  (1987) 525--576.}

\lref\catp{
P.~S. Aspinwall, B.~R. Greene, and D.~R. Morrison, ``{C}alabi--{Y}au moduli
  space, mirror manifolds and spacetime topology change in string theory,''
  Nucl. Phys. B {\bf 416} (1994) 414--480, {\urlfont hep-th/9309097}.}

\lref\clwang{
C.-L. Wang, ``On the incompletenes of the {W}eil--{P}etersson metric along
  degenerations of {C}alabi--{Y}au manifolds,'' Math. Res. Lett. {\bf 4}
  (1997) 157--171.}

\lref\dWN{
B.~de~Wit and H.~Nicolai, ``Extended supergravity with local {$SO(5)$}
  invariance,'' Nucl. Phys. B {\bf 188} (1981) 98--108.}

\lref\dasguptamukhi{
K.~Dasgupta and S.~Mukhi, ``Orbifolds of {M}-theory,'' Nucl. Phys. B {\bf 465}
  (1996) 399--412, {\urlfont hep-th/9512196}.}

\lref\davis{
M.~W. Davis, ``Some group actions on homotopy spheres of dimension seven and
  fifteen,'' Amer. J. Math. {\bf 104} (1982) 59--90.}

\lref\ddg{
D.~E. Diaconescu, M.~R. Douglas, and J.~Gomis, ``Fractional branes and wrapped
  branes,'' J. High Energy Phys. {\bf 02} (1998) 013, {\urlfont
  hep-th/9712230}.}

\lref\dgm{
M.~R. Douglas, B.~R. Greene, and D.~R. Morrison, ``Orbifold resolution by
  {D}-branes,'' Nucl. Phys. B {\bf 506} (1997) 84--106, {\urlfont
  hep-th/9704151}.}

\lref\dm{
M.~R. Douglas and G.~Moore, ``{D}-branes, quivers, and {ALE} instantons,''
  {\urlfont hep-th/9603167}.}

\lref\douglas{
M.~Douglas, ``Branes within branes,'' {\urlfont hep-th/9512077}.}

\lref\drmnineteen{
S.~Mori, D.~R. Morrison, and I.~Morrison, ``On four-dimensional terminal
  quotient singularities,'' Math. Comp. {\bf 51} (1988) 769--786.}

\lref\drmseven{
D.~R. Morrison and G.~Stevens, ``Terminal quotient singularities in dimensions
  three and four,'' Proc. Amer. Math. Soc. {\bf 90} (1984) 15--20.}

\lref\drmtwelve{
D.~R. Morrison, ``Canonical quotient singularities in dimension three,'' Proc.
  Amer. Math. Soc. {\bf 93} (1985) 393--396.}

\lref\dsw{
M.~Dine, N.~Seiberg, and E.~Witten, ``{F}ayet--{I}liopoulos terms in string
  theory,'' Nucl. Phys. B {\bf 289} (1987) 589--598.}

\lref\fine{
J.~Fine, ``On varieties isomorphic in codimension one to torus embeddings,''
  Duke Math. J. {\bf 58} (1989) 79--88.}

\lref\freund{
P.~G.~O. Freund, ``Higher-dimensional unification,'' Physica D {\bf 15} (1985)
  263--269.}

\lref\gallot{
S.~Gallot, ``{\'E}quations diff\'erentielles caract\'eristiques de la
  sph\`ere,'' Ann. Sci. {\'E}cole Norm. Sup. (4) {\bf 12} (1979) 235--267.}

\lref\gomis{
J.~Gomis, ``Anti de {S}itter geometry and strongly coupled gauge theories,''
  Phys. Lett. B {\bf 435} (1998) 299--302, {\urlfont hep-th/9803119}.}

\lref\gromollmeyer{
D.~Gromoll and W.~Meyer, ``An exotic sphere with nonnegative sectional
  curvature,'' Ann. of Math. (2) {\bf 100} (1974) 401--406.}

\lref\gross{
M.~Gross, ``Deforming {C}alabi--{Y}au threefolds,'' Math. Ann. {\bf 308}
  (1997) 187--220, {\urlfont alg-geom/9506022}.}

\lref\gubserklebanov{
S.~S. Gubser and I.~R. Klebanov, ``Baryons and domain walls in an {${\cal
  N}=1$} superconformal gauge theory,'' Phys. Rev. D {\bf 58} (1998) 125025
  {\urlfont hep-th/9808075}.} 

\lref\gukov{
S.~Gukov, ``Comments on {${\cal N}=2$} {{\it AdS\/}} orbifolds,'' 
Phys. Lett. B. {\bf 439} (1998) 23--26, {\urlfont
  hep-th/9806180}.}

\lref\halyo{
E.~Halyo, ``Supergravity on {$AdS_{5/4}\times$} {H}opf fibrations and conformal
  field theories,'' {\urlfont hep-th/9803193}.}

\lref\hayakawa{
Y.~Hayakawa, ``Degeneration of {C}alabi--{Y}au manifold with
  {W}eil--{P}etersson metric,'' {\urlfont alg-geom/9507016}.}

\lref\hirzebruch{
F.~Hirzebruch, ``Singularities and exotic spheres,'' {Exp. 314}, {S}\'eminaire
  Bourbaki, vol. 1966/1967, W. A. Benjamin, New York.}

\lref\instKthree{
P.~S. Aspinwall and D.~R. Morrison, ``Point-like instantons on {K3}
  orbifolds,'' Nucl. Phys. B {\bf 503} (1997) 533--564, {\urlfont
  hep-th/9705104}.}

\lref\joyce{
D.~D. Joyce, ``Compact {$8$}-manifolds with holonomy {${\rm Spin}(7)$},''
  Invent. Math. {\bf 123} (1996) 507--552.}

\lref\kensix{
K.~Intriligator, ``New string theories in six dimensions via branes at orbifold
  singularities,'' Adv. Theor. Math. Phys. {\bf 1} (1997) 271--282, {\urlfont
  hep-th/9708117}.}

\lref\keo{
A.~Kehagias, ``New type {IIB} vacua and their {F}-theory interpretation,''
  Phys. Lett. B {\bf 435} (1998) 337--342, {\urlfont hep-th/9805131}.}

\lref\kervairemilnor{
M.~Kervaire and J.~Milnor, ``Groups of homotopy spheres, {I},'' Ann. of Math.
  (2) {\bf 77} (1963) 504--537.}

\lref\klemmmayr{
A.~Klemm and P.~Mayr, ``Strong coupling singularities and non-abelian gauge
  symmetries in {$N=2$} string theory,'' Nucl. Phys. B {\bf 469} (1996)
  37--50, {\urlfont hep-th/9601014}.}

\lref\ks{
S.~Kachru and E.~Silverstein, ``{4d} conformal field theory and strings on
  orbifolds,'' Phys. Rev. Lett. {\bf 80} (1998) 4855--4858, {\urlfont
  hep-th/9802183}.}

\lref\kw{
I.~R. Klebanov and E.~Witten, ``Superconformal field theory on threebranes at a
  {C}alabi--{Y}au singularity,'' Nucl. Phys. B {\bf 536} (1998) 199--218, 
  {\urlfont hep-th/9807080}.}

\lref\lnv{
A.~Lawrence, N.~Nekrasov, and C.~Vafa, ``On conformal field theories in four
  dimensions,'' Nucl. Phys. B {\bf 532} (1998) 199--209,
  {\urlfont hep-th/9803015}.}

\lref\ls{
R.~G. Leigh and M.~J. Strassler, ``Exactly marginal operators and duality in
  four dimensional {N=1} supersymmetric gauge theory,'' Nucl. Phys. B {\bf 447}
  (1995) 95--136, {\urlfont hep-th/9503121}.}

\lref\mal{
J.~Maldacena, ``The large {N} limit of superconformal field theories and
  supergravity,'' Adv. Theor. Math. Phys. {\bf 2} (1998) 231--252, {\urlfont
  hep-th/9711200}.}

\lref\milnor{
J.~Milnor, ``On manifolds homeomorphic to the 7-sphere,'' Ann. of Math. (2)
  {\bf 64} (1956) 399--405.}

\lref\mukray{
S.~Mukhopadhyay and K.~Ray, ``Conifolds from {D}-branes,'' Phys. Lett. B {\bf
  423} (1998) 247--254, {\urlfont hep-th/9711131}.}

\lref\nahm{
W.~Nahm, ``Supersymmetries and their representations,'' Nucl. Phys. B {\bf 135}
  (1978) 149--166.}

\lref\nearhorizon{
J.~M. Figueroa-O'Farrill, ``Near-horizon geometries of supersymmetric branes,''
  {\urlfont hep-th/9807149}.}

\lref\nonspherII{
D.~R. Morrison and M.~R. Plesser, ``Non-spherical horizons, {II},'' in
  preparation.}

\lref\ofered{
O.~Aharony and E.~Witten, ``Anti-de {S}itter space and the center of the gauge
  group,'' {\urlfont hep-th/9807205}.}

\lref\ot{
Y.~Oz and J.~Terning, ``Orbifolds of {$AdS_5\times S^5$} and {$4d$} conformal
  field theories,'' Nucl. Phys. B {\bf 532} (1998) 163--180,
  {\urlfont hep-th/9803167}.}

\lref\phases{
E. Witten, ``Phases of $N=2$ theories in two dimensions,'' Nucl. Phys. B
  {\bf 403} (1993) 159--222, {\urlfont hep-th/9301042}.}

\lref\pol{
J.~Polchinski, ``Tensors from {K3} orientifolds,'' Phys. Rev. D {\bf 55}
  (1997) 6423--6428, {\urlfont hep-th/9606165}.}

\lref\psa{
P.~S. Aspinwall, ``Enhanced gauge symmetries and {K3} surfaces,'' Phys. Lett. B
  {\bf 357} (1995) 329--334, {\urlfont hep-th/9507012}.}

\lref\reidypg{
M.~Reid, ``Young person's guide to canonical singularities,'' Algebraic
  Geometry, Bowdoin, 1985, Proc. Sympos. Pure Math., vol. 46, part 1, Amer.
  Math. Soc., Providence, RI, 1987, pp.~345--414.}

\lref\reid{
M.~Reid, ``Canonical 3-folds,'' Journ\'ees de G\'eom\'etrie Alg\'ebriqe
  d'{A}ngers (A.~Beauville, ed.) Sitjhoff \& Noordhoof, 1980, pp.~273--310.}

\lref\romans{
L.~J. Romans, ``New compactifications of chiral {$N{=}2, d{=}10$}
  supergravity,'' Phys. Lett. B {\bf 153} (1985) 392--396.}

\lref\sardoinfirrione{
A.~V. Sardo~Infirri, ``Partial resolutions of orbifold singularities via moduli
  spaces of {HYM}-type bundles,'' {\urlfont alg-geom/9610004}.}

\lref\schlessinger{
M.~Schlessinger, ``Rigidity of quotient singularities,'' Invent. Math. {\bf 14}
  (1971) 17--26.}

\lref\seifive{
N.~Seiberg, ``Five dimensional {SUSY} field theories, non-trivial fixed points
  and string dynamics,'' Phys. Lett. B {\bf 388} (1996) 753--760, {\urlfont
  hep-th/9608111}.}

\lref\sethithreeD{
S.~Sethi, ``A relation between {$N=8$} gauge theories in three dimensions,''
  {\urlfont hep-th/9809162}.}

\lref\sixauthors{
M.~Berkooz, R.~G. Leigh, J.~Polchinski, J.~H. Schwarz, N.~Seiberg, and
  E.~Witten, ``Anomalies, dualities, and topology of {$D=6$} {$N=1$}
  superstring vacua,'' Nucl. Phys. B {\bf 475} (1996) 115--148, {\urlfont
  hep-th/9605184}.}

\lref\squash{
S.~S. Gubser, ``{E}instein manifolds and conformal field theories,'' {\urlfont
  hep-th/9807164}.}

\lref\stromcon{
A.~Strominger, ``Massless black holes and conifolds in string theory,'' Nucl.
  Phys. B {\bf 451} (1995) 96--108, {\urlfont hep-th/9504090}.}

\lref\summing{
D.~R. Morrison and M.~R. Plesser, ``Summing the instantons: Quantum cohomology
  and mirror symmetry in toric varieties,'' Nucl. Phys. B {\bf 440} (1995)
  279--354, {\urlfont hep-th/9412236}.}

\lref\svtwo{
M.~A. Shifman and A.~I. Vainshtein, ``On holomorphic dependence and infrared
  effects in supersymmetric gauge theories,'' Nucl. Phys. B {\bf 359} (1991)
  571--580.}

\lref\sv{
M.~A. Shifman and A.~I. Vainshtein, ``Solution of the anomaly puzzle in {SUSY}
  gauge theories and the {W}ilson operator expansion,'' Nucl. Phys. B {\bf 277}
  (1986) 456--486.}

\lref\tianyau{
G.~Tian and S.-T. Yau, ``{K}\"ahler--{E}instein metrics on complex surfaces
  with {$C_1>0$},'' Comm. Math. Phys. {\bf 112} (1987) 175--203.}

\lref\tian{
G.~Tian, ``On {C}alabi's conjecture for complex surfaces with positive first
  {C}hern class,'' Invent. Math. {\bf 101} (1990) 101--172.}

\lref\witbaryons{
E.~Witten, ``Baryons and branes in anti de {S}itter space,'' J. High Energy
  Phys. {\bf 07} (1998) 006, {\urlfont hep-th/9805112}.}

\lref\witfivebrane{
E.~Witten, ``Five-branes and {$M$}-theory on an orbifold,'' Nucl. Phys. B {\bf
  463} (1996) 383--397, {\urlfont hep-th/9512219}.}

\lref\yamagishi{
K.~Yamagishi, ``Supergravity on seven-dimensional homotopy spheres,'' Phys.
  Lett. B {\bf 134} (1984) 47--50.}

Spacetime-filling branes have emerged as an essential feature of string
and M-theory compactifications in at least three contexts:  (1)
new branches of the heterotic string in six dimensions with ``extra'' tensor
multiplets, which can be represented by a Ho\v{r}ava--Witten-type
compactification
of M-theory on $(S^1/\ZZ_2)\times K3$ but with extra spacetime-filling
M5-branes
representing the extra tensor multiplets
\refs{\DuffMinasianWitten,\GanorHanany,\SeibergWittensix};
(2) $F$-theory models in four dimensions
(which can be regarded as compactifications of the IIB string with D7-branes
included)
which in general require spacetime-filling D3-branes to cancel a tadpole
anomaly \SVW;
and (3) M-theory models in three dimensions, which require spacetime-filling
M2-branes
to cancel a similar tadpole anomaly \SVW.
In each of these cases, the spacetime-filling brane meets the compactifying
space at a single point, and the string or M-theory remains finite near the
brane.\foot{This
last requirement excludes consideration of $F$-theory models in eight
dimensions 
with the D7-brane being spacetime-filling.
}

Remarkably, this short list of branes (M5, D3, and M2) is precisely the
list of branes for which a certain scaling limit is expected to lead to a
``boundary'' conformal field theory in the recent
\AdS/\CFT\ conjectures \refs{\mal,\GKPo,\Whol}.
In fact, the scaling limit can be taken even when the space transverse to the
branes
is curved, as in the compactification scenarios above.  The details of the
metric far
from the location $y_0$ of the brane in the transverse space become irrelevant;
for the purposes of studying the scaling limit, the metric on the compactifying
space can be approximated by
some metric on its tangent space $T_{y_0}$ at $y_0$.  In the scaling limit,
the rescaled supergravity metric approaches a metric of the form
$\ads_{p+2}\times S^{k}$
in which the anti-de~Sitter space has been formed out of
the worldvolume of the brane and the radial direction within  $T_{y_0}$,
and $S^{k}$ is the unit
sphere within $T_{y_0}$.

Maldacena's conjecture proposes that the M or string theory on this space
$\ads_{p+2}\times S^{k}$,
with $N$ units of flux of the supergravity $k$-form field strength through
$S^{k}$,
is dual to a specific conformal field theory on the boundary of $\ads_{p+2}$.
The conjecture applies
to the large $N$ limit when a large number of these branes
have been brought together; in the compactification context, fairly large
values
of $N$ can be obtained by bringing together all available branes in a given
model.

Virtually all points\foot{The exception
is the four-dimensional $F$-theory models, where points located along the
D7-branes behave differently;
in particular, the string coupling becomes infinite at such points.  The
behavior of D3-branes
at such points has recently been determined \refs{\FaySpa,\AFM,\aotfth}, and we
will
not consider
them here.} in the compactifying space have identical behavior in this scaling
limit.
That situation changes, however, if we consider a compactifying space which
itself
has singularities.  Such spaces are necessary additions if one wishes to study
the
complete moduli space of string compactifications, since singularities can be
acquired at special values of moduli (occurring at finite distance in the
moduli space).
We have learned much interesting physics in the past few years
by studying what happens at such singularities.
Each of the three compactification scenarios above allows for singularities to
be
acquired, and once the space is singular, there is no apparent obstruction to
moving the branes to the singular point.

What should we expect to happen if we locate $N$ branes at such a singular
point?
The first step is to understand the geometry of the scaling limit.  Instead of
comparing the rescaled metric to a metric on the tangent space $T_{y_0}$,
when $y_0$ is a singular point we should compare to a metric on the {\it
tangent
cone}, which is topologically the product of $\Bbb{R}^+$ and a
$k$-dimensional
``horizon'' $H^{k}$.
The key difference between smooth and singular points is that the horizon is
no longer a sphere.  We propose that
{\it the standard behavior one should expect for branes at a singularity in
the scaling limit is a model of the form $\ads_{p+2}\times H^{k}$},
with the horizon $H$ consisting of all points at some fixed distance from
$y_0$.
Moreover,
{\it the theory on $\ads_{p+2}\times H^{k}$ should have at least as much
unbroken supersymmetry as the original compactified M or string theory.}

Under these conditions, the generalization of Maldacena's conjecture is
clear:
{\it the $M$ or IIB theory
compactified on $\ads_{p+2}\times H^{k}$ with $N$ units of $k$-flux
through
$H^{k}$ should be dual to a conformal field theory ${\cal C}_N$ (depending on
the singularity),
with a large $N$ `t~Hooft-type limit being a field theory limit on the stringy
side.}
In order for this conjecture to be meaningful, however, we need to find ways of
identifying the families
of conformal field theories ${\cal C}_N$.
Note that the M or string moduli (including the locations of the branes) should
determine the
parameters of the conformal field theory.  Since we have good geometric control
of those moduli at weak coupling, we can expect good geometric interpretations
of the
field theory parameters.  We exploit this idea systematically throughout this
paper,
using geometry to aid in our determination  and investigation of the
corresponding
field theories.

The plan of the paper is as follows.
In section 1, we review the scaling limit in detail and explain how it must be
modified in the present context.  We also discuss some generic features of the
singularities we will study.

In section 2, we find a  geometric characterization of the amount of
supersymmetry
carried by the limiting conformal field theories.  Using existing mathematical
theorems,
we show that the $R$-symmetry group of the conformal field theory acts by
isometries on the corresponding horizon manifold $H$.  This discussion is
closely related to the {\it differential geometry}\/ of $H$.

In section 3, we present a very brief discussion of these theories for
M2-branes
and M5-branes.  In the case of M2-branes, we merely give a few examples; in the
case of M5-branes, we sketch the answer, deferring details to a companion
publication
\nonspherII.

In section 4, we give a detailed analysis of the singularities occurring in
D3-brane
theories, focusing on a few particular classes which exhibit many of the
general
features.  The tool here is {\it algebraic geometry}, and we find a nice
correspondence with some of the results obtained earlier via differential
geometry. 

In section 5, we study a particular singularity---a $\ZZ_2\times \ZZ_2$
orbifold---in
considerable detail.  This theory has been studied before in the literature,
but our
focus is on giving more detail and in particular in extracting geometric
interpretations
for field theory parameters and moduli.  This is then applied in section 6,
where we present
a number of new examples, all derived in some way from the $\ZZ_2\times \ZZ_2$
orbifold.

In section 7 we discuss our conclusions.

\bigskip

Extensions of the \AdS/\CFT\ correspondence to supersymmetric
theories with less than maximal supersymmetry have been
extensively studied in the past few months, either by orbifold constructions
\refs{\ks\berkooz\lnv\bkv\FKPZ\gomis\ot\halyo\bj\ahnone\ahntwo{--}\gukov},
by reduction on Hopf fibrations \halyo,
by ``brane box'' constructions \refs{\HSU,\ami},
or by compactification on homogeneous spaces \GmodH\ or other 
Einstein manifolds \DLP.  In fact the general idea of
interpreting branes with less than maximal supersymmetry in terms of
 compactifications on spaces of the form $\ads\times H$ for interesting
manifolds
$H$ goes back at least as far as \DLPS.

Particularly important to the development of our approach were early versions
of it
which appeared in \keo\ and \AFM.  While this work was in progress, we became
aware of
several overlapping projects:  Klebanov and Witten \kw\ independently studied
the field theories for D3-branes
at a conifold singularity---one of the examples we discuss in section 6---and
Gubser \squash\ and
Gubser and Klebanov \gubserklebanov\ studied some
further aspects of those theories, overlapping our results to a
certain extent.  In addition, Acharya, Figueroa-O'Farrill,
Hull, and Spence
\refs{\nearhorizon,\AFHS}
have independently considered the relationship between supersymmetry and the
geometry of the horizon
manifold; there is substantial overlap between their work and section 2 of this
paper.

A preliminary version of the present work was discussed by the first author in a
talk at
the Strings '98 conference in Santa Barbara.

\newsec{The scaling limit}

 We've learned, in studying M and string theory compactifications,
of several interesting examples in which part of the background necessarily
involves some spacetime-filling branes.  We will focus in this
paper on three of these, involving M2-branes, D3-branes, and M5-branes.

The first example is M-theory models in $2+1$ dimensions obtained
by compactifying on
some
eight-manifold with holonomy group contained in $\Spin(7)$.
As shown by Sethi, Vafa
and Witten, there is a tadpole cancellation requirement of an insertion of
$\chi/24$ M2-branes, where $\chi$ is the Euler characteristic of the
eight-manifold \SVW.

Second, consider an $F$-theory model in $3+1$ dimensions, constructed out
of an elliptically fibered Calabi--Yau fourfold.  We will regard this
 as a IIB string theory compactification on the
base of the elliptic fibration (which is a six-manifold), with D7-branes
wrapped
around the four-manifolds along which the fibers in the elliptic fibration have
degenerated.
A similar tadpole cancellation requirement insists that we include
$\chi/24$
spacetime-filling D3-branes in the background \SVW, where $\chi$ is the Euler
characteristic of the Calabi--Yau fourfold.

A third example of this nature is found in the Ho\v{r}ava--Witten
description \HoravaWitten\ of the $E_8\times E_8$ string compactified on K3,
thought of now as
M-theory compactified on $(S^1/\ZZ_2)\times K3$ with some $E_8\times E_8$ gauge
fields on the Ho\v{r}ava--Witten $9$-branes (HW9-branes).  In models where
some M5-branes
associated to small instantons are allowed to move off of the HW9-branes
(becoming spacetime-filling M5-branes),
 interesting new branches of the theory arise
\refs{\DuffMinasianWitten\GanorHanany{--}\SeibergWittensix}.  These new
branches can
be studied from a dual
perspective using $F$-theory, which is one of the reasons that we know that
they exist \refs{\SeibergWittensix,\WitMF,\MVtwo}.
There is a limit of 24 M5-branes which can play a role in these
theories, since 24 is the maximum number of small instantons which can
occur here.

The limitation on the total number of branes in these compactification
scenarios is
an artifact of global features of the compactification space which are
irrelevant in the appropriate scaling limit.  In fact, there is no local reason
to limit the number of branes, and one can study local versions of these
compactification
models with arbitrarily large numbers of spacetime-filling branes.  (This is
analogous to using ALE spaces in place of K3 surfaces to study certain local
aspects of compactification, and to various other similar ``geometric
engineering''
tricks.)  We will do so henceforth whenever it is convenient for our analysis.

When we allow the compactifying space to acquire singularities, the requirement
that the limiting
value of moduli be at finite distance is extremely restrictive on the
singularity
types which can occur.  When the compactifying space can be described by means
of
algebraic geometry (which further restricts the holonomy in $2+1$ dimensional
models
to being contained within $SU(4)$), the only singularities which can appear
\refs{\hayakawa,\clwang} are
those known as {\it Gorenstein canonical singularities}.
 Briefly put, these are singularities in complex dimension
 $n$ for which there exists a non-vanishing holomorphic $n$-form near the
singularity which
 on any smooth blowup of the space extends to a holomorphic form.
 Although not all of these singularities can be embedded in
 Calabi--Yau manifolds, we should still expect to be able to use any space with
 such a singularity to ``compactify'' our M or string theory models.

Returning to the case of smooth compactifying manifolds, the
 low-energy theory in all three of our compactification scenarios has a
supergravity description,
in which the spacetime-filling branes are included by means of a ``warped
product''
metric.  Consider a flat $p$-brane with worldvolume $M^{p+1}$
($(p{+}1)$-dimensional
Minkowski space) in a spacetime of dimension $D$, in which the
transverse
space $I^{k+1}$ need not be flat.
The values of $(p,D)$ of interest here are $(2,11)$, $(3,10)$, and $(5,11)$,
and the dimension of the internal space
$k+1=D-p-1$ takes the values $8$, $6$ and $5$. The most general form
of a metric which preserves $(p{+}1)$-dimensional Poincar\'e invariance is
\eqn\warpmet{
ds^2 = \phi(y) \, dx_M^2 + \widetilde{dy}_I^2,
}
where $\phi(y)$ is a positive function on $I$ called the
warping factor.  We use the notation $M\timesw I$ to denote
such a warped product.
We treat $x$ and $y$ as dimensionful coordinates on $M$ and $I$,
but note that the warping factor $\phi(y)$ is dimensionless and so $y$ must be
measured with respect to a specific length scale in order to define $\phi$.  We
will use the $D$-dimensional Planck length $\ell=\ell_D$
so that $\phi$ is actually a function of $y/\ell$.

To get a solution to the supergravity equations of motion, we follow
\refs{\BBMeight,\keo,\AFM}\foot{The metric in \BBMeight\ is written in terms of
$\Delta(y)=f(y)^{2\over3}$.}
and use a special form
of eq.~\warpmet\ in which the warping factor has been written as a power of
another function,
and the metric on $I$ has been rewritten with a conformal factor which is
 a different power of the warping factor:
\eqn\warpmetsp{
ds^2 = f(y)^{{(3-k)/(k-1)}} \, dx_M^2 + f(y)^{2/(k-1)} \, dy_I^2 .
}
This metric must be
accompanied by an expectation value for the supergravity $k$-form field
strength
\eqn\kform{
\left[d( f(y)^{-1}) \wedge \dvol_M + \star\left( d( f(y)^{-1})\wedge
\dvol_M\right)\right]_{k}
}
where $[\cdots]_{k}$ denotes the projection onto the $k$-form piece.
The equation of motion for
$f(y)$  demands that it be a harmonic function on $I$ away from the location of
the branes.
(In the case of the M2-brane,
there is a more general form of this solution in which the $k$-form field
strength
also has a component in the internal space, and the equation of motion for the
warping
factor is suitably modified \BBMeight.)

The original supergravity solutions for the D3-brane \HorowitzStrominger,
the M2-brane \DuffStelle, and the M5-brane \Guven\ took exactly this form
\warpmetsp, using
the specific function
\eqn\warpfac{
f(y)=1+{c_k\,g^{11-D}N\over (r/\ell)^{k-1}}
}
which depends only on the distance $r$ from the brane; there must also be a
source term in the
 equation of motion for the function $f(y)$, which satisfies:
\eqn\feom{
\Delta f = \tilde{c}_k\,g^{11-D} N\, \delta(y-y_0) .
}
In these formulas, $N$ denotes the number of branes which are located at $y_0$,
the
(dimensionful) constant $\tilde{c}_k$
is determined by flux quantization rules for the particular type of branes
being studied, and
 $-\tilde{c_k}/(k-1)c_k$ is the volume of a unit sphere of dimension $k$.
 The string coupling $g$ only enters when $D=10$.

Incorporating branes into more general solutions of the form \warpmetsp\ is
then straightforward:
if we have $N_j$ branes located at $y_j\in I$ then
the warping factor should satisfy
\eqn\feomgen{
\Delta f = \tilde{c}_k\,g^{11-D}\sum_j N_j\, \delta(y-y_j) .
}
When we include branes into more general compactifications, we need not demand
that
the metric be asymptotically flat, so our solutions to eq.~\feomgen\ will be
more general
than eq.~\warpfac.

In order for a supergravity solution on the warped product
$M^{p+1}\timesw I$ to retain some supersymmetry
in the effective $(p{+}1)$-dimensional theory, there must be some appropriate
spinors
defined on $M\timesw I$.   The component of
such a spinor lying in the internal space $I$ must be covariantly constant
with respect to the metric $dy_I^2$. (This was the primary reason for including
an
appropriate power of the warping factor multiplying $dy_I^2$ in
eqn.~\warpmetsp.)
If there is an ${\cal N}$-dimensional space of covariantly constant spinors,
then the resulting theory has ${\cal N}$ supersymmetries.
In particular, whenever ${\cal N}>0$, the metric $dy_I^2$ must be Ricci-flat,
and must therefore have a holonomy group
which is a {\it proper}\/ subgroup of $SO(k+2)$.

Consider $N$ branes at $y_0$, measuring $r$ from $y_0$.
We locally rewrite the metric on $I$
in the form
\eqn\Imetric{
dy_I^2=r^2\left({dr^2\over r^2}+d\Omega_H^2 + O({r\over\ell})\right),
}
where $d\Omega_H^2$ is a dimensionless metric on the ``horizon'' manifold
$H^{k}$---the
unit sphere.
Since $f$ satisfies eqn.~\feom, it has precisely the same local singularity as
eqn.~\warpfac, i.e., there is some regular function $f_0(y)$ with
\eqn\formoff{
f(y)={c_k\,g^{11-D}N\over (r/\ell)^{k-1}}+f_0(y)
={c_k\,g^{11-D}N\over (r/\ell)^{k-1}}\left(1+O({r\over\ell})\right).
}
It follows that the conformal factor in the metric on $I$ can be written in
the form
\eqn\conffactor{
f(y)^{2/(k-1)}={\ell^2\over r^2}\, (c_k\,g^{11-D}N)^{2/(k-1)}
\left(1+O({r\over\ell})\right).
}

We now, following \mal,
introduce a new variable $u$ satisfying
\eqn\defofu{
{r\over\ell}=(u\ell)^{2/ (k-3)},
}
and consider a scaling limit with $u$ held fixed and $\ell\to0$.
We are interested in the behavior of the dimensionless two-tensor $ds^2/\ell^2$
(the metric in Planck units) in this scaling limit.  We write
\eqn\dsoverl{
{ds^2\over\ell^2} = {1\over\ell^2}\,f(y)^{{(3-k)/(k-1)}} \, dx_M^2
+ {r^2\over\ell^2}\,f(y)^{2/(k-1)} \, \left({dr^2\over r^2}+d\Omega_H^2 +
O({r\over\ell})\right)
}
and then change variables (using also eqns.~\formoff\ and \conffactor) to get
\eqn\dsoverlu{
{ds^2\over\ell^2} = {(c_k\,g^{11-D}N)}^{{(3-k)/(k-1)}} \, u^2\, dx_M^2
+ (c_k\,g^{11-D}N)^{2/(k-1)} \, \left(({2\over k{-}3})^2\,{du^2\over
u^2}+d\Omega_H^2\right)
+O(u\ell)
}
so that in the limit, $ds^2/\ell^2$ approaches
\eqn\dsoverllim{
 {(c_k\,g^{11-D}N)}^{{(3-k)/(k-1)}} \, u^2\, dx_M^2
+ {4\,(c_k\,g^{11-D}N)^{2/(k-1)}\over (k{-}3)^2}\,{du^2\over u^2}+
(c_k\,g^{11-D}N)^{2/(k-1)} \, \, d\Omega_H^2 .
}

This limiting metric \dsoverllim\ has two interpretations:  on the one hand, it
is a warped product of Minkowski
space and the cone $C(H)$ over the horizon $H$, where the cone metric is
defined to be
\eqn\conemetric{
dy_{C(H)}^2 = du^2 + u^2 \, d\Omega_H^2
}
and the warped product structure is given by
\eqn\warpcone{
u^2 \, dx_M^2 + u^{-2} dy_{C(H)}^2
}
(suppressing constants for clarity).
On the other hand, \dsoverllim\ can be interpreted as a product metric on
$\ads_{p+2}\times H^{k}$.
The cone $C(H)$ with its limiting metric should be considered as a kind of
linearization of
the original
Ricci-flat
metric near $y_0$, the location of the brane.

We now consider what happens if the original internal space $I$ had
singularities, and we are
attempting to place $N$ branes at a singular point $y_0$.  As discussed in the
introduction,
we propose that the standard behavior one should expect in this situation is
that there is
a horizon $H^{k}$ (points at ``unit distance'' from $y_0$) whose cone $C(H)$
provides a good linearization
of the limiting Ricci-flat metric near $y_0$.  More precisely, we will assume
that $H$ has
a (dimensionless) ``metric''\foot{As noted below, in certain circumstances $H$
has
singularities and we will need a generalized form of metric.} $d\Omega_H^2$
such that near $y_0$,
\eqn\ansatz{
dy_I^2=r^2\left({dr^2\over r^2}+d\Omega_H^2 + O({r\over\ell})\right)
}
for some radial coordinate $r$ centered at $y_0$.

When \ansatz\ is satisfied, the analysis above can be repeated verbatim:
nowhere did we assume
that $H$ is a sphere.  The conclusion is that the scaling limit of
$ds^2/\ell^2$ can
be interpreted as either a warped product $M^{p+1}\timesw C(H)$, or as a direct
product
$\ads_{p+2}\times H$.  In particular, for any Gorenstein canonical singularity for which
\ansatz\ holds, there must exist an Einstein metric on $H$ whose corresponding
cone $C(H)$
serves as the leading approximation to limiting Ricci-flat metrics from nearby
smooth
compactifications.

We can thus formulate a general program as follows.  For
each type of Gorenstein canonical
 singularity satisfying \ansatz, we should first carefully study the
horizon $H$, 
determining both its topology and the set of 
Einstein metrics on it whose cones could arise as
limits from Ricci-flat metrics.  For each such metric, we expect that
 $M$ or IIB theory compactified
on $\ads_{p+2}\times H$ with $N$ units of flux through $H$
will determine a conformal field theory on the 
boundary of $\ads_{p+2}$.  Ultimately, we would like to find  more direct
descriptions 
of these conformal field theories.

There is an important subtlety which arises when the singularity at $y_0$ is
not isolated:
the horizon will intersect the singular locus and will not itself be smooth.
Our program should be modified in such cases to allow for singularities
on $H$, and to search for Einstein ``metrics'' with a specified type
of singular behavior.
In the case of M5-brane and D3-brane theories, the singularity of the horizon
is an
ADE singularity (in real codimension four), and it is natural to expect an
orbifold-type
``metric'' at the singularity.  For M2-branes, the horizon can have a
Gorenstein canonical threefold
singularity in real codimension six and it is not clear what kind of singular
behavior
of the metric to expect there.

An important aspect of our general program is determining how much unbroken
supersymmetry each
of these models have; it is to this issue that we turn next.

\newsec{Supersymmetry and holonomy}

In the scaling limit, we found a compactification on a space of the form
$\ads\times H$.  To have a supergravity solution, we need an Einstein metric on
$H$.  Moreover, if the effective theory on $\ads$ is to retain some
supersymmetry,
we need appropriate spinors defined on $\ads\times H$.  The condition on the
$H$ component of the spinor is that it should be a so-called {\it Killing
spinor},
so the dimension of the space of Killing spinors counts the number of
supersymmetries
of the associated theory.

Metrics for which Killing spinors exist have been extensively studied in the
mathematics literature.  The key result for our purposes is a theorem of
B\"ar \baer, which establishes a natural isomorphism between the space of
Killing spinors on $H$ and the space of covariantly constant spinors on the
cone $C(H)$.  This is exactly what should be expected if our basic scenario
is valid:  the limit of the Ricci-flat metric
on the original internal space $I$ should approach a Ricci-flat metric on
the cone $C(H)$ with at least as many covariantly constant spinors as the
original
metric on $I$.  These then translate directly into Killing spinors on $H$,
as we should expect when comparing the amount of unbroken supersymmetry in
the two interpretations $M\timesw C(H)$ and $\ads\times H$ of our limiting
compactification.

The great utility of B\"ar's theorem is that it can be combined with
 Berger's classification of holonomy groups \refs{\berger,\bryant} to yield a
 classification of what supersymmetries are possible and what geometric
structures
are associated to them.  We have collected the results of this classification
into
two tables, which we now explain.

\topinsert
$$\vbox{\offinterlineskip\halign{
\strut # height 15pt depth 8pt& \quad$#$\quad\hfill\vrule&
\quad$#$\quad\hfill\vrule&
\quad$#$\quad\hfill\vrule\cr
\noalign{\hrule}
\vrule &{\cal N}& \hbox{Superconformal algebra}&\hbox{Maximal holonomy
representation}\cr
\noalign{\hrule}
\vrule &(2,0)&(\go(6,2) \oplus \gsp(2),\ ({\bf 8},{\bf 4}))&\hbox{trivial
$5$-dim'l rep.}     \cr
\vrule &(1,0)&(\go(6,2) \oplus \gsp(1),\ ({\bf 8},{\bf 2}))&(SU(2),\ {\bf
2}\oplus{\bf 2}\oplus {\bf 1}) \cr
\noalign{\hrule}
\vrule &4&(\go(4,2) \oplus \gsu(4),\ ({\bf 4},{\bf 4})
\oplus({\bf\bar{4}},{\bf\bar{4}}))&\hbox{trivial $6$-dim'l rep.}\cr
\vrule &2&(\go(4,2) \oplus \gu(2) ,\ ({\bf 4},{\bf 2})
\oplus({\bf\bar{4}},{\bf\bar{2}}))&(SU(2),\ 2\times ({\bf 2}\oplus{\bf 1}))
\cr
\vrule &1&(\go(4,2) \oplus \gu(1) ,\ ({\bf 4},{\bf
1}_1)\oplus({\bf\bar{4}},{\bf 1}_{-1}))&(SU(3),\ {\bf 3}\oplus {\bf\bar{3}})
  \cr
\noalign{\hrule}
\vrule &8&(\go(3,2) \oplus \gso(8),\ ({\bf 4},{\bf 8}))&(\Bbb{Z}_2,\ \{\pm
\boldone_8\})       \cr
\vrule &6&(\go(3,2) \oplus \gso(6),\ ({\bf 4},{\bf 6}))&(U(1) ,\ 4\times({\bf
1}_1 \oplus  {\bf 1}_{-1}))  \cr
\vrule &5&(\go(3,2) \oplus \gso(5),\ ({\bf 4},{\bf 5}))&(SU(2),\ 4\times{\bf
2}) \cr
\vrule &4&(\go(3,2) \oplus \gso(4),\ ({\bf 4},{\bf 4}))&(SU(2){\times}SU(2),\
2\times (({\bf 2},{\bf 1})\oplus({\bf 1},{\bf 2}))) \cr
\vrule &3&(\go(3,2) \oplus \gso(3),\ ({\bf 4},{\bf 3}))&(Sp(2),\ {\bf
4}\oplus{\bf 4})                     \cr
\vrule &2&(\go(3,2) \oplus \gso(2),\ ({\bf 4},{\bf 2}))&(SU(4),\ {\bf 4}\oplus
{\bf\bar{4}})               \cr
\vrule &1&(\go(3,2)             ,\  {\bf 4}         )&(\Spin(7),\ {\bf 8})
                         \cr
\noalign{\hrule}
}}$$
\bigskip
\centerline{Table 1. Superconformal algebras and holonomy representations.}
\bigskip
\endinsert

For theories on $\ads_{p+2}$, the superconformal algebras must be particular
ones
from Nahm's list \nahm, with a bosonic part of the form
$\go(p{+}1,2)\oplus\frak{g}$
(where $\frak{g}$ is the $R$-symmetry algebra), and a fermionic part which is a
representation of the bosonic algebra.  The algebras which occur in our
theories
are labeled by the number of supersymmetries, and are
listed in the middle column of Table 1 in the form
\centerline{(bosonic algebra, fermionic representation).}

The holonomy group on the cone $C(H)$ is a subgroup $\Hol_{C(H)}\subseteq
SO(D-p-1)$.
The covariantly constant spinors are then determined as fixed elements of
$\Hol_{C(H)}$ in the
 spin representation of $SO(D-p-1)$.  Now for
each fixed number ${\cal N}$ of spinors there is a group ${\cal H}_{\cal
N}\subseteq SO(D-p-1)$
which fixes precisely that number of spinors.  If the holonomy group of $C(H)$
is known, the
amount of supersymmetry can then be determined by the condition
$\Hol_{C(H)}\subseteq {\cal H}_{\cal N}$, $\Hol_{C(H)}\not\subseteq {\cal
H}_{{\cal N}+1}$
(up to conjugacy).

The third column of Table 1 lists these possible ``maximal holonomy''
groups ${\cal H}_{{\cal N}}$, together with
their $(D{-}p{-}1)$-dimensional representation.\foot{In the third section
of the table,
the maximal holonomy groups appearing after the first line are
$\Spin(2)$, $\Spin(3)$, \dots, $\Spin(7)$ but we have given most of these a
more familiar name.}
Note that the number of spinors which is
fixed can never be just one less than the dimension of the spinor
representation (since
fixing all but one of the spinors forces the last one to be fixed as well), but
that there is otherwise no restriction on the ${\cal N}$'s which occur.

Not all of the ``maximal holonomy representations'' can actually occur as
holonomy
groups; however, as we will see shortly, for each ``maximal holonomy'' group
${\cal H}_{\cal N}$
in the table there is some holonomy group $C(H)$ which lies in ${\cal H}_{\cal
N}$
but not in ${\cal H}_{{\cal N}+1}$.

\topinsert
$$\vbox{\offinterlineskip\halign{
\strut # height 15pt depth 8pt& \quad$#$\quad\hfill\vrule&
\quad # \quad\hfill\vrule&\quad # \quad\hfill\vrule\cr
\noalign{\hrule}
\vrule & \hbox{Holonomy representation}      &\hbox{Cone geometry}
&\hbox{Horizon geometry}            \cr
\noalign{\hrule}
\vrule &\hbox{trivial $5$-dim'l rep.}        &$\Bbb{R}^5$             &$S^4$
                          \cr
\noalign{\hrule}
\vrule &\hbox{trivial $6$-dim'l rep.}        &$\Bbb{R}^6$             &$S^5$
                          \cr
\vrule &(SU(3),\ {\bf 3}\oplus{\bf\bar{3}})  &Calabi--Yau threefold
&Einstein--Sasaki $5$-manifold      \cr
\noalign{\hrule}
\vrule &\hbox{trivial $8$-dim'l rep.}        &$\Bbb{R}^8$             &$S^7$
                          \cr
\vrule &(Sp(2),\ {\bf 4}\oplus {\bf 4})      &hyper-K\"ahler fourfold
&$3$-Sasaki $7$-manifold            \cr
\vrule &(SU(4),\ {\bf 4}\oplus {\bf\bar{4}}) &Calabi--Yau fourfold
&Einstein--Sasaki $7$-manifold      \cr
\vrule &(\Spin(7),\ {\bf 8})                  &$\Spin(7)$-manifold
&nearly-parallel $G_2$-manifold \cr
\noalign{\hrule}
}}$$
\bigskip
\centerline{Table 2. Geometry of cones over smooth, simply-connected horizons.}
\bigskip
\endinsert

If the horizon manifold $H$ is smooth and simply-connected and the cone $C(H)$
is not flat, then the holonomy of the
metric on $C(H)$ must
act irreducibly in the sense that 
$\Hol_{C(H)}$ cannot be written as 
$G'\times G''$ in such a way that the
corresponding representation is a direct sum of a representation of $G'$
with a representation of $G''$ \gallot.
This puts quite severe restrictions on the
possible holonomies which can occur for such cones $C(H)$, limiting them to
those in the first column of Table 2.
The associated geometry for each of these holonomy groups is
 fairly well known; we list it in the second column
of Table 2.  Less familiar is the fact that the horizon manifolds also have a
special geometric
structure, determined by the existence of Killing spinors.
These special geometric structures on the horizons are indicated in the final
column of Table 2.

We will give the precise definitions for these geometric structures on the
horizon manifolds
in an appendix; here, we simply note
that each geometric structure is a direct consequence of the existence of
a certain number of Killing spinors.  There are some quite remarkable
mathematical theorems
which have been proven about these geometric structures
\refs{\FKone,\FKtwo,\baer}.

First, every $3$-Sasaki manifold which is not a sphere
has three distinguished vector fields generating an infinitesimal
action of $\gsu(2)$ (as part of the definition).  The theorem is that this can
be integrated to
 an $SU(2)$ group
action on the manifold, by isometries.  Thus, in the case of a $3$-Sasaki
seven-manifold, the $R$-symmetry group
$SU(2)$ or $SO(3)$ acts by isometries.  (The center of $SU(2)$ will act
faithfully in some cases, and
trivially in other cases.)

Second, every Einstein--Sasaki manifold which is not $3$-Sasaki has a
distinguished vector field
(again part of
the definition).  The theorem is that  this can be integrated to a $U(1)$
action on the manifold.
Thus, for an Einstein--Sasaki five-manifold or seven-manifold, the
$R$-symmetry group $U(1)$ or
$SO(2)$ acts by isometries.  Furthermore, for Einstein--Sasaki five-manifolds
which are not $3$-Sasaki, the Einstein--Sasaki structure is unique; hence,
so is the corresponding $U(1)$ action.

In the $3$-Sasaki case, when the $SU(2)$ action is ``regular'', then the
quotient by the
$SU(2)$ action is a quaternionic K\"ahler manifold.
The metric on the $3$-Sasaki manifold
can be recovered by using an appropriately scaled Hermitian metric on a
quaternionic
line bundle over the
K\"ahler--Einstein manifold, restricted to the unit three-sphere bundle
within that 
line bundle
\refs{\BGM}. For simply-connected
$3$-Sasaki seven-manifolds with regular $SU(2)$ action,
the conclusion is that the manifold must either be $S^7$ or the space
$SU(3)/S^1_{1,1}=N(1,1)$
described in \DNKK.

Similarly, in the Einstein-Sasaki case, when the $U(1)$ action is regular, the
quotient by
the $U(1)$ action is a K\"ahler--Einstein manifold of positive curvature,
and the Einstein--Sasaki manifold can be
recovered as the unit circle bundle in a certain Hermitian line bundle over the
K\"ahler--Einstein manifold.
For Einstein--Sasaki five-manifolds, this yields a complete classification
for the 
case of regular $U(1)$-actions, since all positively curved K\"ahler--Einstein
four-manifolds
are known \refs{\tianyau,\tian}.
The possible cases are as follows \FKone\ (see also \refs{\DLP,\keo}).

\item{1)} $H/U(1)\cong \Bbb{C}\Bbb{P}^2$ and either $H\cong S^5$ or $H\cong
S^5/\Bbb{Z}_3$.
These metrics are unique up to rescaling---in each case, they are induced by
 the ``round'', or constant curvature, metric
on $S^5$.

\item{2)} $H/U(1)\cong \Bbb{C}\Bbb{P}^1\times \Bbb{C}\Bbb{P}^1$ and either
$H\cong V_{4,2}$ or $H\cong V_{4,2}/\Bbb{Z}_2$, where $V_{4,2}\cong
SO(4)/SO(2)$ is known as
the ``Stiefel manifold of framed two-planes in four-space''.  (These
horizons and the associated supergravity solutions have also been studied
under the names $T^{11}$ and $T^{112}=T^{11}/\BZ_2$ \romans.)  The metric in
each case is again unique up
to scaling, due to a theorem of B\'erard Bergery \berardbergery.  Writing
$SO(4)\cong S^3\times S^3$,
the metric on $H$ is induced from the product of round metrics on the
three-spheres of equal volume.

\item {3)} $H/U(1)$ is a del Pezzo surface, the blowup of $\Bbb{C}\Bbb{P}^2$ at
$k$ points
with $3\le k\le 8$, and $H$ is a simply-connected $S^1$-bundle over this del
Pezzo surface.
Note that in this case, K\"ahler--Einstein metrics are known to exist on
$H/U(1)$ \refs{\tianyau,\tian},
but they are not known explicitly.

\noindent
In \FKone\ it is also pointed out that for free group actions of $\Gamma\subset
SU(3)$ on
$S^5$ other than $\Gamma=\Bbb{Z}_3$, the $U(1)$ action is {\it not}\/ regular.
Thus, there are many other interesting possibilities for Einstein--Sasaki
five-manifolds.

\subsec{Detailed classification of holonomy groups}

We can obtain a more detailed classification of holonomy groups for M5-branes
by considering
the connected component $\Hol^0$ of the holonomy
group of the cone $C(H)$.  The possible choices for this connected component
again follow
from Berger's classification \berger: it must be $SU(2)$ or trivial.

Suppose there is an M5-brane with $\Hol^0(C(H))=SU(2)$. After taking a finite
unramified
cover $\widetilde{H}\to H$ we get a cone $C(\widetilde{H})$ whose holonomy is
reducible, with $C(\widetilde{H})$ locally the product of $\Bbb{R}$ (with a
flat metric)
and a four-dimensional space whose holonomy is $SU(2)$.  By \gallot, $H$
cannot be smooth; on
the other hand, the Gorenstein canonical singularities which can occur on
$SU(2)$ spaces are only
the ADE singularities, for which the holonomy is finite.  The conclusion is
that the
holonomy group of $C(H)$ must be finite for M5-branes, with the horizon taking
the form
 $H=S^4/\Gamma$ for some finite group
$\Gamma\subset SU(2)$.  Notice that $\Gamma$ always fixes two points, so that
if the holonomy
is nontrivial (that is, if we get an ${\cal N}=(1,0)$ theory rather than an
${\cal N}=(2,0)$
theory) then the horizon is singular.

Similarly, in the case of D3-branes, the possible choices for $\Hol^0(C(H))$
are $SU(3)$,
$SU(2)$ and trivial \berger, and we find several possibilities for the holonomy
(assuming that there
is {\it some}\/ supersymmetry).

\item{1.}  If $\Hol^0(C(H))=SU(3)$ then $\Hol(C(H))=SU(3)$ and we get ${\cal
N}=1$
supersymmetry.  We will see several examples of
this case later in the paper.  The horizon is an Einstein--Sasaki
five-manifold with a $U(1)$ action which geometrically realizes the $\CN=1$
$R$-symmetry group by isometries of $H$.

\item{2.} The case $\Hol^0(C(H))=SU(2)$ is unlikely to occur.  As in the case
of M5-branes, an
unramified cover of $H$ must have reducible holonomy, and so by \gallot\ $H$
cannot be smooth.
The singularities on $H$ all derive from the complex codimension two
singularities on $C(H)$,
which must again be ADE singularities.  However, since we now have a {\it
family}\/ of ADE singularities
we cannot immediately conclude that the holonomy is broken to a finite group.
(It
seems likely that it will be.\foot{This conclusion is based on our analysis of
the ``suspended
pinch point'' below.  That singularity is associated to a family of ADE
singularities, but
its holonomy does not appear to break to a finite group.  However, due to
complications
within the family of ADE singularities we in fact get ${\cal N}=1$
supersymmetry
for this example and we suspect that the holonomy is all of $SU(3)$ in this
case.
Unfortunately, the metric is not known explicitly.})

\item{3.}  If $\Hol^0(C(H))$ is trivial, then $\Hol(C(H))=\Gamma$ is a finite
group, with
$H=S^5/\Gamma$.  If $\Gamma$ is not trivial but $\Gamma\subset SU(2)$, then $H$
has singularities
along a circle
and we get ${\cal N}=2$ supersymmetry.  (This is consistent with the discussion
in case 2.) In this case, the horizon has a natural $SU(2)\times U(1)$
isometry group, again geometrically realizing the $R$-symmetries.
\hfill\break 
If $\Gamma\subset SU(3)$, $\Gamma\not\subset SU(2)$,
then we get ${\cal N}=1$ supersymmetry (and again have a $U(1)$ action
which realizes the $R$-symmetry group).  Both isolated and non-isolated 
singularities can
occur in this latter case.  (For a classification of non-isolated singularities
with $\Gamma$
a cyclic group, see \drmtwelve.)

An analysis similar to this could also be carried out for M2-branes, but it is
quite involved
so we shall defer it to some future occasion on which a more detailed studied
of M2-branes
is made.

\newsec{M2-branes and M5-branes}

To study M2-branes at singularities following our program, we would need some
kind of
classification of the singularities which can occur.  In the case of general
${\cal N}=1$ theories in which the holonomy group of the cone is $\Spin(7)$,
there
do not appear to be any techniques for making such a classification---the
geometry
of degenerate $\Spin(7)$ manifolds is an almost completely unexplored
subject.\foot{There are
some examples known, however---the first constructions of complete
\bryantsalamon\
and of compact \joyce\ $\Spin(7)$-manifolds are both related to degenerations
of
the sort considered here.} Of course,
one could attack this from the other point of view, and attempt to classify all
Einstein seven-manifolds with a ``nice'' three-form, but this would also
appear to be
quite difficult.  Many examples of such manifolds are known (see \DNKK\ for a
list of
known examples circa 1985\foot{Additional examples have recently been
discussed in \AFHS.}).  A third approach---equally difficult---would be to try
to
classify all families
of ${\cal N}=1$ conformal theories in three dimensions with a large $N$ limit.

Even for theories with ${\cal N}=2$ supersymmetry, the singularities involved
are Gorenstein
canonical singularities of
Calabi--Yau fourfolds, and their complete classification seems an extremely
challenging
problem.\foot{See \drmnineteen\ for a hint of some of the difficulties.}
We hope to return to a more detailed study of some classes of these
singularities
in a future work; for the present, we will focus on a few examples.

\subsec{M2-branes at orbifolds}

\newcount\ftref

First, let us note that a variety of supersymmetries can be obtained just from
orbifold
constructions---horizons in $\Bbb{C}^4/\Gamma$---even assuming that the action
of
$\Gamma$ on $S^7$ is free.
When the group $\Gamma$ is cyclic, there is a  characterization from algebraic
geometry---the singularity
is said to be ``terminal''---which singles out the freely acting
$\Gamma$'s which are contained in $SU(2)\times SU(2)$
\drmseven.  In fact, any such group is generated by
$\diag(e^{2\pi i/k},e^{-2\pi i/k},e^{2\pi ia/k},e^{-2\pi ia/k})$
for some relatively prime integers $a$ and $k$.
If $a=1$, $k=2$, we get ${\cal N}=8$ supersymmetry with horizon manifold
$\Bbb{R}\Bbb{P}^7$ \refs{\AOY,\sethithreeD}.
If $a=\pm1$, $k>2$ then $\Gamma$ lies in $U(1)$ and we get ${\cal N}=6$
supersymmetry \halyo.\foot{The standard lore in the supergravity literature
which excludes ${\cal N}=5, 6$
supersymmetry from $\ads_4\times X^7$ compactifications applies only when $X^7$
is simply connected
\refs{\CRW}.}
\ftref=\ftno
If $a\ne\pm1$, the resulting theory has ${\cal N}=4$ supersymmetry.
  All of these singularities have the interesting
property of having neither (complex structure)
deformations \refs{\schlessinger}, nor (Calabi--Yau) resolutions.\foot{Any
Calabi--Yau resolution must have a ``small'' exceptional set and so by \fine\
must be
toric; but it's
easy to show that these singularities do not have toric Calabi--Yau
resolutions.}

As a slight variant of this example, if we take $\Gamma$ to be a binary
dihedral group,
so that $\Bbb{C}^2/\Gamma$
is a $D_k$ singularity, and diagonally embed $\Gamma$ into
$SU(2)\times SU(2)$, then the corresponding orbifold theory has
${\cal N}=5$ supersymmetry.$^{\the\ftref,}$\foot{${\cal N}=5$ supergravity
theories were constructed in \dWN.} A similar construction produces ${\cal
N}=5$ supersymmetry
from $E_6$, $E_7$, and $E_8$ singularities.

Many other orbifolds are possible, with both free and non-free
actions on
$S^7$, leading to a variety of different amounts of supersymmetry.

\subsec{M2-branes and exotic spheres}

Another type of example is produced by hypersurface singularities.  For
instance, the
hypersurface in
$\Bbb{C}^5$ defined by
\eqn\hypers{
v^2+w^2+x^2+y^3+z^{6k-1}=0
}
has an isolated Gorenstein canonical singularity at the origin for any integer
$k$.  These
singularities can be realized on Calabi--Yau fourfolds for low values of
$k$---e.g.,
on a degree six hypersurface in $\Bbb{C}\Bbb{P}^5$ in the case of $k=1$.  The
singularities
\hypers\ were studied many years ago by Brieskorn
\refs{\brieskorn,\hirzebruch}, who showed
that the corresponding horizon manifold is topologically isomorphic to $S^7$,
but
not in general diffeomorphic to $S^7$---in particular, in the case $k=1$ it is
not
diffeomorphic to $S^7$.  These horizons thus give examples
of {\it exotic seven-spheres},\foot{For earlier discussions
of exotic seven-spheres in the physics literature, see
\refs{\yamagishi,\freund}. 
The reappearance of this mathematical topic in physics---in the context of
string
theory---was predicted by S.~Shenker in 1996.}
and we should expect to be able to
put M2-branes at the relevant singularities and find Einstein metrics on the
exotic spheres. 

The set of diffeomorphism classes of topological seven-spheres is known to
form a group of
order $28$ \kervairemilnor, and Brieskorn showed that the horizons for the
hypersurfaces
\hypers\ with $k=1, 2, \dots, 28$ include all of these exotic spheres.  For
addressing the
problem of finding Einstein metrics on these spaces, Milnor's original
description \milnor\ of
exotic seven-spheres as $S^3$-bundles over $S^4$ is probably more useful.  In
fact,
Gromoll and Meyer used that description to construct a metric with nonnegative
sectional
curvature on one of the
exotic seven-spheres \gromollmeyer\ (one of the generators of the group),
 but it is apparently still unknown whether or not any of the
 exotic seven-spheres admit  Einstein metrics.

  The Gromoll--Meyer metric
has an isometry group of $SO(3)\times O(2)$; moreover, any exotic seven-sphere
has a
differentiable action of $SO(3)\times O(2)$ \davis, which is the largest its
isometry
group could be for any metric.  It is unclear whether we should expect the
$SO(3)$ or the
$SO(2)$ factor to play the role of the $R$-symmetry group, i.e., whether we
should
expect ${\cal N}=3$ or ${\cal N}=2$ supersymmetry for the resulting
superconformal
theories.  We expect {\it at least}\/ ${\cal N}=2$ supersymmetry since this is
coming
from a singularity on a Calabi--Yau manifold.

\subsec{M5-branes}

Turning now to M5-branes, as we observed in section 2.1,
in order to obtain a model with ${\cal N}=(1,0)$ supersymmetry we must consider
a horizon of the form $H=S^4/\Gamma$ with $\Gamma\subset SU(2)$ a finite group.
In the global context of the Ho\v{r}ava--Witten heterotic string compactified
on $(S^1/\ZZ_2)\times K3$, this means
that we must allow the K3 surface to acquire a rational double point at
$Q$---so that $(S^1/\ZZ_2)\times K3$ is singular all along
$S^1\times\{Q\}$, choose a point $P\in S^1\times\{Q\}$, and place $N$
M5-branes at $P$.  It is important that the branes be located at the
same position in the eleventh dimension.

In order to find a description of the resulting theory, it is useful to employ
the dual model in $F$-theory, worked out in \instKthree\ with some important
additions in \kensix.\foot{The six-dimensional $E8\times E8$
theories studied in \refs{\instKthree,\kensix} also
have an interesting duality relating them to models with $SO(32)$ heterotic
$5$-branes
\PRmirror.}  We will briefly sketch this, deferring details to a companion
publication \nonspherII.  Our task is to
place $N$ small instantons at $Q$ and study the phase transition to the
model in which the corresponding M5-branes have left the boundary, keeping
all $N$ of the branes located at the same position in the eleventh dimension.
In \instKthree\ it was explained how to do this, by studying the behavior
of the discriminant locus in the $F$-theory dual model.  Placing $N$ small
instantons at a singular point $Q$ creates a singularity in the base of
the elliptic fibration of the $F$-theory model, which must then be resolved.

The sizes of the exceptional divisors in the resolution translate to distances
among the M5-branes and the boundary HW9-branes in the Ho\v{r}ava--Witten
description.  In order to keep the M5-branes together in the eleventh
dimension, it is thus necessary to blow down all but one of the exceptional
divisors.  (The one remaining is the one adjacent to the boundary, which
measures
the distance between the collective M5-brane location and the boundary.)
This sends all gauge groups appearing in the effective field theory to
strong coupling.  A precise description of the strong-coupling fixed points
thus obtained can be found in \refs{\instKthree,\kensix,\nonspherII}.

Note that it will be difficult to make direct checks of the generalized
conjecture
in this case due to the presence of orbifold singularities in the horizon.
We do not yet know how to treat such singularities in M-theory.  (But see
\refs{\dasguptamukhi,\witfivebrane} for orientifolds.)

This proposal for a dual field theory for M5-branes at orbifolds differs from
one which
has previously been studied in the literature \refs{\FKPZ,\ahnone}.  In that
earlier
proposal, the $N$ M5-branes were placed at {\it different}\/ locations in
the eleventh dimension.

{}From the global analysis above, it is clear that there is one further type of
point
which must be considered in order to have a complete story for M5-branes:
the {\it transition points}\/ at which the M5-branes are
placed at a singular point, but with that point located on the HW9-brane.
We do not currently have a proposal for the effective theory in such
cases.  The analysis of these transition points will presumably involve the
type of orientifold description
pioneered by Berkooz \berkooz\ for M5-branes at the HW9-brane at a smooth
point of K3 (see also \AKY).

\newsec{Gorenstein canonical singularities in three complex dimensions}

For the remainder of this paper, we focus our attention on
D3-branes in four-dimensional $F$-theory models, which we regard
as
type IIB string compactifications on backgrounds which include D7-branes.
We restrict our attention
to singular points of the compactifying space away from the D7-branes.
(A study of what can happen {\it at}\/ the D7-branes was initiated in
\refs{\FaySpa,\AFM,\aotfth}.)  The relevant singularities are the {\it
Gorenstein canonical
singularities in three complex dimensions}.  When we place $N$ D3-branes
at such a singularity, we will find a four-dimensional theory with ${\cal N}=1$
or ${\cal N}=2$ supersymmetry.\foot{A general analysis of such theories and
their connection to \ads\ supergravity theories was given in \FZ.}

Gorenstein canonical singularities in complex dimension three have been studied
extensively
in the algebraic geometry literature (see \reid\ for the definitions and
earliest results, and \refs{\reidypg} for a thorough review).
There is a kind of inductive classification of these singularities, obtained by
blowing them up one step at a time; there are also detailed descriptions
available
for special classes of these singularities such as the toric ones.  We will
consider
in detail two special types of Gorenstein canonical singularities.

\subsec{Singularities resolved by only one blowup}

Gorenstein canonical threefold singularities which arise by contracting a
single extremal
ray (i.e., those which can be resolved with a single blowup) can be classified
fairly
easily.  These singularities form three types \Wilson, depending on the
behavior of the
contraction mapping:  we can have curves contracting to points (type I), a
surface contracting
to a point (type II), or a surface contracting to a curve (type III).
If the complex structure is generic, these singularities can be more explicitly
described as follows:

\item{$\bullet$}
  type I: a collection of $n\ge1$ smooth ${\Bbb CP}^1$'s, each with normal
bundle
  ${\cal O}(-1)\oplus {\cal O}(-1)$, is contracted to $n$ ``conifold'' points

\item{$\bullet$}
  type III: a rational ruled surface whose ruling has $n\ge0$ degenerate fibers
is contracted
  to a curve

\item{$\bullet$}
  type II: a del Pezzo surface of Picard number $n+2$ is contracted to a point
($-1\le n\le 7$)

This same list of singularities is familiar from another application.  When
 M-theory is compactified on a Calabi--Yau
threefold with such a singularity
 \refs{\MS,\DKV}, there is a five-dimensional field theory associated to the
singularity which decouples
 from gravity \seifive.  Most of these field theories can be described in
 gauge theory terms, as follows:

 \item{$\bullet$}
 type I:  $U(1)$ gauge theory with $n\ge1$ ``electrons'', which can also be
described as
  $n$ D8-branes in Type IA string theory on
 $S^1/\ZZ_2$

 \item{$\bullet$}
 type III: $SU(2)$ gauge theory with $n\ge0$ ``quarks'', which can also be
described as
  $n$ D8-branes at an orientifold plane in Type IA
 string theory on $S^1/\ZZ_2$

 \item{$\bullet$}
 type II: strong-coupling limit of $SU(2)$ gauge theory with $n$ ``quarks''
($0\le n\le7$).
 There is also a variant with a $\ZZ_2$ $\theta$-angle when $n=0$, and a theory
corresponding to
 $n=-1$ which cannot be obtained as a strong-coupling limit of a gauge theory.

\noindent
(In \MS, these theories were labeled by their global symmetry groups:
$A_{n-1}$, $D_n$ and
$E_{n+1}$ in the three cases above.)  Compactifying on a circle to obtain
a IIA theory in four dimensions, the type I singularities correspond to
the familiar conifold points at which there are massless hypermultiplets
\stromcon\
and a conifold transition \GMS; the type III singularities correspond to
enhanced
gauge symmetry and can also exhibit a transition \refs{\klemmmayr,\KMP};
and the type II singularities are associated to ``extremal transitions''
whose physics is less well understood.

Returning to the discussion at hand, when we consider $N$ of the space-filling
D3-branes present in an $F$-theory model whose
base has acquired such a singularity, and move the branes to the singular
locus, there
are several things which can happen.
In the case of a type I contraction, the compactifying space has several
singular points
but they are all locally isomorphic to a conifold singularity, so we must study
branes
at a conifold to cover this case.  In the case of a type
III contraction, the singularity is not isolated, but there are only two
isomorphism classes of
local singularity types which occur, corresponding to the contraction at a
nonsingular fiber
of the ruling on the surface (the generic case), or at a singular fiber.  In
the case of
a type II contraction, there is only one singular point so there is no
ambiguity about
where to place the branes.

In short, to cover all of the cases above we must consider the following kinds
of singular
points and their associated horizons:

\item{$\bullet$}
type I: a conifold point with local equation $xy=zt$, whose horizon is
$H_{con}=V_{4,2}=(S^3\times S^3)/U(1)$ \CdO.

\item{$\bullet$}
type III generic: a quotient singularity $\Bbb{C}^3/\ZZ_2$ (in which $\ZZ_2$
acts on only two of
the three complex coordinates) with local equation $xy=z^2$, whose horizon is
$S^5/\ZZ_2$

\item{$\bullet$}
type III special: a ``suspended pinch point'' with local equation $xy=z^2t$,
whose horizon is a circle bundle over
a product of weighted projective spaces
$\Bbb{W}\Bbb{C}\Bbb{P}^{1,2}\times \Bbb{W}\Bbb{C}\Bbb{P}^{1,2}$
(as we will see in section 4.3)

\item{$\bullet$}
type II: the complex cone over a del Pezzo surface, whose horizon is a circle
bundle over
the del Pezzo surface.
The complex cone and the circle bundle are built from the anti-canonical line
bundle of the del Pezzo surface.

There are two kinds of del Pezzo surfaces:  the surface
$\Bbb{F}_0=\Bbb{C}\Bbb{P}^1\times
\Bbb{C}\Bbb{P}^1$,
or the blowup of $\Bbb{C}\Bbb{P}^2$ at $k$ points with $0\le k\le8$.  Note that
the circle bundle
in the anti-canonical bundle of $\Bbb{C}\Bbb{P}^2$ yields $S^5/\Bbb{Z}_3$
as its horizon manifold, and the circle bundle
in the anti-canonical bundle of $\Bbb{F}_0$ yields
$V_{4,2}/\Bbb{Z}_2$ as its horizon manifold.

Thus, we find an almost perfect match between isolated Gorenstein canonical
singularities obtained by contracting
a single extremal ray (classified via algebraic geometry), and
Einstein--Sasaki five-manifolds with a regular $U(1)$-action (classified via
differential geometry).
There is a small puzzle, however: the cones over del Pezzo surfaces which are
the blowups
of $\Bbb{C}\Bbb{P}^2$ at one or two points are missing form the list of regular
Einstein--Sasaki horizons.  It is not even known if there exist
Einstein--Sasaki metrics
on these spaces.  One possibility is that they have such metrics, but---like
many orbifolds---the
$U(1)$ action is not regular.  Another possibility is that the limiting metrics
have a
more complicated asymptotic behavior than our basic ansatz \ansatz.  We note in
passing that
these same two singularities exhibited exceptional behavior in the earlier
application to
construction of five-dimensional field theories \MS: on the one hand, these
were the only two
whose deformation space has extra first-order directions that do not extend to
full
deformations \refs{\altmann,\gross}; on the other hand, the global symmetry
groups of the
corresponding five-dimensional field theories were the only two which are not
semisimple groups---each
one them has a $U(1)$ factor.  (These are the groups denoted $\widetilde{E}_1$
and $E_2$ in \MS.)

\subsec{Toric singularities}

Another class of Gorenstein canonical singularities which can be easily
classified is the class
of toric
singularities;\foot{For a review of toric geometry for physicists, see
\refs{\catp,\summing}}
this includes all singularities of the form $\Bbb{C}^n/\Gamma$ for finite
abelian groups $\Gamma\subset U(n)$.
According to Reid's criterion \reid, the data needed to define a toric
Gorenstein canonical
singularity of complex dimension $n$ is a convex polygon in $\Bbb{R}^{n-1}$
whose vertices have
integer coordinates.  Given that data, one construction of the singularity is
as follows:\foot{As
we will see shortly, there are other possible constructions.} let
$v_1$, \dots, $v_k$ be all of the vectors in $\Bbb{R}^{n-1}$ which have integer
coordinates,
and which lie in either the interior or the boundary of the polygon.  Consider
the linear relations among these vectors
\eqn\linrel{
\sum_{j=1}^k Q^jv_j=0
}
with integer coefficients $Q^j$, subject to
\eqn\linrelcond{
\sum_{j=1}^k Q^j=0.
}
Let  $\vec{Q}_1$, \dots,
$\vec{Q}_{k-n}$ be a basis for the set of all such relations, and use the
matrix $(Q_i^j)$ to specify
the charges in a representation of $U(1)^{k-n}$ on $\Bbb{C}^k$.  The singular
space
is then the symplectic reduction $\Bbb{C}^k\catquot U(1)^{k-n}$, using the moment map
centered at the
origin.  (This mathematical construction should be familiar from studies of
$\CN=(2,2)$ abelian gauge theory in two dimensions, with gauge group
$U(1)^{k-n}$ and chiral matter in the 
representation specified
by $(Q_i^j)$ and no Fayet--Iliopolous terms---see 
\refs{\phases,\summing}, for example.)  Clearly, this singularity 
doesn't change if
we shift the location of the polygon by adding an integer vector to it.

Varying the center of the moment map then produces partial or complete
resolutions of this
singularity.\foot{By including all integer vectors $v_j$ lying within
the polygon, we have included all possible toric blowups with a non-vanishing
holomorphic $n$-form.}  (In the $\CN=(2,2)$ gauge theory, this corresponds
to turning on certain Fayet--Iliopoulos $D$-terms.)  The location of the
center of the moment map determines the sizes of holomorphic curves within
the exceptional set of the resolution, i.e., the blowup moduli of the
singularity.  Particular values of these moduli will correspond to partial
rather than complete resolutions.
In general, any convex sub-polygon can represent the residual singularity in
one of the partial resolutions.

An alternate description of toric singularities is often very useful: they can
be described
by a collection of polynomial equations of the form ``one monomial equals
another monomial.''
To produce such a description from the data above, introduce homogeneous
coordinates
$x_1$, \dots, $x_k$, and find a minimal generating set $z_1$, \dots $z_\ell$
for
\eqn\monos{
\{\hbox{\rm $U(1)^{k-n}$-invariant monomials}\ z=x_1^{a_1}\cdots x_k^{a_k}\ |\
a_i\ge0\}.
}
The $z_j$'s give coordinates in the ambient space of the toric singularity, and
the polynomial
equations relating various monomials in the $z_j$'s can be found in a
straightforward manner.

For example,  if we let $\Gamma=\Bbb{Z}_m\times \Bbb{Z}_m$ act on $\Bbb{C}^3$
via the generators
$\diag(e^{2\pi i/m},e^{-2\pi i/m},1)$ and $\diag(e^{2\pi i/m},1,e^{-2\pi
i/m})$, then
the quotient singularity $\Bbb{C}^3/\Gamma$ is described by the polygon with
vertices
$(0,0)$, $(m,0)$ and $(0,m)$.  The set of vectors $v_j$ consists of all integer
vectors
$(k,\ell)$ with $k\ge0$, $\ell\ge0$, and $k+\ell\le m$.  Notice that the convex
polygon
associated to {\it any}\/ three-dimensional
toric Gorenstein canonical singularity can be embedded as a sub-polygon of
$\left<(0,0), (m,0), (0,m)\right>$ for some $m$, after shifting its location by
an appropriate
vector.

\bigbreak

\noindent{\it Partial resolutions of the ${\Bbb Z}_2\times {\Bbb Z}_2$
orbifold}\par\nobreak\medskip\nobreak

To be more explicit in the case of $\Bbb{Z}_2\times \Bbb{Z}_2$, let us label
the vectors
as
\eqn\labelvec{
v_0=(0,0) \quad v_1=(2,0) \quad v_2=(0,2) \quad w_0=(1,1) \quad
w_1=(0,1) \quad w_2=(1,0) .
}
(These vectors are illustrated in Figure~1(a).)
A basis for the relations subject to \linrelcond\ is then given by
\eqn\relbas{
v_0+w_0-w_1-w_2=0, \quad v_1-w_0+w_1-w_2=0, \quad v_2-w_0-w_1+w_2=0.
}
We introduce homogeneous coordinates $x_0$, $x_1$, $x_2$, $y_0$, $y_1$, $y_2$
and consider the $U(1)^3$ action determined by \relbas; it has charge matrix
\eqn\chrgmat{
\pmatrix{x_0&x_1&x_2&y_0&y_1&y_2\cr 1&0&0&1&-1&-1\cr 0&1&0&-1&1&-1\cr
0&0&1&-1&-1&1}.
}
The moment map, using arbitrary $D$-terms, is therefore
\eqn\mommap{
\matrix{
|x_0|^2+|y_0|^2-|y_1|^2-|y_2|^2=\zeta_1\cr
|x_1|^2-|y_0|^2+|y_1|^2-|y_2|^2=\zeta_2\cr
|x_2|^2-|y_0|^2-|y_1|^2+|y_2|^2=\zeta_3}
}
Another equation which follows from these is
\eqn\anothereq{
|x_0|^2+|x_1|^2-2|y_2|^2=\zeta_1+\zeta_2.
}

\midinsert
$$\matrix{{\epsfxsize=1in\epsfbox{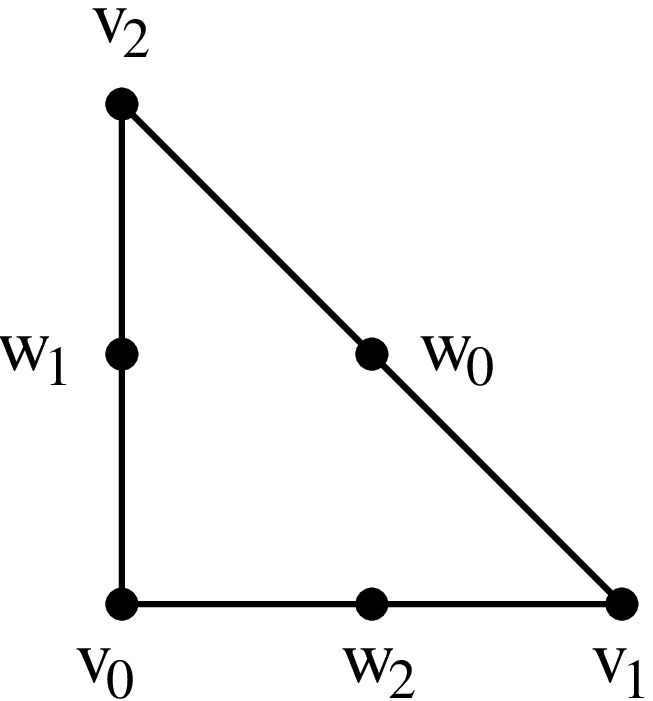}}&
{\epsfxsize=1in\epsfbox{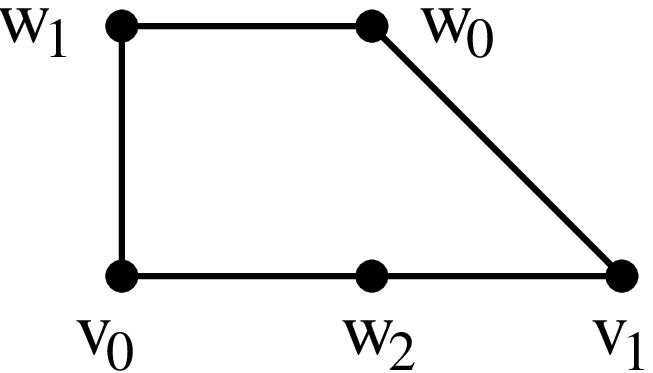}}\cr
\hbox{Figure 1(a). $\Bbb{Z}_2\times\Bbb{Z}_2$ orbifold.}& \hbox{Figure 1(b).
Suspended pinch point.}\cr
{\epsfxsize=.75in\epsfbox{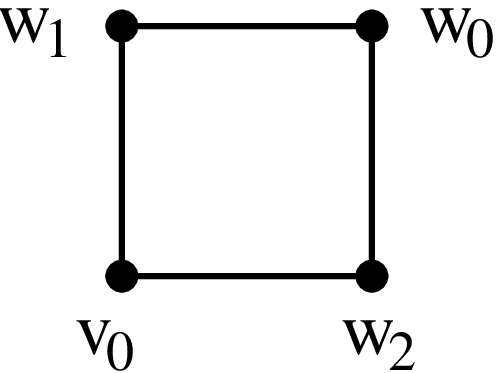}}&
{\epsfxsize=.9in\epsfbox{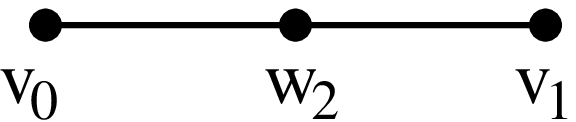}}\cr
\hbox{Figure 1(c). Conifold.}& \hbox{Figure 1(d). $\Bbb{Z}_2$ orbifold.}}$$
\endinsert

When $\zeta_1$, $\zeta_2$ and $\zeta_3$ are generic, the symplectic reduction
is smooth.
There are several singularity types which can be obtained for special values of
the
$\zeta_i$.  These are illustrated in Figure~1, and described as follows.
Each singularity is associated to a subgroup $U(1)^k$ of $U(1)^3$ with the
property
that at least $3-k$ of the homogeneous coordinates are uncharged under
$U(1)^k$.
The singular space is then described by symplectic reduction of the space
$\Bbb{C}^{k+3}$
spanned by the remaining $k+3$
homogeneous coordinates by $U(1)^k$.  For each partial resolution, we have
listed in
Table~3 the condition on $D$-term coefficients which selects out $U(1)^k$, and
the corresponding charge matrix on the remaining $k+3$ homogeneous coordinates.
The polygons corresponding to the relevant homogeneous coordinates are
illustrated in
Figure~1.

\midinsert
$$\vbox{\offinterlineskip\halign{
\strut # height 27pt depth 23pt&
\quad#\quad\hfill\vrule&\quad$#$\quad\hfill\vrule& \quad$#$\quad\hfill\vrule\cr
\noalign{\hrule}
\vrule&$\Bbb{Z}_2\times\Bbb{Z}_2$ orbifold&
\zeta_1=\zeta_2=\zeta_3=0&
\pmatrix{x_0&x_1&x_2&y_0&y_1&y_2\cr 1&0&0&1&-1&-1\cr 0&1&0&-1&1&-1\cr
0&0&1&-1&-1&1}\cr
\noalign{\hrule}
\vrule &suspended pinch point&\zeta_1=\zeta_2=0&\pmatrix{x_0&x_1&y_0&y_1&y_2\cr
1&0&1&-1&-1\cr 0&1&-1&1&-1}\cr
\noalign{\hrule}
\vrule &conifold&\zeta_1=0&\pmatrix{x_0&y_0&y_1&y_2\cr1&1&-1&-1}\cr
\noalign{\hrule}
\vrule &$\Bbb{Z}_2$
orbifold&\zeta_1+\zeta_2=0&\pmatrix{x_0&x_1&x_2&y_2\cr1&1&0&-2}\cr
\noalign{\hrule}
}}$$
\bigskip
\centerline{Table 3. Charge matrices for partial resolutions of the
$\Bbb{Z}_2\times\Bbb{Z}_2$ orbifold.}
\bigskip
\endinsert

For each of these singularities, we can compute the $U(1)^k$-invariant
monomials
and the equations they satisfy, giving the algebraic realization of the
singularity.
This is carried out in Table~4.  The equations enable us to recognize some of
these
singularities as familiar ones, such as the ${\Bbb Z}_2$ orbifold
$z_1z_3=z_2^2$ (``type III generic'' in section 4.1), 
the conifold singularity
$z_1z_4=z_2z_3$ (``type I'' in section 4.1),
and the ``suspended pinch point''\foot{This singularity is the
suspension, in the sense of singularity theory, of the pinch points which occur
in the theory of ordinary singularities of surfaces \refs{\GH}.}
$z_1z_3=z_2^2z_4$ (``type III special'' in section 4.1).

\midinsert
$$\vbox{\offinterlineskip\halign{
\strut # height 23pt depth 19pt&
\quad#\quad\hfill\vrule&\quad$#$\quad\hfill\vrule& \quad$#$\quad\hfill\vrule\cr
\noalign{\hrule}
\vrule&$\Bbb{Z}_2\times\Bbb{Z}_2$ orbifold&
\matrix{z_1=x_0^2y_1y_2,& z_2=x_1^2y_0y_2,\hfill\cr z_3=x_2^2y_0y_1,& 
z_4=x_0x_1x_2y_0y_1y_2\hfill}&z_1z_2z_3=z_4^2\cr
\noalign{\hrule}
\vrule &suspended pinch point&
\matrix{z_1=x_0^2y_1y_2,&z_2=x_0x_1y_2,\hfill\cr
z_3=x_1^2y_0y_1,&z_4=y_0y_1\hfill}&z_1z_3=z_2^2z_4\cr
\noalign{\hrule}
\vrule &conifold&
\matrix{z_1=x_0y_1,\hphantom{x_1^2}&z_2=x_0y_2,\hfill\cr
z_3=y_0y_1,\hphantom{x_1^2}&z_4=y_0y_2\hfill}&z_1z_4=z_2z_3\cr 
\noalign{\hrule}
\vrule &$\Bbb{Z}_2$ orbifold&
\matrix{z_1=x_0^2y_2,\hphantom{x_1}&z_2=x_0x_1y_2,\hfill\cr
z_3=x_1^2y_2,\hphantom{x_1}&z_4=x_2\hfill}&z_1z_3=z_2^2\cr 
\noalign{\hrule}
}}$$
\bigskip
\centerline{Table 4. Invariant monomials and equations for partial resolutions
of the
$\Bbb{Z}_2\times\Bbb{Z}_2$ orbifold.}
\bigskip
\endinsert

The presentations of these singularities which we have given include all
possible
``crepant'' toric blowups by varying the $D$-terms.  (The {\it crepant}\/
blowups
are the ones for which the pullback of the holomorphic $n$-form near the
singularity
is non-vanishing.)  However, in many cases there is a more ``efficient''
description
of the singularity, involving fewer fields and a smaller group.  In terms of
the
convex polygon, the most efficient description is obtained by using only the
vertices
of the polygon and omitting the other vectors $v_j$.  One modification which
must
be made to the construction is that if the vertices do not span a primitive
sublattice
of $\Bbb{Z}^n$, then after performing the symplectic reduction we must
implement
a further orbifold by a finite group $G$ (the torsion subgroup of the quotient
of
$\Bbb{Z}^n$ by the span of the $v_j$'s).

\midinsert
$$\matrix{{\epsfxsize=.9in\epsfbox{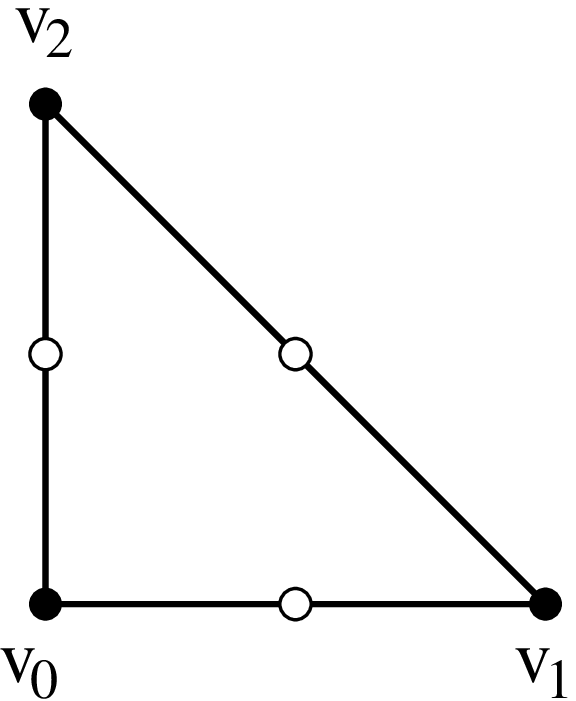}}&\qquad&
{\epsfxsize=.9in\epsfbox{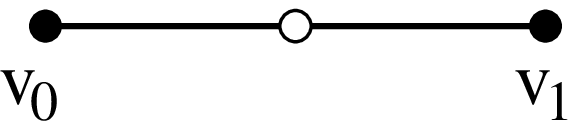}}}$$
\centerline{Figure 2. The combinatorial data for $\Bbb{Z}_2\times\Bbb{Z}_2$ and
$\Bbb{Z}_2$ orbifolds.}
\endinsert

For two of our examples---the $\Bbb{Z}_2\times \Bbb{Z}_2$ orbifold and
$\Bbb{Z}_2$ orbifold---this
procedure leads back directly to an orbifold description.  The combinatorics
for the
most efficient description are shown in Figure~2.  In the first case, the
invariants
for the $\Bbb{Z}_2\times \Bbb{Z}_2$ action $(x_0,x_1,x_2)\mapsto
((-1)^{a+b}x_0,
(-1)^ax_1,(-1)^bx_2)$ are $z_1=x_0^2$, $z_2=x_1^2$, $z_3=x_2^2$ and
$z_4=x_0x_1x_2$, with
equation $z_1z_2z_3=z_4^2$, so we recover our original description.  In the
second case,
the invariants for the $\Bbb{Z}_2$ action $(x_0,x_1,x_2)\mapsto(-x_0,-x_1,x_2)$
are $z_1=x_0^2$, $z_2=x_0x_1$, $z_3=x_1^2$, and $z_4=x_2$ with equation
$z_1z_3=z_2^2$
and we again recover our original description.

\midinsert
$$\matrix{{\epsfxsize=.75in\epsfbox{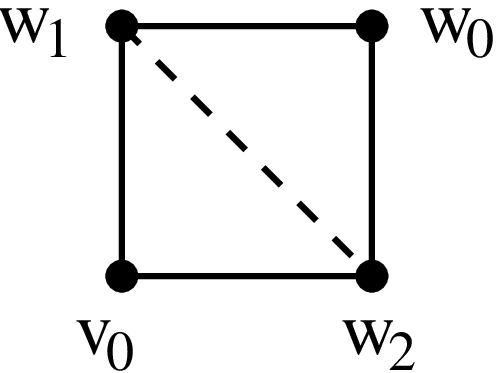}}&\qquad&
{\epsfxsize=.75in\epsfbox{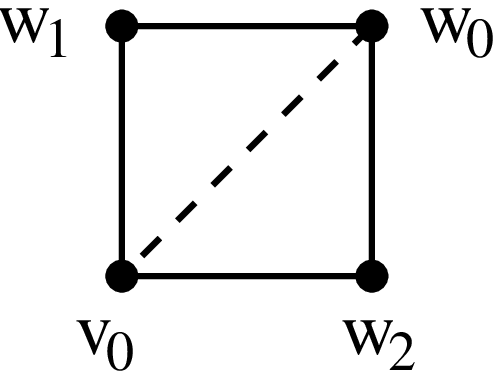}}}$$
\centerline{Figure 3. The two resolutions of the conifold.}
\endinsert

For the other two examples---the suspended pinch point and the conifold---the
most efficient
description involves a single $U(1)$.  In the case of the conifold, this is the
description
which we have already given.  Varying the $D$-term for the $U(1)$  produces the
two small
resolutions of the conifold, related by a flop, as illustrated in Figure~3.

\midinsert
$$\matrix{{\epsfxsize=1in\epsfbox{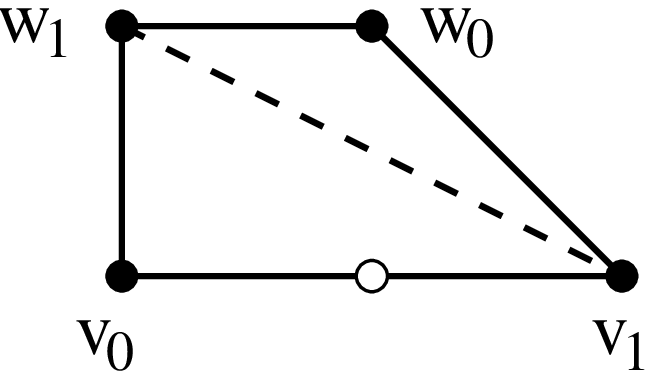}}&\qquad
&{\epsfxsize=1in\epsfbox{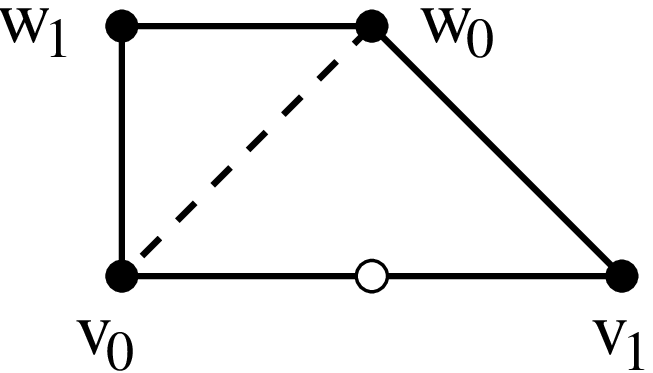}}}$$
\centerline{Figure 4. The partial resolutions of the suspended pinch point in
its minimal presentation.}
\endinsert

In the
case of the suspended pinch point, the efficient description involves the
vectors
$v_0$, $v_1$, $w_1$ and $w_2$, related by
\eqn\spprel{
v_0-v_1+2w_0-2w_1,
}
which leads to the charge matrix
\eqn\sppchg{
\pmatrix{x_0&x_1&y_0&y_1\cr1&-1&2&-2}.
}
The invariants monomials are $z_1=x_0^2y_1$, $z_2=x_0x_1$, $z_3=x_1^2y_0$ and
$z_4=y_0y_1$
with equation $z_1z_3=z_2^2z_4$ so that we recover our original description.
Varying the $D$-term for the $U(1)$ produces two small partial resolutions,
with exceptional
set $\Bbb{C}\Bbb{P}^1$ but still having singularities along a curve.  The toric
data for
these is illustrated in Figure~4.

\bigbreak

\noindent{\it Complex cones over del Pezzo
surfaces}\par\nobreak\medskip\nobreak 

Two of the complex cones over del Pezzo surfaces (i.e., singularities of
``type II'' in the terminology of section 4.1) have particularly simple
toric descriptions:  they are the complex cones over $\BC\BP^2$ and
over ${\Bbb F}_0$.  The corresponding convex polygons are illustrated
in Figure 5; they involve the vectors
\eqn\labelvecbis{
v_0=(0,0) \quad  w_0=(1,1) \quad
w_1=(0,1) \quad w_2=(1,0) \quad u_1=(2,1) \quad u_2=(1,2) .
}
The relations subject to \linrelcond\ are generated by
\eqn\relbaspeetwo{
v_0+u_1+u_2-3w_0=0
}
for the complex cone over $\BC\BP^2$, and by
\eqn\relbasfzero{
w_1+u_1-2w_0=0, \quad w_2+u_2-2w_0=0
}
for the complex cone over ${\Bbb F}_0$.

\midinsert
$$\matrix{{\epsfxsize=1in\epsfbox{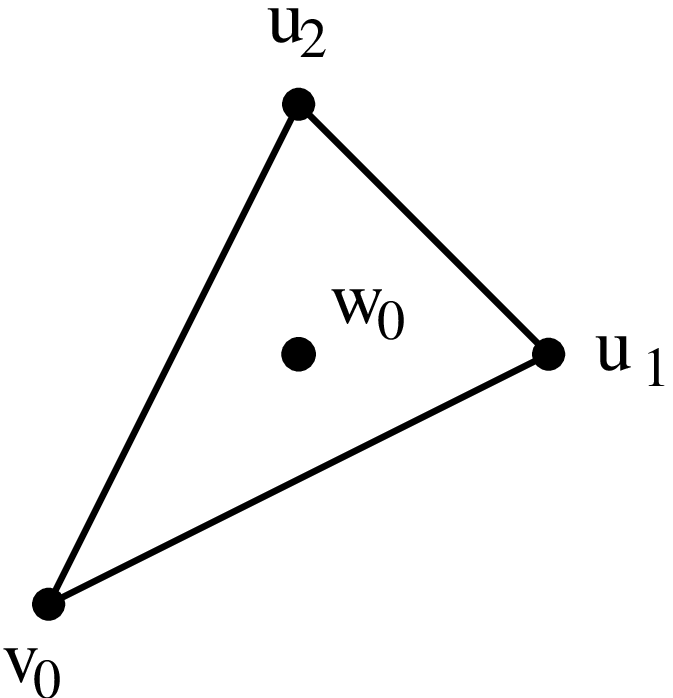}}&
{\epsfxsize=1.1in\epsfbox{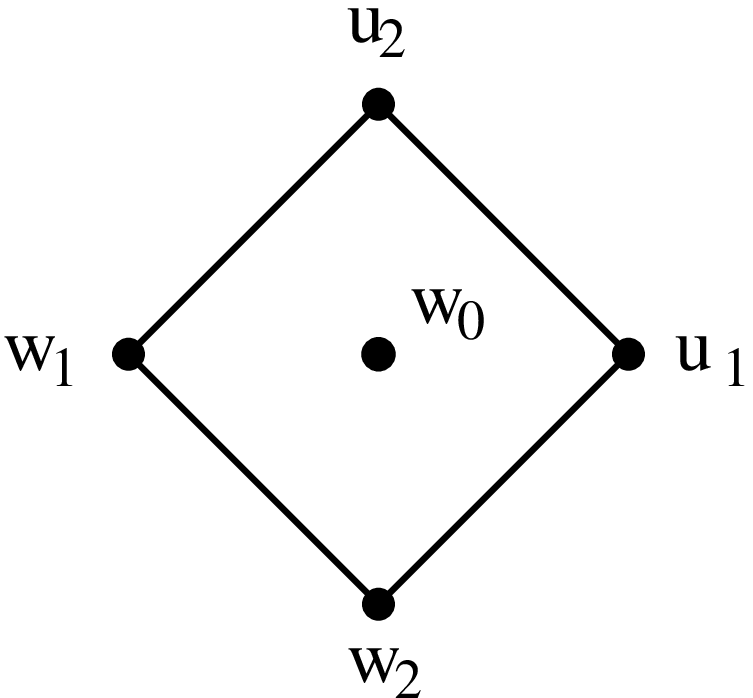}}\cr
\hbox{Figure 5(a). Complex cone over $\BC\BP^2$.}& \hbox{Figure 5(b).
Complex cone over ${\Bbb F}_0$.}\cr
}$$
\endinsert

In the case of the complex cone over $\BC\BP^2$, the singularity can
be described by means of homogeneous coordinates $x_0$,  $y_0$, $z_1$, $z_2$
with $U(1)$ action
\eqn\chrgmatpeetwo{
\pmatrix{x_0&y_0&z_1&z_2\cr 1&-3&1&1}.
}
The moment map with $D$-term $\zeta$ is given by
\eqn\mommap{
\matrix{
|x_0|^2-3|y_0|^2+|z_1|^2+|z_2|^2=\zeta .}
}
Varying the $D$-term leads to two possible partial resolutions
(``unresolved'' and ``fully resolved'') as illustrated in Figure~6.

\midinsert
$$\matrix{{\epsfxsize=1in\epsfbox{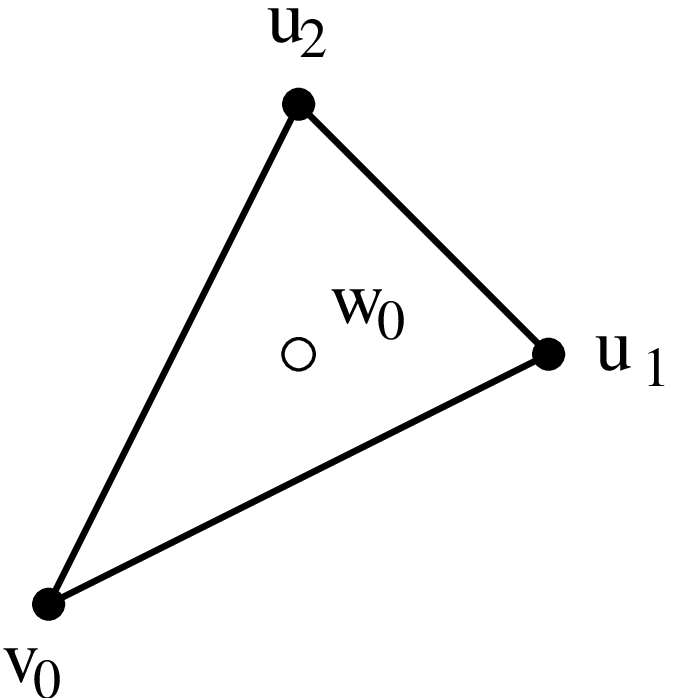}}&\qquad&
{\epsfxsize=1in\epsfbox{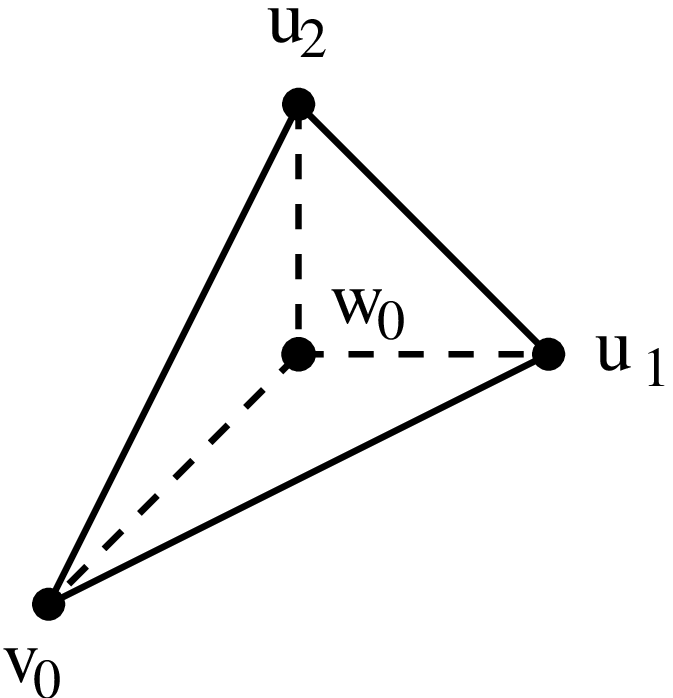}}\cr
}$$
\centerline{Figure 6.  Partial resolutions of the complex cone over
$\BC\BP^2$.} 
\endinsert

The ``most efficient'' description omits the vertex $w_0$ as illustrated in
the left side of Figure~6.  Notice that if we gauge fix to eliminate the
corresponding homogeneous coordinate $y_0$, there is a residual ${\Bbb Z}_3$
gauge symmetry.  That is, {\it the complex cone over $\BC\BP^2$ is a
${\Bbb Z}_3$ orbifold.}

In the case of the complex cone over ${\Bbb F}_0$, the singularity can
be described by means of homogeneous coordinates   $y_0$, $y_1$, $y_2$,
$z_1$, $z_2$ with $U(1)^2$ action
\eqn\chrgmatfzero{
\pmatrix{y_0&y_1&y_2&z_1&z_2\cr -2&1&0&1&0\cr -2&0&1&0&1}.
}
The moment map with arbitrary $D$-terms $\zeta_1$, $\zeta_2$ is given by
\eqn\mommapfzero{
\matrix{
-2|y_0|^2+|y_1|^2+|z_1|^2=\zeta_1 \cr
-2|y_0|^2+|y_2|^2+|z_2|^2=\zeta_2 .}
}
Notice that the difference between these equations is
\eqn\difffzero{
|y_1|^2-|y_2|^2+|z_1|^2
-|z_2|^2=\zeta_1 -\zeta_2.
}
Varying the $D$-terms leads to four possible partial resolutions
(including ``unresolved'' and ``fully resolved'') as illustrated in Figure~7.
Each of the two partially resolved cases has a curve of ${\Bbb Z}_2$
singularities. 

The ``most efficient'' description omits the vertex $w_0$ as in the left
portions of Figure~7.  If we gauge fix the second $U(1)$ in \chrgmatfzero\ to
eliminate the homogeneous coordinate $y_0$, there is a residual ${\Bbb Z}_2$
group action on a conifold point (recognizable from the remaining
moment map \difffzero).  That is,
{\it the complex cone over ${\Bbb F}_0$ is a ${\Bbb Z}_2$ orbifold of a
conifold.}

\midinsert
$$\matrix{{\epsfxsize=1.1in\epsfbox{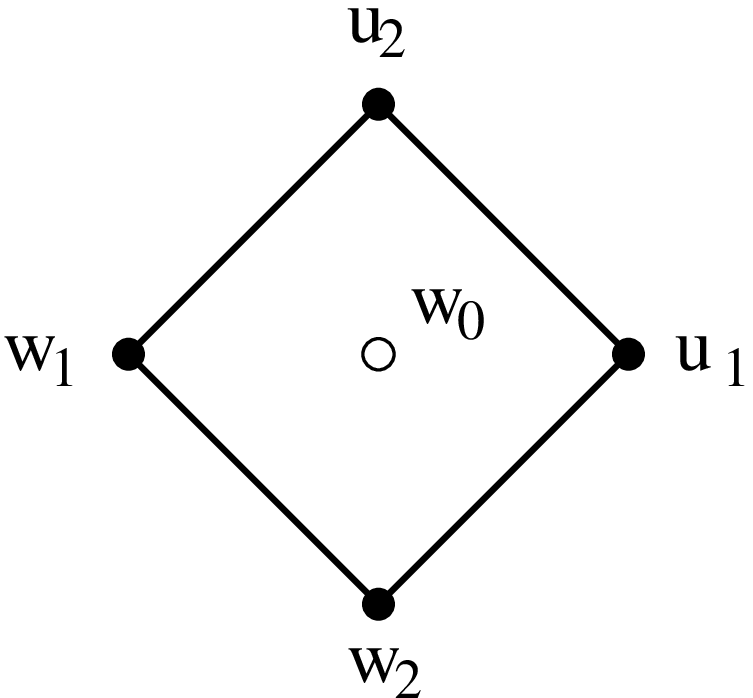}}&
{\epsfxsize=1.1in\epsfbox{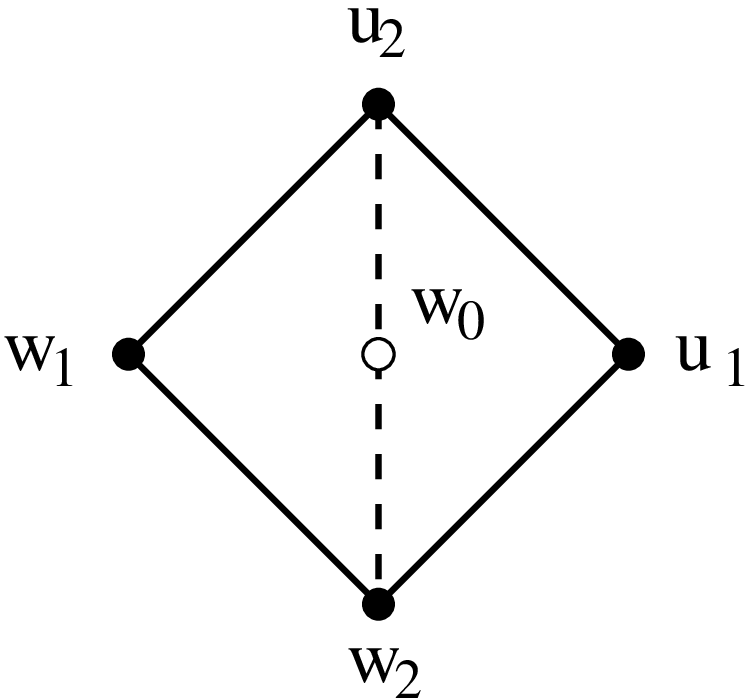}}&
{\epsfxsize=1.1in\epsfbox{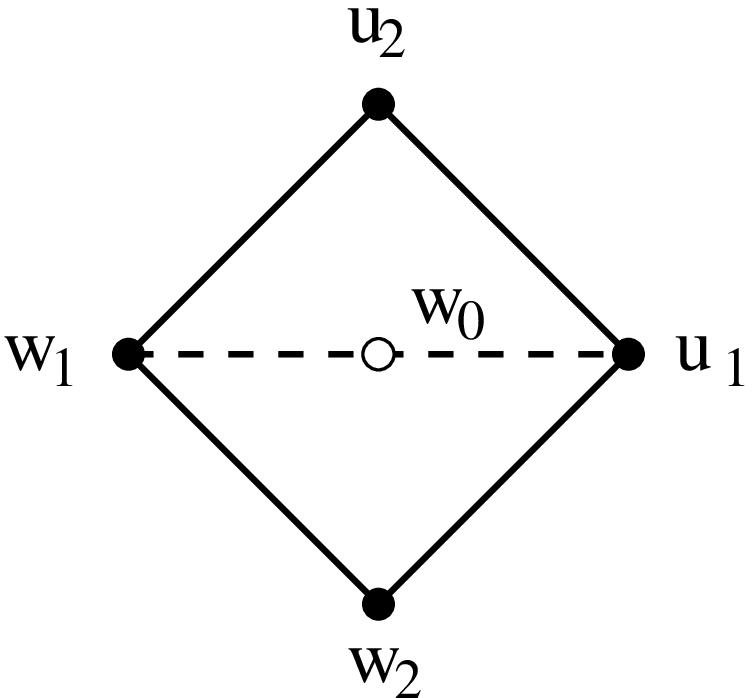}}&
{\epsfxsize=1.1in\epsfbox{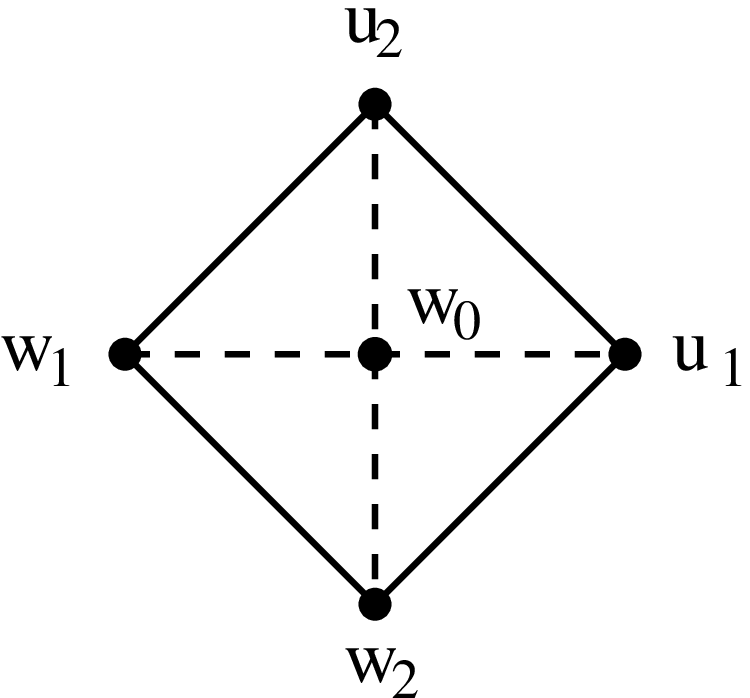}}
}$$
\centerline{Figure 7.  Partial resolutions of the complex cone over ${\Bbb
F}_0$.} 
\endinsert

\subsec{Horizon manifolds}

For each of the toric singularities discussed in the previous section,
we will now study the corresponding horizon $H$ in some detail, determining
its isometry group, its homology, and the parameter space for minimum
volume cycles in certain homology classes.  (All of these properties will be
useful when we subsequently compare \AdS\ compactifications to field theory
models.)  As we learned in 
section $2$, when the singularity leads to an ${\cal N}=1$ theory
the horizon should have an Einstein--Sasaki structure with
an associated $U(1)$ action (the $R$-symmetry in the field theory)
so that $H/U(1)$ has the structure of a
K\"ahler--Einstein orbifold.  The horizon can then be reconstructed as the
unit circle bundle in a certain holomorphic line bundle ${\cal L}$ over the 
K\"ahler--Einstein space.

When the $U(1)$ action is regular, $H/U(1)$ is a K\"ahler--Einstein
manifold (in fact, a del Pezzo surface) and we can use the Leray spectral
sequence\foot{For an introduction to spectral sequences, see \botttu.} 
for the fibration $H\to H/U(1)$ to calculate the cohomology and
homology of 
$H$.  The del Pezzo surface $H/U(1)$ has vanishing first and third
cohomology groups, and its second cohomology group is isomorphic to $\ZZ^b$
for some $b$, the second Betti number of $H/U(1)$.  The  first Chern class
of ${\cal L}$ determines a map $\lambda:\ZZ\to\ZZ^b$, and a dual map
$\lambda^*:\ZZ^b\to\ZZ$, which appear in the spectral sequence.
The nonzero entries in the $E_2$ term of the Leray spectral
sequence are then:
{
$$\vbox{\offinterlineskip\halign{
\hfill\quad$#$\quad\hfill&\strut # & 
\quad$#$\ &\hfill$#$\hfill&\ $#$\ &\hfill$#$\hfill&\ $#$\quad\cr
1&\vrule height 15 pt depth 5 pt &\ZZ&&\ZZ^b&&\ZZ\cr
0&\vrule height 31 pt depth 8 pt &\ZZ 
&\raise8pt\hbox{\epsfbox{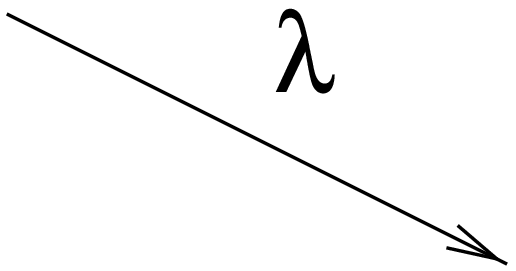}}&\ZZ^b
&\raise8pt\hbox{\epsfbox{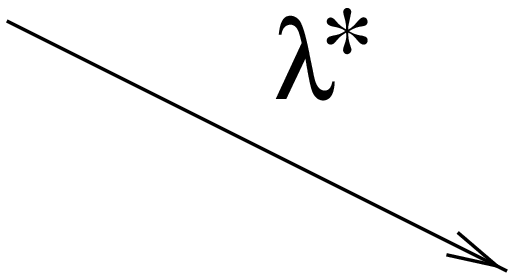}}&\ZZ\cr 
\noalign{\hrule}
E_2^{p,q}&\vrule height 15 pt depth 8 pt &0&1&2&3&4\cr
}}$$
}
Suppose that $c_1({\cal L})$ is divisible by $m$ in $H^2(H/U(1))$, so that
the cokernel of $\lambda$ contains the $m$-torsion group $\ZZ_m$.  Then
the $E_3=E_\infty$ term of the Leray spectral sequence is
{
$$\vbox{\offinterlineskip\halign{
\hfill\quad$#$\quad\hfill&\strut # & 
\quad$#$\ &\hfill$#$\hfill&\ $#$\ &\hfill$#$\hfill&\ $#$\quad\cr
1&\vrule height 15 pt depth 5 pt &0&0&\ZZ^{b-1}&0&\ZZ\hphantom{{}_m}\cr
0&\vrule height 15 pt depth 8 pt &\ZZ 
&\qquad0\qquad
&\ZZ^{b-1} \oplus \ZZ_m
&\qquad0\qquad
&\ZZ_m\cr 
\noalign{\hrule}
E_\infty^{p,q}&\vrule height 15 pt depth 8 pt &0&1&2&3&4\hphantom{{}_m}\cr
}}$$
}
We can read off the cohomology of $H$ from the $E_\infty$ term as
$H^i(H)=\bigoplus_{p+q}E_\infty^{p,q}$, and determine the homology
by Poincar\'e duality or the universal coefficient theorem.  The result is
shown in Table~5.

\midinsert
$$\vbox{\offinterlineskip\halign{
\strut # height 15pt depth 8pt& \hfill\quad$#$\quad\hfill\vrule&
\hfill\quad$#$\quad\hfill&
\hfill\quad$#$\quad\hfill&
\hfill\quad$#$\quad\hfill&
\hfill\quad$#$\quad\hfill&
\hfill\quad$#$\quad\hfill&
\hfill\quad$#$\quad\hfill\vrule\cr
\noalign{\hrule}
\vrule&i&0&1&2&3&4&5\cr
\noalign{\hrule}
\vrule&H^i(H) &\ZZ & 0 & \ZZ^{b-1}\oplus \ZZ_m
& \ZZ^{b-1} & \ZZ_m & \ZZ\cr
\noalign{\hrule}
\vrule&H_i(H)&\ZZ & \ZZ_m & \ZZ^{b-1} & \ZZ^{b-1}\oplus\ZZ_m
&  0 & \ZZ\cr
\noalign{\hrule}
}}$$
\bigskip
\centerline{Table 5. Cohomology and homology of horizon manifolds.}
\bigskip
\endinsert

To compute the isometry groups of our horizons, we must work with the
explicit descriptions of the horizon manifolds.  The ${\cal N}=1$
supersymmetry in the
associated compactification of the IIB string on $\AdS_5\times H$ requires
a complex Killing spinor on $H$, and the isometry group of $H$ contains a
distinguished $U(1)$ which simply shifts the phase of that Killing
spinor---this becomes the $U(1)$ $R$-symmetry in the associated conformal
field theory.  To determine the symmetries of the supersymmetric theory, we
need to compute the group of all isometries of $H$ which commute with this
$U(1)$.  We will call this the group of 
{\it supersymmetric isometries.}\foot{Similar definitions could be made
for horizons corresponding to theories with extended supersymmetry,
using the larger $R$-symmetry groups relevant in such cases.}

\bigbreak

\noindent{\it Orbifolds}\par\nobreak\medskip\nobreak

For orbifold singularities $\Bbb{C}^3/\Gamma$, with $\Gamma\subset SU(3)$,
we consider the induced action of
$\Gamma$ on $\Bbb{C}\Bbb{P}^2$.  The space $\Bbb{C}\Bbb{P}^2/\Gamma$ will
inherit
the structure of a K\"ahler--Einstein orbifold, from the K\"ahler--Einstein
metric on $\Bbb{C}\Bbb{P}^2$.  Similarly, the horizon $S^5$ of the origin in
$\Bbb{C}^3$ inherits an action of $\Gamma$, and $S^5/\Gamma$ will be the
corresponding
horizon for $\Bbb{C}^3/\Gamma$.  In the ${\cal N}=1$ case, the
 $U(1)$ action descends to $S^5/\Gamma$ in
such
a way that $(S^5/\Gamma)/U(1)=\Bbb{C}\Bbb{P}^2/\Gamma$.  (In other words,
$U(1)$ acts with charge $1$ on each of the three complex variables in
$\Bbb{C}^3$.) The horizon will be singular
if the action of $\Gamma$ on $\Bbb{C}^3$ has a non-isolated fixed point locus.
When present, the singularities occur along $S^1$'s (i.e., in real
codimension four within $H^5$).

The isometries of $S^5/\Gamma$ lift to isometries of $S^5$ and so lie in
$SO(6)$.  The commutator in $SO(6)$ of the $U(1)$ $R$-symmetry group is
$U(3)$, so in the ${\cal N}=1$ case
the supersymmetric isometries of $S^5/\Gamma$ are given by
\eqn\susyisomorbi{
\{g\in U(3)\ |\ g\Gamma=\Gamma g \}/\Gamma.
}
For example, when $\Gamma=\BZ_3$ acting as the center of $SU(3)$,
then \susyisomorbi\ becomes $U(3)/\BZ_3$, whereas when
$\Gamma=\BZ_2\times\BZ_2$ acting with generators
$(x,y,z)\to(-x,-y,-z)$ and $(x,y,z)\to(-x,y,-z)$
then \susyisomorbi\ becomes $U(1)^3\rtimes S_3$.

The $U(1)$ action on $S^5/\Gamma$ is regular only in the case
$\Gamma=\Bbb{Z}_3$ \FKone.  As remarked earlier, this is the horizon
of the complex cone over $\BC\BP^2$.
In this case, the second Betti number $b$ of
$H/U(1)=\Bbb{C}\Bbb{P}^2$ is $1$, and the torsion in the cokernel of
$\lambda$ has order $m=3$.  The nonzero homology groups are
$H_0=H_5=\Bbb{Z}$ and $H_1=H_3=\Bbb{Z}_3$.

More generally, when $\BC^3/\Gamma$ has an isolated singularity so that
$H=S^5/\Gamma$ is a manifold, the homology and cohomology of $H$ can be
computed directly with the results given in Table~6.  The computations
proceeds as follows:  first, the free part of
homology and cohomology is concentrated in degrees zero and five, since all
$p$-forms can be pulled back to $S^5$.  Next, since
$\pi_1(H)=\Gamma$, the first homology group  $H_1(H)$ is the Abelianization
$\Gamma/[\Gamma,\Gamma]$ of $\Gamma$, which we denote by $\Gamma'$.
Poincar\'e duality and the universal coefficient theorem then determine
$H^2$, $H_3$ and $H^4$.  Finally,  by the
Hurewicz theorem, $H_2(H)=0$, and it follows by Poincar\'e duality or the
universal coefficient theorem that $H^3(H)=0$ as well.

\midinsert
$$\vbox{\offinterlineskip\halign{
\strut # height 15pt depth 8pt& \hfill\quad$#$\quad\hfill\vrule&
\hfill\quad$#$\quad\hfill&
\hfill\quad$#$\quad\hfill&
\hfill\quad$#$\quad\hfill&
\hfill\quad$#$\quad\hfill&
\hfill\quad$#$\quad\hfill&
\hfill\quad$#$\quad\hfill\vrule\cr
\noalign{\hrule}
\vrule&i&0&1&2&3&4&5\cr
\noalign{\hrule}
\vrule&H^i(S^5/\Gamma) &\ZZ & 0 &  \Gamma'
& 0 & \Gamma' & \ZZ\cr
\noalign{\hrule}
\vrule&H_i(S^5/\Gamma)&\ZZ & \Gamma' & 0 & \Gamma'
&  0 & \ZZ\cr
\noalign{\hrule}
}}$$
\bigskip
\centerline{Table 6. Cohomology and homology of manifolds $S^5/\Gamma$.}
\bigskip
\endinsert

In the case that $\Gamma=\BZ_k$ is a cyclic group acting diagonally on the
coordinates $(x,y,z)$, there are natural
three-cycles which represent nontrivial classes in $H_3$ given by $x=0$,
$y=0$, and $z=0$.  Each is the image of some $S^3$ from $S^5$, itself
modded out by $\Gamma$.  Each of these is a minimum volume
representative of its homology class.  (The precise classes which are
represented depend on the action of $\BZ_k$.)

\bigbreak

\noindent{\it Conifold}\par\nobreak\medskip\nobreak

Turning next to the conifold, the space $H/U(1)$ is
$\Bbb{F}_0=\Bbb{C}\Bbb{P}^1\times \Bbb{C}\Bbb{P}^1$,
as already noted in section 2.
The K\"ahler--Einstein metric on this space is compatible with a toric
construction of the space; the simplest toric construction uses homogeneous
coordinates $x_0, x_1, y_0, y_1$ with a $U(1)^2$ action given by the charge
matrix
\eqn\fzerochg{
\pmatrix{x_0&x_1&y_0&y_1\cr1&1&0&0\cr0&0&1&1}.
}
To have equal-volume round metrics on the two-spheres, we must insist that the
two $D$-terms for the two $U(1)$'s be equal.

Line bundles over this space can be be described in toric geometry by including
a fifth homogeneous variable $p$.  The line bundle relevant to the conifold
has toric data
\eqn\conilbchg{
\pmatrix{x_0&x_1&y_0&y_1&p\cr1&1&0&0&-1\cr0&0&1&1&-1}.
}
This time we use $D$-terms of zero, in order to get a cone-type toric variety.
In order to more easily recognize this singularity, we can use a modified
basis for the $U(1)$'s, with charge matrix
\eqn\conilbchgbix{
\pmatrix{x_0&x_1&y_0&y_1&p\cr1&1&-1&-1&0\cr0&0&1&1&-1}.
}

The horizon will be the unit circle bundle inside the line bundle, defined by
$|p|^2=1$.  Thus, using this condition, we can describe the horizon by two
$D$-term equations
\eqn\conilbeqn{
|x_0|^2+|x_1|^2-1=0,\quad |y_0|^2+|y_1|^2-1=0,
}
and mod out by two $U(1)$'s.  One of the $U(1)$'s can be used to gauge-fix
$p$, setting it equal to $1$.  This leaves another $U(1)$, acting on the
remaining variables as
\eqn\conilbuoneaction{
(x_0,x_1,y_0,y_1)\mapsto
(e^{i\theta}x_0,e^{i\theta}x_1,e^{-i\theta}y_0,e^{-i\theta}y_1).
}
Thus we see a description of the horizon of the conifold in the form
$(S^3\times S^3)/U(1)$
with action \conilbuoneaction.

The $U(1)$ $R$-symmetry acts by rotating the phase of $p$ in the original
description; after our gauge fixing, and with an appropriate choice of
gauge for the remaining $U(1)$, it acts with charge $1$ on the complex
variables $x_0$, $x_1$, $y_0$, $y_1$.  The commutator of this $U(1)_R$ in
$Isom(S^3\times S^3)$ is $U(2)^2\rtimes \Bbb{Z}_2$ (with the $\Bbb{Z}_2$
acting by exchanging the factors).  Since the $U(1)$ given by
\conilbuoneaction\ is a normal subgroup of  $U(2)^2\rtimes \Bbb{Z}_2$,
we find that the supersymmetric isometries of the conifold horizon are
given by
\eqn\susyisomconi{
(U(2)^2\rtimes \Bbb{Z}_2)/U(1),
}
with the $U(1)$ action given by \conilbuoneaction.

The homology in this case is determined by the second Betti number $b=2$
of $H/U(1)=\Bbb{C}\Bbb{P}^1\times \Bbb{C}\Bbb{P}^1$, and the fact that the
cokernel of $\lambda$ is torsion-free (i.e., $m=1$).  Thus,
the nonzero homology groups are 
$H_0=H_2=H_3=H_5=\Bbb{Z}$.  (In fact, $H$
is topologically the product of $S^2$ and $S^3$ \CdO.)

Once again we can describe minimum-volume representatives for certain
generators of $H_3(H)$.  Representing $H$ as $(S^3\times S^3)/U(1)$, there are
two natural projections to $S^3/U(1)=S^2$ (the left factor and the right
factor).  Each gives a description of $H$ as being swept out by an $S^2$ of
minimum-volume $S^3$'s (of homology class $A_i$).  However, the two
families of $S^3$'s belong to 
{\it opposite}\/ homology classes, i.e., $A_1 + A_2 = 0$ in homology.

\bigbreak

\noindent{\it Complex cone over $\BF_0$}\par\nobreak\medskip\nobreak

The case of the complex cone over $\Bbb{F}_0$ has many
similar aspects to the conifold.
This time we use the anti-canonical bundle of $\Bbb{F}_0$ as ${\cal L}$,
whose toric data is 
\eqn\conefzerochg{
\pmatrix{x_0&x_1&y_0&y_1&p\cr1&1&0&0&-2\cr0&0&1&1&-2}.
}
The unit circle bundle is $|p|^2=1$, which modifies the $D$-term equations to
\eqn\conefzeroeqn{
|x_0|^2+|x_1|^2-1=0,\quad |y_0|^2+|y_1|^2-1=0,
}
as before.  This time, however, when we use one of the $U(1)$'s to gauge-fix
$p$,
there is a residual $\Bbb{Z}_2$ symmetry.  Thus, in addition to modding out
$S^3\times S^3$ by $U(1)$ as specified in \conilbuoneaction, we need to mod out
by an additional $\Bbb{Z}_2$, acting as
\eqn\conefzerozeetwoaction{
(x_0,x_1,y_0,y_1)\mapsto (-x_0,-x_1,y_0,y_1).
}
The horizon is thus $(S^3\times S^3)/(U(1)\times\Bbb{Z}_2)$.

Since $U(1)\times\Bbb{Z}_2$ is normal in $U(2)^2\rtimes\Bbb{Z}_2$, the
group of supersymmetric isometries is
\eqn\susyisomconefzero{
(U(2)^2\rtimes \Bbb{Z}_2)/(U(1)\times\Bbb{Z}_2),
}
with the $U(1)\times\Bbb{Z}_2$ action given by \conilbuoneaction\ and
\conefzerozeetwoaction. 

The homology in this case is again determined by the second Betti number $b=2$,
and the torsion ${\Bbb Z}_2$ in the cokernel of $\lambda$ (i.e., $m=2$).  Thus,
the nonzero homology groups are 
$H_0=H_2=H_5=\Bbb{Z}$, $H_1={\Bbb Z}_2$, and $H_3=\BZ\oplus {\Bbb Z}_2$.  
This horizon $H$ is topologically the quotient of
$S^2\times S^3$ by a freely acting ${\Bbb Z}_2$.  (So in particular,
$\pi_1(H)=\BZ_2$.) 

Again, there are minimum-volume representatives for certain classes in
$H_3$, consisting of two families
of three-spheres, such that their sum is the torsion element (i.e.,
$A_1+A_2$ represents the two-torsion in $H_3$).  The parameter
space for each of these families is a two-sphere.

\bigbreak

\noindent{\it Suspended pinch point}\par\nobreak\medskip\nobreak

Finally, in the case of the suspended pinch point, we can again make a
similar construction. 
We begin with a product of two copies of $\Bbb{W}\Bbb{C}\Bbb{P}^{1,2}$,
whose toric description is
\eqn\wcpchg{
\pmatrix{x_0&x_1&y_0&y_1\cr1&2&0&0\cr0&0&1&2}.
}
(As before, we keep the $D$-terms equal.)  The line bundle of interest is $1/3$
of the anti-canonical bundle, which has a toric description
\eqn\wcplbchg{
\pmatrix{x_0&x_1&y_0&y_1&p\cr1&2&0&0&-1\cr0&0&1&2&-1}.
}
(using zero for both $D$-terms).  In another basis for the $U(1)$'s, we find
\eqn\wcplbchgbis{
\pmatrix{x_0&x_1&y_0&y_1&p\cr1&2&-1&-2&0\cr0&0&1&2&-1}.
}
In this form, we easily recognize the toric data for the suspended pinch point.

To find the horizon, we again impose $|p|^2=1$ and find equations
\eqn\wcplbeqn{
|x_0|^2+2|x_1|^2-1=0,\quad |y_0|^2+2|y_1|^2-1=0.
}
We use one $U(1)$ to gauge fix $p$ to $1$, leaving a second $U(1)$ with action
\eqn\wcplbuoneaction{
(x_0,x_1,y_0,y_1)\mapsto
(e^{i\theta}x_0,e^{2i\theta}x_1,e^{-i\theta}y_0,e^{-2i\theta}y_1).
}
This $U(1)$ action is not regular: the orbits with $x_0=y_0=0$ are half the
size of
the generic orbits.  There is an $S^1$ of such orbits on the horizon manifold
itself.

So the horizon again has a description as $(S^3\times S^3)/U(1)$, but with the
$U(1)$-action
given by \wcplbuoneaction, which leads to singularities along an $S^1$.

To find the group of supersymmetric isometries in this case, note that the
normalizer in $U(2)^2\rtimes\Bbb{Z}_2$ of the $U(1)$ action
\wcplbuoneaction\ is $U(1)^4\rtimes\Bbb{Z}_2$; thus, the group of
supersymmetric isometries is
\eqn\susyisomspp{
(U(1)^4\rtimes\Bbb{Z}_2)/U(1)
}
with the $U(1)$ action given by \wcplbuoneaction.

Since this space has singularities, the usual homology might not be the right
thing to calculate.  We can, however, still study the parameter spaces for
minimum volume representatives of certain kinds of cycles.
As in the conifold case, there are two natural projections, in this case
projections  to $\BW\BC\BP^{1,2}$, which express $H^5$ as a family of
three-cycles.  However, in this case there is one three-cycle which is
distinguished as having smaller volume---the one which lies over the unique
orbifold point in $\BW\BC\BP^{1,2}$.  We thus use those two three-cycles as
minimum-volume representatives.  The corresponding classes $A_i$ satisfy
$A_1+A_2=0$ in the appropriate homology theory.

\newsec{An orbifold model}

We have conjectured that for each of the singularities discussed in the
previous section, and for each $N$, there should be a dual conformal field
theory (thought of as living on the boundary of $AdS_5$).  In order
to test this conjecture, we need to be able to identify 
the low-energy \CFT\ on the worldvolume, and perform computations there as
well as in the 
compactification on $AdS_5\times H^5$.  Our approach to the former
problem is based
on the construction \dm\ of the theory of branes at orbifolds
as an appropriate projection of the theory of branes on the covering space.
This method was used in \dgm\ to compute
the worldvolume theory for branes at an
orbifold singularity when the transverse space is locally described as
$\BC^3/\Gamma$.  
One of the useful products of this analysis, applied for example in
\refs{\brg,\mukray}, is that it yields a map between resolutions of the
singularity
and deformations of the field theory.  Using this we can find the
field-theoretic description of the partial resolutions of a given
quotient singularity.  As discussed in section {\it 4.2}, by studying
the case $\Gamma = \BZ_m\times\BZ_m$ this method in
principle allows us to treat all toric Gorenstein canonical
singularities.  Singularities of this type were studied from a
different point of view in \ami .  In this work we will focus on the
simplest such model, the case $\Gamma=\BZ_2\times\BZ_2$  and its partial
resolutions.  We will also discuss the complex cones over $\BC\BP^2$
and ${\Bbb F}_0$.

The \AdS /\CFT\ correspondence for D3-branes at quotient singularities in
three complex dimensions was studied in
\refs{\ks,\lnv,\bkv,\ot,\halyo,\bj,\gukov}.  From the point of view of
the \AdS\ compactification, the quotient acts on the $S^5$, leaving
the \AdS\ space unchanged.  Thus we find in these examples horizons of
the form $H^5=S^5/\Gamma$ as discussed in section {\it 4.3\/} above.
On the brane worldvolume, the prescription of \dgm\ leads for
$\Gamma\subset SU(3)$ to an $\CN{=}1$ model, with a gauge group which
is a product of $U(N)$ factors and chiral matter in bifundamental
representations.  The one-loop $\beta$-functions for the $SU(N)$
couplings in these theories vanish \refs{\lnv,\bkv,\bj}.  In the
large $N$ limit, these theories are in fact finite, as shown in the
latter two references.  Before launching into the computation for the
$\BZ_2\times\BZ_2$ model, we make some general remarks regarding
orbifold models.  The goal here is to set out our understanding of
these systematically with a view towards the generalizations we will
attempt in section 6.  Since the examples we discuss here have been
studied in detail in the references given, we will be brief in
presenting them and focus on those aspects which are most relevant to 
our generalizations.

\subsec{Orbifolds and the \AdS /\CFT\ correspondence}

For simplicity we initially frame our discussion for a cyclic group
$\Gamma=\BZ_k$ acting on three complex coordinates as $X^I\to
\omega^{a_I} X^I$ with $\omega$ a $k^{\rm th}$ root of unity. 
The worldvolume theory will be supersymmetric if $\Gamma\subset
SU(3)$, i.e. $\sum_I a_I = 0$.
The extension to other groups is not difficult.  The essential idea
behind the construction of \dm\ is that the theory of $N$ branes near the
singularity can be constructed by a projection from the theory of $Nk$
branes on the covering space $\BC^3$.  The resulting worldvolume
theory is then found to be an $\CN{=}1$ gauge theory with gauge group
$U(N)^k$ and chiral multiplets $X^I_{l,l+a_I}$ in bifundamental
representations $(\bk,\bkb)$ under $U(N)_l\times U(N)_{l+a_I}$ where
$l=1,\ldots k$.  These interact via a cubic superpotential that is
simply the restriction to these modes of the cubic superpotential of
the original $\CN{=}4$ theory.  Note that there are $3N$ fundamental
and anti-fundamental chiral multiplets under each $SU(N)$ factor, so
that the one-loop $\beta$-functions vanish.  Also, since all matter is
in bifundamental representations, the diagonal $U(1)$ subgroup
decouples completely from the theory and describes a free photon.  As
discussed in \refs{\mal,\Whol,\ofered} this photon is absent from the \AdS\
description.  We will ignore this issue in the sequel, but note that
in our theories there remain $k-1$ Abelian factors in the gauge group,
under which some of the massless matter is charged.  This means of
course that these decouple in the infrared and have no interesting
dynamics in the conformal limit.  In spite of this decoupling,
the fate of these gauge symmetries
is still of interest and will be discussed further below.

The closed-string sector of a IIB compactification on a quotient space
(or, in the low-energy limit, on a space with a quotient singularity)
bears some distinctive marks of its orbifold character.  Notable among
these are the existence of twisted sectors and a ``quantum'' symmetry
which acts on states in these sectors.  Twisted-sector states describe
strings which, in $\BC^3$, close only up to the action of an element
$g\in\Gamma$ (which labels the sector). The quantum symmetry can be
summarized geometrically as the statement that the identity of $g$ is
conserved by interactions (where for multi-string states the group
elements should be multiplied to obtain the total ``twist'').
Massless states in this sector arise from strings localized near
loci fixed by $g$.  In Calabi--Yau compactifications, the
massless states in twisted sectors of an orbifold singularity are known to
correspond to the moduli of the singularity.  Namely, for each
two-cycle in a resolution of the singularity we have a 
massless (${\cal N}=2$) hypermultiplet.
The four real scalars are given by the period of the
K\"ahler class about the two-cycle; the periods of the two two-form
fields $b_{NS}$ and $b_{RR}$; and the period of the four-form field
about the dual four-cycle. In terms of the $SU(2)_R$ symmetry these
form a complex doublet.  
There are additional light states in four dimensions associated to the
degenerating two-spheres.  D3-branes wrapping these two-spheres describe
strings 
in the four-dimensional theory.  These are BPS states, and the string
tension vanishes when the two-sphere degenerates to zero volume and
the corresponding two-form periods vanish as well. 

When the supersymmetry is broken to $\CN{=}1$ by the presence of the
transverse D3-branes (or in the \AdS\ picture by the curvature and
five-form flux) the hypermultiplet of twisted-sector
moduli breaks up into two chiral
multiplets with, in principle, different masses.  The two two-form
periods pair up to form the complex scalar $B = b_{RR}+ib_{NS}$ in one
chiral multiplet $B$, and the K\"ahler mode $r$ is complexified by the
four-form period $a$ to form the complex scalar $R = a + ir$
in a second chiral
multiplet $R$.  The couplings of this latter multiplet are
perhaps most naturally described if we dualize the real part; the
result is a {\it linear chiral multiplet}, containing a real scalar $r$
and a two-form potential $c$ (the integral of the self-dual RR
four-form over the two-cycle).  

\bigbreak

\noindent{\it Non-isolated singularities}\par\nobreak\medskip\nobreak

The discussion we wish to make will
depend upon the nature of the quotient singularity.  In particular,
quotient singularities can occur in codimension two (along complex
curves) or three (at points, the complement of the singular point
being smooth).
We begin with the case of a non-isolated singularity.
As an example for this case, we consider the $\BZ_k$ action
given by $\vec{a} = (1,-1,0)$.  The structure transverse to the fixed $Z$
axis is then $\BC^2/\BZ_k$.  Resolving this singularity leads to $k-1$
homology two-cycles $\Sigma_i$ with an intersection matrix given by
(minus) the 
Cartan matrix of $A_{k-1}$.  In the scaling limit, this
quotient leads to an $\CN{=}2$
theory with gauge group $U(N)^k$ (combining the vector multiplets of
$\CN{=}1$ with the projections of $Z$) and bifundamental
hypermultiplets transforming in the $(\bk,\bkb)$ representation of
$U(N)_l\times U(N)_{l+1}$ \dm .  The quantum $\BZ_k$ symmetry acts in
these models by cyclic permutations of the $k$ factors of the gauge
group, and corresponding permutations of the charged hypermultiplets. 
In general, the theory also has an $SU(2)\times U(1)$ $R$-symmetry.

The four-dimensional theory in this case can be obtained by
dimensional reduction of a six-dimensional theory, and the 
spectrum and couplings of the twisted sectors were
computed in \dm .  The couplings of twisted RR fields are computed by
an explicit worldsheet calculation following \douglas.  The couplings
of the other fields then follow by supersymmetry.  These calculations
show that the  scalar $a_j$ arising from the RR four-form in the $j$th
twisted sector has
nontrivial gauge transformation properties under the Abelian factors
in the gauge group
\eqn\atrans{
a_j\to a_j + \sum_{l=1}^k \omega^{jl}\epsilon_l\ .
}
Thus the $U(1)$ factors in the gauge group are broken, and the
low-energy gauge group is $SU(N)^k$.  The supersymmetric completion of
the coupling implied by \atrans\ shows that the metric modes $r_j$
parameterize the FI $D$-terms.  Note that the Abelian factors in the
gauge group (except for the trivial diagonal) are still present as
global symmetries in the worldvolume field theory.  We refer to
the conserved charges as baryon numbers under the various $SU(N)$
factors.  Thus the global symmetry group in general contains
$U(1)^{k-1}$ in addition to the $R$-symmetry.

The chiral primary operators charged under these symmetries are 
formed by antisymmetrizing the product of $N$ 
bifundamental fields on
both gauge indices.  These would actually be absent if the gauge
symmetry included the Abelian factors, so their existence verifies the
arguments above that the latter are broken.  These baryons, unlike the baryon
vertex studied in \witbaryons , represent actual particles in the
theory.  
Identifying the spectrum of baryons and identifying these states in
the dual \AdS\ model provides strong evidence for the conjecture; in
particular, because these states become infinitely massive (in \AdS )
or have divergent conformal weights (in the \CFT ) in the large $N$
limit, this comparison suggests that some aspects of the theories
agree at finite $N$.  In the model at hand, each of the bifundamental
hypermultiplets leads according to this prescription to a
hypermultiplet of baryon states.  The quantum symmetry
permutes these baryons as it does the bifundamental fields, and we can
construct a basis of linear combinations 
with charges corresponding to the twisted sectors by discrete Fourier
transform.  Thus, there is one baryon (and one antibaryon) state in each
sector, with conformal weight $N$ (which is determined by the $R$-charge, 
and is thus protected from quantum corrections).

The supersymmetric completion of the RR two-form coupling shows that
the two-form periods appear in the worldvolume action as
\eqn\bisg{
{1\over 8\pi}{\rm Im} \int d^2\theta \sum_{j,l}\omega^{jl} B_j
\tr(W_l^2)\ . 
}
Thus, the two-form periods enter the worldvolume theory as gauge
couplings and $\theta$ angles.  It is interesting in this context that
the two-form periods naturally parameterize a two-torus, since
$b_{NS}\sim b_{NS}+1$ while $b_{RR}\sim b_{RR} + \tau_S$ where $\tau_S
= {\chi\over 2\pi} + {2\pi i\over g}$ is the complexified string
coupling.  A striking confirmation of this identification was found in
\lnv , where this periodicity was shown to correspond to the known
$S$-duality group for an $\CN{=}2$ orbifold model.  The (bare) gauge
couplings are thus given by $\tau_l = \tau_S + \sum_j \omega^{jl}
b_j$.  The point $b_j=0$, which we interpret as the undeformed
quotient theory, will have all two-form periods equal to one-half
their maximal value \psa .  We define our variables so that this point
corresponds to $b_j=0$.  Note that the nonzero two-form periods give
the wrapped D3-brane strings mentioned above a tension of order
$\alpha'$ , so they are expected to be absent from the low-energy
theory.  Indeed, the model we have described has no candidate
tensionless strings, consistent with our interpretation of this point as
the {\it orbifold point}\/ at  which the CFT is nonsingular.

{}From the point of view of
the dual \AdS\ compactification, the horizon $H^5 = S^5/\Gamma$ is
singular in these cases, as explained in section {\it 4.3}.  The 
singularity occurs along the image of the circle $|Z|=1$, itself a
circle in the quotient (of $1/k$ the size), with the transverse space
described locally by $\BC^2/\BZ_k$.  A supergravity theory
compactified on this space would be singular, but as first pointed out
in \ks\ we expect the IIB theory to be well-behaved.  In particular,
we expect the same general structure of twisted sectors and twisted
moduli for the compactification.  The spectrum of states in the
untwisted sector can be found by projecting the known spectrum of the
$AdS _5\times S^5$ compactification to $\Gamma$-invariant states.
Restricting attention to the Kaluza--Klein states of the supergravity
theory reproduces the spectrum of chiral primary operators in the
untwisted sector of the worldvolume theory whose conformal dimensions
remain finite in the large $N$ limit \ot .
(In the field theory, untwisted states can be easily recognized
by their invariance under the global ``quantum'' symmetry).  The
twisted sector states were first studied in detail by Gukov \gukov .  The
background curvature and five-form flux change the spectrum in a
calculable way, and applying this correction Gukov was able to
reproduce the spectrum of chiral primary fields charged under the
quantum symmetry from the spectrum of twisted states in the \AdS\
compactification in $\CN{=}2$ orbifolds.  In particular, this analysis
demonstrates that the quantum symmetry is realized in these models as
the quantum symmetry associated to the quotient singularity in $H^5$.

The work of \refs{\ot,\gukov} did not include the baryons in the
spectra of the models. (These were, however, subsequently
discussed in \gubserklebanov .)  Along the lines described by Witten
\witbaryons , one expects that the baryons are represented by
D3-branes wrapped around three-cycles in $H^5$.
Witten considers related theories, and points out that a
D3-brane wrapping a nontrivial three-cycle in $H^5$ would lead to a
BPS particle with mass $m\sim V_3/g$.  Since $V_3\sim R^3$ with $R\sim
(gN)^{1/4}$, the conformal dimension of the operator in the
\CFT\ coupling to this field is
\eqn\bmass{
\Delta\sim mR\sim N
}
independent of $g$.  This is in accordance with the baryon-like
particles we predict.

Thus, we would like to find that the
horizon in this case has a homology group $H_3\supseteq\BZ^{k-1}$.
Since the horizon is singular, this is somewhat delicate to compute
explicitly and we give a heuristic description, motivated by the fact
that it yields physically reasonable answers.  As noted above, the
local structure transverse to the singular circle is $\BC^2/\BZ_k$.
This singularity, when resolved, leads to a chain of exceptional
${\BC\BP}^1$'s $\Sigma_i$ with an intersection matrix given by (minus) the
Cartan matrix of $A_{k-1}$.  We can construct a basis of three-cycles
dual to this, given by $C_j = \Sigma_j^*\times S^1$.  One can show that
these transform under the quantum $\BZ_k$ symmetry in all nontrivial
one-dimensional representations, in accord with our expectation that
we find one baryonic state in each twisted sector (the antibaryons are
of course represented by D3-branes wrapped with the opposite
orientation).  In the quotient
space, of course, all of these have vanishing volume and hence the
scaling argument above is suspect.  However, at the orbifold point we
have nonzero two-form periods about these cycles, and we expect this
to lead to a nonzero mass for these particles, scaling by dimensional
analysis as in \bmass .  Constructing the untwisted baryon state is
more difficult, since the supergravity approximation breaks down for
states as massive as this.  

These same three-cycles lead in the \AdS\ theory to additional gauge
symmetries, whose gauge fields 
$A_j = \int_{C_j} A^{(4)}$
are the reduction of the self-dual RR
four-form on the three-cycles.  The baryon states described above are
charged under these symmetries, which hence should 
correspond to the baryon number global symmetries in the \CFT .
The charges of
the baryons suggest that $A_j$ couples to $\sum_l \omega^{jl} J_l$
where $J_l$ is the worldvolume current for baryon number under the
$l^{\rm th}$ factor in the gauge group.  In section 4 we saw that the
$R$-symmetry of the \CFT\ is realized as a gauge symmetry in the \AdS\ 
compactification related to isometries of $H^5$.  In general, the
Abelian symmetry described here will complete the continuous global
symmetries of the worldvolume theory, so that we have a complete
correspondence.  In particular cases there may be an enhanced global
symmetry on the worldvolume.  As we will see in the examples we treat
in detail, this will be matched by enhanced isometry groups for the
horizon. 

The discrete global symmetries of the worldvolume theory will in
general include the quantum symmetry mentioned above.  This will not
act on the geometric horizon $H^5$, as we might expect for quantum
symmetries.  On the 
other hand, there will be an action on the cycles $C_j$ and
hence on the baryon spectrum, as discussed above.  There is one
additional discrete symmetry that will arise in all of these models. 
In the field theory this is implemented as charge conjugation,
exchanging the two chiral multiplets in a hypermultiplet.  The
interpretation of this in the \AdS\ model was discussed in \kw .
Charge conjugation reverses the orientation of the open strings, and
so corresponds to the action of the center of the $SL(2,\BZ)$ duality
group of the IIB theory.  This acts on the two-form fields as
$b_{NS}\to -b_{NS}$ and $b_{RR}\to -b_{RR}$.  It thus reverses the
sign of the two-form periods.  This symmetry is thus unbroken for
$B=0$, but because of the periodicity, also for the orbifold value
$B=\half(1+\tau_S)$, which describes the origin of our deformation
space, as described above.  

\bigbreak

\noindent{\it Isolated singularities}\par\nobreak\medskip\nobreak

We now turn to the case of isolated quotient singularities occurring in
codimension three.  We consider
the quotient $\BC^3/\BZ_k$ with $a_I$ all nonzero. (This is
isolated for most choices of the $a_I$ when $k$ is odd,
which we assume).  Resolving
this singularity leads to $(k-1)/2$ new homology two-cycles.  A new
feature is that the dual four-cycles are in this case localized near
the singularity.  The $k-1$ twisted sector fields are constrained by a
reality condition 
\eqn\real{
B_j = B^*_{k-j}\qquad\qquad R_j = R^*_{k-j}\ .
}
The construction of \dgm\ 
leads here to an $\CN{=}1$ theory with gauge group $U(N)^k$,
bifundamental chiral multiplets $X^I_{l,l+a_I}$ and a cubic
superpotential (the restriction to the surviving fields of the
cubic superpotential from 
${\cal N}=4$ super Yang--Mills theory). As always, the diagonal $U(1)$
decouples and is not included in the dual \AdS\ model.  Once more, the
quantum $\BZ_k$ symmetry acts by permuting the factors in the gauge
group and permuting the chiral multiplets as dictated by their
representation content.   In addition, the model has generically a
$U(1)^3$ global symmetry,\foot{The global symmetry group is larger if
the $a_I$'s are not all distinct.} the $I^{\rm th}$ factor of which 
acts by phases on all the fields descended
from $X^I$.  One combination of these is anomalous and can be combined
with the na\"\i ve $U(1)_R$ symmetry to form a non-anomalous
$U(1)'_R$.  

An important observation \IRU\ is that for these singularities the
worldvolume theory is chiral and the Abelian factors in the gauge
group as written above suffer from anomalies.
This can be easily seen in the example
above in which the $U(1)$ factors have mixed anomalies with the
nonabelian factors.  Under Abelian transformations with parameters
$\epsilon_l$ the effective action changes by 
\eqn\anom{
\sum_{ll'} {\cal A}_{ll'} \epsilon_l \,\tr(F_{l'}\wedge F_{l'})\ ,
}
with 
\eqn\anomaly{
{\cal A}_{ll'} = N(\delta_{l',l+a_I} - \delta_{l,l'-a_I})\ .
}
This is puzzling, since we have derived the model
in what seems like a consistent manner from an obviously well-defined
string theory.  The resolution of this puzzle was described in \IRU
.\foot{The mechanism described here is essentially an extension of the
ideas of \dsw .  These were applied in a similar context to
six-dimensional type I compactifications in \sixauthors.
} 
It requires an additional coupling (which can be expressed either
in terms of the two-form potentials $c_j$ or the scalars $a_j$)
\eqn\more{
{1\over k}\sum_{j=1}^{k-1} e_j\, c_j\wedge *\tr(\gamma^{-j}I_G) = 
{1\over k}\sum_{j=1}^{k-1} e_j \sum_{l=1}^k \omega^{-jl}
a_j\,\tr(F_l\wedge F_l)\ ,}
where $I_G$ is the Chern--Simons form for the gauge fields and $e_j$ is
a constant given by
\eqn\funny{e_j = \prod_{I=1}^3 2\sin (\pi j a_I)\ .
}
We are not aware of a direct calculation of this
coupling along the lines of \refs{\douglas,\dm} but it easy to
verify that it indeed cancels the anomaly \anom\ when we take into
account the transformation properties of $a_j$.  In \dm\ a version of
this argument was used to demonstrate the need for the Chern--Simons
couplings.  In the $\CN{=}2$ examples above $e_j=0$.\foot{In general, for
a non-isolated singularity leading to an $\CN{=}1$ theory
the coefficient $e_j$ corresponding to the subgroup
fixing the singular curve will vanish.}

As in the previous case, the Chern--Simons couplings show that the
$U(1)$ factors in the gauge group (apart from the diagonal) are
broken, and the $D$-terms are given by the background moduli as \dgm
\eqn\edgm{
\zeta_l = \sum_{j=1}^{k-1} \omega^{jl} R_j\ .
}
These are real by \real\ and sum to zero since we omit the $j=0$ term
in \edgm.  Note that in this case the baryon number symmetries are
absent as global symmetries as well.  From the point of view of the
field theory they are broken by anomalies.   (Note that the closely related
gauge symmetries which act as baryon number on the worldvolume but also act
on the RR fields are free of anomalies but, as discussed above, are broken
by RR expectation values.)

Once more, the absence of the Abelian factors predicts the presence of
baryons in the spectrum.  In general, there will be three families of
$k$ baryons arising from the projections of $X^I$, forming three
$k$-dimensional representations of $\BZ_k$.  We thus predict three
baryons in each sector.  Since the baryon number symmetries are broken
by anomalies, baryon number will not in general be conserved.   The
three baryonic states in each sector will however be distinguished by
their charges under the $U(1)^3$ symmetry mentioned above, since the
baryon formed as $(X^I)^N$ will have charge $N$ under the $I^{\rm th}$
$U(1)$. 

The \AdS\ dual theory is obtained \ks\ by letting the quotient group
act on the horizon.  
In this case, as described in section {\it 4.3}, since the singularity
is isolated the horizon is smooth.  
For the Abelian $\Gamma$ we are discussing here, the supersymmetric
isometries given by \susyisomorbi\ will extend the $R$-symmetry to at
least the $U(1)^3$ Cartan subgroup of $U(3)$, in agreement with the
continuous global symmetry of the generic quotient model as discussed
above.  In particular cases, this may be enhanced.  Note that in this isolated
case the self-dual RR four-form gives rise to no gauge fields in the
\AdS\ compactification because the homology is pure torsion; this
reflects the breaking by anomalies of the corresponding global
symmetries in the worldvolume theory.  
In addition, charge conjugation, together with the
permutation $i\to k-i$ on the gauge groups and the corresponding
permutation on the chiral matter multiplets, acts as a $\BZ_2$
symmetry.  In the \AdS\ theory this will as usual be realized as the
center of the $SL(2,\BZ)$ duality group.  Once more, this is a
symmetry of the orbifold theory with both two-form periods equal to
one-half their maximal value. In these models as well we find no trace
of the tensionless strings expected to arise when the two-form periods
vanish.  This is a somewhat stronger statement here than in the
non-isolated case.  Since in this case the strings would be
constrained by a potential to move on the brane worldvolume we would
expect them to represent fluctuating degrees of freedom in the field
theory. 

The projection to $\Gamma$-invariant states
of the supergravity spectrum on $S^5$ reproduces the chiral primary
untwisted states whose conformal weight remains finite at large $N$
in the worldvolume
theory \ot .  The origin of the twisted states is not so clear.
Because the group action is free, twisted states involve strings with
a minimal length of order $R$, so na\"\i vely these cannot correspond to
the charged vertex operators of low dimension present in the \CFT.  An
example of the latter would be
\eqn\puzl{
\sum_{l=1}^k \omega^{jl}\tr (W^2_l)\ ,
}
for $j=1,\ldots k$, with conformal dimension 2.  We do not know a
satisfactory resolution to this problem.

We can also consider the \AdS\ counterparts to the baryon states. 
We would like to describe these as
D3-branes wrapped on three-cycles in $H^5$; the only available
cycles are torsion classes.  In fact, as shown
in {\it 4.3\/} the relevant homology group is 
$H_3(H^5)=\BZ_k$.  The $k-1$ torsion cycles have three-volumes of
order $R^3$ so D3-branes wrapping them will lead to BPS particles with
mass as in \bmass .  Note that these torsion cycles do not lead to
additional gauge symmetry in the \AdS\ compactification, in agreement
with the fact that the corresponding global symmetries on the
worldvolume are broken by anomalies.

As discussed in {\it 4.3\/}, there are explicit geometric representatives
for some of these cycles.  Consider an $S^3\subset
S^5$ determined by setting $X^I=0$ for some $I$.  This (homologically
trivial) submanifold is preserved by $\Gamma$, so projects to a
submanifold in the quotient, given by $S^3/\Gamma$.  The triviality
upstairs, however, only demonstrates that the cycle so obtained is a
torsion cycle.  We thus find three representatives (labelled by $I$)
for each homology class in $H_3$, in agreement with the baryon
spectrum in the \CFT .  The three states are characterized by their
charges under the $U(1)^3$ symmetry. But the representation theory of
$U(1)^3\rtimes \BZ_k$ together with the fact that the charge under the
diagonal $U(1)$ is determined by the five-form flux will fix these to
agree with those found in the worldvolume theory. 
Note that the fact that
these are associated to D3-branes wrapping torsion cycles implies that
their number is conserved only mod $k$.  This is in accord with the
fact that baryon number is broken (note that the mod $k$ conservation
is not related to an unbroken subgroup of this but rather to the
quantum symmetry).

\subsec{Branes at a $\BZ_2\times\BZ_2$ singularity}

We now turn to a detailed study of a particular quotient singularity,
which we will use as a tool to generate new examples in section 6.
We first discuss the worldvolume theory, comparing in the next
subsection with the \AdS\ picture. 
We consider the theory of $N$ D3-branes at the origin of a
space locally described as $\BC^3/\BZ_2\times \BZ_2$.  We can take
the $\BZ_2\times \BZ_2$ action
\eqn\g{\eqalign{
(X,Y,Z) &\to (-X,-Y,Z)\cr
(X,Y,Z) &\to (-X,Y,-Z)\ .\cr
}}

Following \dgm\ we model $N$ branes near this by using $4N$ branes on
the covering space, leading to an $\CN{=}4$ theory with gauge group
$U(4N)$ on the worldvolume.  This is projected to the orbifold by
letting the discrete group act on Chan--Paton indices via the regular
representation in addition to its action on the spacetime indices.
The projection leaves an $\CN{=}1$ theory with gauge group $U(N)^4$
(of which the diagonal $U(1)$ decouples completely)
and the chiral multiplets surviving the projection are
\eqn\xs{\eqalign{
&X_{14}, X_{23}, X_{41}, X_{32}\cr
&Y_{13}, Y_{31}, Y_{24}, Y_{42}\cr
&Z_{12}, Z_{21}, Z_{34}, Z_{43}\cr
}}
where $X^I_{ij}$ transforms in the representation $(\bk,\bkb)$
of $U(N)_i\times U(N)_j$.
These interact via a superpotential descended from the $\CN{=}4$ theory
\eqn\w{\eqalign{
W &= \tr\Big(
Z_{12}(X_{23}Y_{31} - Y_{24}X_{41}) +
Z_{21}(X_{14}Y_{42} - Y_{13}X_{32}) + \cr
&\qquad Z_{34}(X_{41}Y_{13} - Y_{42}X_{23}) +
Z_{43}(X_{32}Y_{24} - Y_{31}X_{14})\Big)\ .\cr}}

As discussed above, the gauge symmetry is in fact broken to $SU(N)^4$.
The model has a non-anomalous global symmetry group
\eqn\globo{
G= \left( U(1)^3\times U(1)^3\rtimes S_4\right)\times \BZ_2\ .
}
The first $U(1)^3$ factor is the (effective) baryon number
symmetry (recall that the diagonal acts trivially).  As discussed above,
this arises from $RR$ symmetries in the \AdS\ model.
The second $U(1)^3$ factor acts by phases on the chiral multiplets,
and we can choose a basis of generators acting on chiral multiplets  
arising from one of the complex coordinates with charge one.

One linear combination of the latter three $U(1)$'s (the diagonal) is
anomalous, and combines with the anomalous $U(1)_R$ symmetry of the
model (under which all lowest components of superfields are fixed,
gluinos have charge $1$ and quarks charge $-1$) to a non-anomalous
$U(1)'_R$ subgroup of $G$, under which the squarks all have charge
$2/3$.  The $S_4$ permutes the four factors of the gauge group.  It
also acts on the twelve chiral multiplets, permuting these as dictated
by the index structure.  For example, the permutation (12) exchanges
$X_{14}$ with $Y_{24}$ and $X_{23}$ with $Y_{13}$ (as well as the
fields in the conjugate representations), and $Z_{12}$ with $Z_{21}$.
The superpotential is not invariant under $S_4$, so odd permutations
must be combined with a $U(1)_R$ rotation by $\pi\over 2$.\foot{A
potential anomaly in this symmetry, along with the reason for its
cancellation, was pointed out in \kw .  We are being somewhat
imprecise here; we restrict attention to the action on bosons.  Thus
we ignore the fact that $e^{i\pi R}=(-1)^F$ is not the identity.}
The
final $\BZ_2$ factor in $G$ acts on the vector multiplets by charge
conjugation and exchanges $X^I_{ij}$ with $X^I_{ji}$.  This too
changes the sign of $W$ so must be combined with a $U(1)_R$ rotation
by $\pi\over 2$. Thus, a $U(1)^5\rtimes A_4$ subgroup are not
$R$-symmetries.  The quantum symmetry discussed above is a
$\BZ_2\times\BZ_2$ subgroup of $G$.  As we shall see, this is
precisely the subgroup of $S_4$ which permutes the chiral fields in
\xs\ along rows only.  More abstractly, it is the kernel of the
natural map $S_4\to S_3$ giving the action on the spacetime indices
and its nontrivial elements are products of disjoint two-cycles.

The moduli space of classical vacua is given by the solutions to the
$F$-term equations following from \w\ and the $D$-term equations,
modulo $SU(N)^4$ gauge transformations.  It is parameterized
by the holomorphic gauge-invariants in the chiral superfields modulo
the Jacobian ideal of \w .  These
include meson-like operators constructed by
finding products which transform in the adjoint representation of some
$SU(N)$ factor and then taking traces of products of these.  Since we are
taking the trace, 
products which are related by cyclic permutations of the factors will lead to
identical invariants.  Modulo the equations of motion following from
\w, these invariants are generated by
\eqna\invnts{
$$\eqalignno{
\eqalign{
x_1 &= X_{14}X_{41}\cr
x_2 &= X_{23}X_{32}\cr}
\qquad &\eqalign{
y_1 &= Y_{13}Y_{31}\cr
y_2 &= Y_{24}Y_{42}\cr}
\qquad\eqalign{
z_1 &= Z_{12}Z_{21}\cr
z_2 &= Z_{34}Z_{43}\cr}&\invnts{}\cr
\noalign{and}
a &= X_{14}Y_{42}Z_{21}\ ,\cr}
$$}

\vskip-12pt

\noindent
where $a$ is the unique cubic adjoint in the sense that when traces
are taken all cubics are equal modulo the equations of motion.  In
addition, there are baryon-like operators formed, as discussed above,
by antisymmetrizing on both gauge indices a product of $N$
bifundamental fields.  This yields twelve baryonic invariants
$B_{ij}\sim X_{ij}^N$, subject to relations of the form (suppressing
numerical factors)
\eqn\brelns{
\eqalign{
B_{14} B_{41} &= x_1^N\cr
B_{23} B_{32} &= x_2^N\cr
}
\qquad\eqalign{
B_{13} B_{31} &= y_1^N\cr
B_{24} B_{42} &= y_2^N\cr
}
\qquad\eqalign{
B_{12} B_{21} &= z_1^N\cr
B_{34} B_{43} &= z_2^N\ .\cr
}}
Together with \invnts{} these give coordinates on the full moduli
space. 

To forge the connection to the construction of \dgm\ we note that the
baryonic $U(1)^3$ symmetry acts on the classical moduli space $\CM(N)$
preserving the symplectic structure. 
We can thus construct, if we wish, the symplectic reduction by this
group action (or equivalently a holomorphic quotient by the
complexified group action).  The symplectic reduction is determined by
a choice of the values of the moment map $\mu:\CM(N)\to {\Bbb R^3}$.
This reduction is equivalent to computing the
moduli space of a theory with the same matter content but a larger
gauge group $U(N)^4/U(1)$.  The components in ${\Bbb R^3}$ of the
moment map correspond to the values $\zeta_i$ of the three 
independent Fayet--Iliopoulos $D$-terms
which enter this latter construction.  
Thus, $\CM(N)$ fibers over $\Bbb R^3$ with the fiber over $\zeta$ 
being a $U(1)^3$ bundle over the reduced space 
$\CM_0(N;\zeta)$.  In view of our discussion above, the
transition from $\CM_0$ to the full moduli space can be reinterpreted
as reinstating the baryonic invariants.  We thus restrict attention to
\invnts{}, and notice that these commute (when traces are taken) using
\w , and satisfy  the relations
\eqn\rel{x_i y_j z_k = a^2\ .}
Note that if we treat the variables as scalars then the solution
of $xyz=a^2$ is $\CM_0(1;0) = \BC^3/\BZ_2\times\BZ_2$, precisely our
transverse space. 

The moduli space $\CM(N)$ has several branches.\foot{This phenomenon
was anticipated by Sardo Infirri \sardoinfirrione, who pointed out
that for non-isolated quotient singularities, the construction of
\sardoinfirrione\ (the one applied in \dgm) will yield a space
satisfying the $F$-flatness conditions which has several components.}
At generic (nonzero) $x,y,z$, the equations of motion imply that
$x^I_i=x^I_j$ and we can drop the subscripts.  Thus on this branch of
the moduli space the meson-like 
invariant traces parameterize the space of $S_N$ invariants of $N$
numbers (roots of the characteristic polynomial) satisfying \rel, i.e.,
parameterize the space $\CM_0(1;0)^N/S_N$ which we can identify with
$\CM_0(N;0)$.  The other level sets of $\mu$ (which together fill out the
full moduli space) will be $U(1)^3$ bundles over $\CM_0(N;\zeta)$
which---as is familiar from toric geometry---will be related to partial
resolutions of $\CM_0(N;0)$.  (The $\zeta$'s serve to label the blowup
moduli.)  In fact, given the structure of the
meson-like invariant traces we can identify $\CM_0(N;\zeta)$ with
$\CM_0(1;\zeta)^N/S_N$.  This is
precisely the configuration space for motion of $N$ branes on the
(partially resolved) space $\CM_0(1;\zeta)$.   At generic points on this
branch, the gauge symmetry is broken to $U(1)^{N-1}$, and the only
massless matter consists of $3(N-1)$ neutral chiral multiplets, with no
superpotential interaction.  Thus the model has an accidental
$\CN{=}4$ supersymmetry and is precisely the low-energy description of
the motion of $N$ branes at generic (smooth) points.

The other branches of 
the moduli space are found by setting, say, all
$y_i=z_i=0$.  Then the meson-like invariants restricted to this branch
are generated by the two
$x_i$ coordinates and parameterize a space  $\left(
\BC^N/S_N\right)^2$, meeting the previous branch along the diagonal
in the square.  Along this branch $y$ and $z$ are massive. The gauge
symmetry is broken to $U(1)^{2(N-1)}$.  The interpretation of this branch
following \refs{\pol,\dgm,\lnv,\ddg} is that the $N$ branes have split
into $2N$ ``fractional'' branes, each with one-half the tension of a
D3-brane.  These are in fact D5-branes wrapped about a vanishing cycle
(there is precisely one such cycle at generic points along the $x$
axis).  The two D5-branes wrap the two-cycle with opposite
orientations; the total six-form charge thus vanishes.  The D5-branes
do, however, acquire a four-form charge through the Chern--Simons
coupling, equal to 1/2 the charge of a fundamental D3-brane (and a
tension in four dimensions determined by this charge and the BPS
condition they satisfy).  Notice that these wrapped branes are not
able to move off in the $y$ and $z$ directions, in accordance with the
fact that $y$ and $z$ are massive along this branch.  Also, the
restriction on the $\zeta_i$ above is in line with our interpretation
of these as blowup moduli for the singularity.  Along the
``fractional'' branch the cycle wrapped by the D5-branes is
constrained to remain at zero size.  Of course, two
more such branches exist along the $y$ and the $z$ axes.  Also, there
are mixed branches in which some of the $N$ branes have split along
each of the axes and some have remained unsplit and hence free to move
off in any direction.  At the origin of moduli space the nature of the
objects is ambiguous, since they are free to move off on either
branch.  They are perhaps best described as bound states of D3-branes
and wrapped D5-branes \ddg .

{}From these pieces we can now assemble our total moduli space of
vacua, as a fibration over the space of $\zeta$ values.  At generic
values we find a $U(1)^3$ bundle over the space $\CM_0(N;\zeta)$ which is
the configuration space of $N$ points moving on the resolved
singularity with blowup parameters $\zeta$.  In codimension one,
whenever the sum of two $\zeta_i$ vanishes, the fiber has an
additional branch, meeting the previous branch along a complex curve
in the base of the $U(1)^3$ bundle.  Along this new branch we find a
$U(1)^2$ fibration\foot{Geometrically one of the $S^1$ directions in the
fiber is absent along this branch because the cycle about which it
measured the four-form period is absent.  More directly, when the
modulus $\zeta$ of the complex scalar vanishes its phase becomes
irrelevant.}  over
$(\BC^N/S_N)^2$.  The moduli space of vacua thus contains both the
moduli of the singularity and those describing motion of the branes
along it.

The utility of this indirect construction is the following.  We have
seen that for $N=1$ \invnts{} parameterize the transverse space.  
The toric methods of \dgm\ allow us
to pursue this identification further; we can in fact identify directly
a map from the FI parameters $\zeta_i$ appearing in the construction
and the blowup moduli of the singularity (see appendix B for details
of this as applied to this example, following \brg .)  
In this paper, our interest in this quotient singularity is in finding
the loci in its classical moduli space at which nonabelian gauge
groups are unbroken, and associating these to special partial
resolutions of the singularity which contain nontrivial singular
points.  We then identify these gauge theories as the worldvolume
theory for branes at the corresponding singularities.  The point is,
that in the toric construction one finds singularities for those
values of $\zeta_i$ for which the moduli space $\CM_0(1;\zeta)$ contains
a point at which some Abelian gauge group remains unbroken.  For these
values, $\CM_0(N;\zeta)$ will contain points with unbroken $U(N)$
symmetries.  In particular, this corresponds to a point in $\CM(N)$
with unbroken $SU(N)$ gauge symmetry.  This argument allows us to
forge a link between the toric analysis of $\CM_0(1,\zeta)$ using the
trick of \dgm\ and the classical moduli space of the field theory.
The worldvolume theory for branes at the singularities arising from
the partial resolution determined by some particular value of
$\zeta_i$ will be the low-energy theory determined by the
corresponding point in the moduli space of the theory we are treating
here. 

The model we have described is indeed a conformal field theory.
The one-loop $\beta$-functions for the $SU(N)$ couplings
all vanish, and the cubic superpotential is a marginal
deformation at weak coupling.  We can confirm this
result beyond weak coupling using the methods of \ls .  The existence of
marginal operators can be tested non-perturbatively by studying the
``scaling coefficients'' determining the $\beta$-functions for various
couplings in terms of the anomalous dimensions \refs{\sv,\svtwo}
.  Making some assumptions on the genericity of the latter
as functions of the couplings we can deduce the dimension of the critical
subspace in coupling space from linear dependencies among the
coefficients.

Eq.~\w\ is the unique superpotential
preserving the entire global symmetry group.  If we preserve the
symmetry, the model has only two independent couplings---the gauge
coupling and the superpotential coefficient.  These are written in
terms of the anomalous dimension of the chiral fields (when the
symmetry is unbroken all the dimensions are of course equal)
\eqn\as{\eqalign{
A_g &= -6N\gamma\cr
A_h &= 3\gamma/2\ .\cr
}}
Clearly, $\gamma=0$ leads to a conformal theory.
This imposes one condition on the two couplings
leaving a line of fixed points enjoying the full global symmetry.  We
can parameterize these theories by $\tau = {\theta\over 2\pi} +
{4\pi i\over g^2}$; the superpotential coupling will now be determined
by the requirement of conformal invariance and are not an additional
marginal coupling.

We can look for additional marginal operators by relaxing the symmetry
requirements.  If we require only that $U(1)_R$ and charge conjugation
(together with the exchange of $X^I_{ij}$ with $X^I_{ji}$) be
symmetries, we can parameterize the most general gauge-invariant
superpotential.  This has eight possible terms (all of which appear in
our $W$).  These pair up into four pairs related by charge conjugation
as above.  Together with the four gauge couplings we thus have eight
coefficients written as functions of the six independent anomalous
dimensions
\eqn\a{
\eqalign{
A_{g_1} &= -2N(\gamma_{12}+\gamma_{13}+\gamma_{14})\cr
A_{g_2} &= -2N(\gamma_{12}+\gamma_{23}+\gamma_{24})\cr
A_{g_3} &= -2N(\gamma_{13}+\gamma_{23}+\gamma_{34})\cr
A_{g_4} &= -2N(\gamma_{14}+\gamma_{24}+\gamma_{34})\cr}
\qquad
\eqalign{
A_{h_1} &= \half(\gamma_{12}+\gamma_{23}+\gamma_{13})\cr
A_{h_2} &= \half(\gamma_{12}+\gamma_{24}+\gamma_{14})\cr
A_{h_3} &= \half(\gamma_{13}+\gamma_{14}+\gamma_{34})\cr
A_{h_4} &= \half(\gamma_{23}+\gamma_{24}+\gamma_{34})\ ,\cr
}}
where $\gamma_{ij}$ is the anomalous dimension of $X^I_{ij}$ and
$X^I_{ji}$.  The notation is such that $S_4$ acts on the indices of
both $g$ and $h$.  These are easily seen to satisfy four
linear relations.  Thus, the vanishing of all of \a\ imposes only
four conditions on the six anomalous dimensions.  In terms of the eight
original couplings we thus predict a four (complex) dimensional space
of fixed points, of which a one-dimensional subspace preserves the full
$S_4$ symmetry, while at a symmetric point we have three truly marginal
deformations permuted by $S_3$.  The $S_4$ structure then guarantees
that these are charged under the quantum $\BZ_2\times\BZ_2$ symmetry
with charges $+-$, $-+$, and $--$.
We can parameterize the entire fixed space by the four gauge couplings
$\tau_i$, which are permuted by the action of $S_4$.  The symmetric
subspace on which all of the couplings are equal is parameterized by
$\tau_{++} = \sum\tau_i$, and we can form combinations with the
weights listed above
\eqn\taus{\eqalign{
\tau_{+-} &= \tau_1+\tau_2-\tau_3-\tau_4 \cr
\tau_{-+} &= \tau_1-\tau_2+\tau_3-\tau_4 \cr
\tau_{--} &= \tau_1-\tau_2-\tau_3+\tau_4 \ ,\cr
}}
where the subscripts indicate the charges with which these transform
under our chosen generators for the quantum symmetry, represented in
$S_4$ by
\eqn\gens{
S_1 = (12)(34)\qquad{\rm and}\qquad S_2 = (13)(24)\ .
}

For comparison with the $AdS$ model we mention one more property of
the model that is easily probed.  Chiral primary operators are given
by gauge-invariant polynomials in the chiral fields modulo
descendants; the latter are expressed as the Jacobian ideal of
$W$. For these operators the scaling dimension is determined by the
$R$-charge $\Delta={3\over 2} R$, hence protected from quantum
corrections (provided we use the correct $R$-symmetry).  Below we list
some of these operators.  For comparison to the $AdS$ picture, we also
give their transformation properties under the global symmetry group.
We label the operators by their charges under the continuous global
symmetry of \globo\ and under the 
quantum symmetry group.   We restrict
attention to scalars with conformal dimension $\Delta\le 4$, except
for the baryonic states which are not part of the towers of
operators constructed over the low-lying ones.

\item{1.} The kinetic energy operators for the four gauge fields $\tr
F_i^2$ are
marginal ($\Delta=4$) operators as discussed above.  They are permuted
by $S_4$ and are all invariant under $U(1)^3$.  Since they are
permuted by $S_4$ we can form one invariant combination, and three
others transforming in one-dimensional representations of the discrete
symmetry.  Explicitly, these are
\eqn\sectors{\eqalign{
f_{++} &= {1\over 4}\sum_{i=1}^4 \tr F_i^2\cr
f_{--} &= {1\over 4}\tr(F_1^2 - F_2^2 - F_3^2 + F_4^2)\cr
f_{-+} &= {1\over 4}\tr (F_1^2 - F_2^2 + F_3^2 - F_4^2)\cr
f_{+-} &= {1\over 4}\tr (F_1^2 + F_2^2 - F_3^2 - F_4^2)\ ,\cr
}}
coupling to the combinations of $\tau_i$ as in \taus . 
These couplings are distinguished, as found above, as being exactly
marginal.

\item{2.} The meson operators $\tr (X^I_{ij} X^I_{ji})$, with
$\Delta= 2$.  There are six operators of this type.  The
operators of charge (2,0,0) under $U(1)^3$ transform in a
two-dimensional representation of $S_4$. Forming linear combinations
we find
\eqn\xsq{\eqalign{
x_{++} &= \tr(X_{14}X_{41}+X_{23}X_{32})\cr
x_{--} &= \tr(X_{14}X_{41}-X_{23}X_{32})\ .
}}
Similarly, we find the operators $y_{++}$ and $y_{-+}$ with charge
(0,2,0) and $z_{++}$ and $z_{+-}$ with charge (0,0,2).

\item{3.} The operators $\tr(W_\alpha W^\alpha)_i$ with $\Delta = 3$.
These are permuted by $S_4$, so we form linear
combinations as in \sectors\ above.  These all have $R$-charge $2$ but
their $U(1)^3$ charges must follow from anomaly cancellation, and one
finds easily
\eqn\wcharges{
\eqalign{w_{++}\quad &(1,1,1)\cr
	w_{--}\quad &(3,0,0)\cr
}
\qquad
\eqalign{w_{-+}\quad &(0,3,0)\cr
	w_{+-}\quad &(0,0,3)\ .\cr
}
}

\item{4.} The operator $d = \tr(X_{ij}Y_{jk} Z_{ki})$, with dimension
$\Delta = 3$.  Note that modulo descendants there is precisely
one such operator.  It is invariant under the quantum symmetry and
carries charge (1,1,1).

\item{5.} The operators
$\tr(X^I_{ij}X^J_{jk}X^I_{kl}X^J_{li})$, with $\Delta =
4$.  Modulo descendants, there is one operator of type $++$ for each
pair of superscripts. To form nontrivial representations of the
discrete group we find that we must have $I=J$ and thus we have
\eqn\xfourth{\eqalign{
q_{++}^{IJ}\quad
&(4,0,0),(0,4,0),(0,0,4),(2,2,0),(2,0,2),(0,2,2)\cr
q_{+-}\quad &(0,0,4)\cr
q_{-+}\quad &(0,4,0)\cr
q_{--}\quad &(4,0,0)\ .\cr
}}

\item{6.} Finally, because the gauge group is $SU(N)^4$, we can
construct baryon operators of the form
\eqn\bars{
B_{ij} = \epsilon^{a_1\cdots a_N}\epsilon_{b_1\cdots b_N}
\left(X^I_{ij}\right)_{a_1}^{b_1}\cdots
\left(X^I_{ij}\right)_{a_N}^{b_N}\ ,
}
with $I$ determined by $ij$.
The $S_4$ transformation properties, like those of the $X^I$, are
determined by the index structure.  Thus, for each $I$ we can form
two baryons and two antibaryons. Since baryon number is broken in this
theory the latter term requires clarification.  It is appropriate in
the sense that a baryon-antibaryon state can decay to mesonic states. 
Constructing linear combinations
with definite charges under the quantum symmetry, we find 
in each sector a baryon and an antibaryon.  In the twisted sectors
these have charges $(N,0,0)_{--}$, $(0,N,0)_{-+}$, and
$(0,0,N)_{+-}$.  In the untwisted sector we find all three charge
configurations. 

\subsec{The \AdS\ compactification}

The dual \AdS\ theory is found following \ks .  We utilize the
construction of \dgm\ once more, constructing the theory as a quotient
of the theory with $4N$ branes.  But in this case, we first use the
original \AdS /\CFT\ duality to map this latter to a IIB
compactification on $AdS_5\times S^5$.  We then implement the quotient
in this dual theory.  It is clear that since we are taking the
quotient by a subgroup of $SO(6)$, this acts purely on the $S^5$,
leaving the \AdS\ space untouched.  Thus the resulting theory is indeed
of the form $AdS_5\times H^5$, predicting a conformal field theory on
the branes.
The horizon is in this case simply
\eqn\h{H^5 = S^5/(\BZ_2\times\BZ_2)\ ,}
which is singular along three circles.  As discussed in {\it 5.1}, we
expect to find three twisted sectors corresponding to these fixed loci.

In section 4, the group of supersymmetric isometries of the quotient
was computed to be $U(1)^3\rtimes S_3$. The
$U(1)$ factors correspond to rotation by a phase of each of the three
complex coordinates, while the $S_3$ permutes these.  Each of the
generators of the continuous group acts as a rotation on one of the
fixed circles; the discrete group permutes the three. These are gauge
symmetries in the \AdS\ compactification and will couple to the global
symmetry currents in the boundary \CFT .  We can ask which of these are
$R$-symmetries.   To answer this, note that our one supersymmetry
generator, 
out of the four in the covering theory, is the invariant spinor in the
decomposition of the $SO(6)$ spinor representation under $SU(3)$
${\bf 4} = {\bf 3} \oplus {\bf 1}$.  The $R$-symmetries are then those
transformations which act (by phases) on this representation.  This
immediately shows that the diagonal $U(1)$ and all odd permutations
in $S_3$ are $R$-symmetries, and indeed predicts the correct action
on the supercharges.

Thus the isometry group of $H^5$ contains the $R$-symmetry of the
\CFT\ as shown in general in section 3,
as well as a part of its global symmetry, but not all of the latter. 
First of all, the baryonic symmetry is not realized as an
isometry.  As in our general discussion of quotient singularities,
this corresponds to a gauge symmetry in the \AdS\ theory carried by
periods of the RR four-form about three-cycles in $H^5$.  These are
constructed along the lines of the discussion for non-isolated
singularities.  The structure of $H^5$ transverse to each of the three
singular circles is $\BC^2/\BZ_2$, containing in its resolution a
two-cycle $\Sigma_i$.   These lead to nontrivial three-cycles in $H^5$ 
of the form $C_i = \Sigma_i\times S^1_i$.  The periods of the RR
four-form about these yield the gauge fields for a $U(1)^3$ symmetry
corresponding precisely to the baryon number symmetry in the
worldvolume, as in our general discussion of quotients above.

Second,there are still discrete group actions we have not accounted for
by isometries.  The
first of these, as expected, is the quantum $\BZ_2\times\BZ_2$ action
under which the untwisted sector states are uncharged, while
the twisted sector states have charges $(--)$, $(-+)$, and $(+-)$.  It
is gratifying to find that the extension of $S_3$ by this action is
precisely $S_4$, as discussed above.  Having now found this, we now
note that if we find three objects permuted by $S_3$ then their
charges under the quantum symmetry must be the ones listed above,
since this is the only situation that arises from $S_4$ representations.
The
additional $\BZ_2$ associated to charge conjugation is also not evident as
an isometry.  Charge conjugation is associated to orientation reversal
for the open strings in the theory, and this symmetry is mapped under
the correspondence to the action of the center of the $SL(2,\BZ)$
duality group of IIB string theory.  It thus commutes with all
other actions.  As shown in \kw, this action must be accompanied by a
nontrivial $R$-transformation.  Thus the \AdS\ picture reproduces
exactly the global symmetry group found in the \CFT\ description.

Having matched the global symmetries we move on to compare the moduli
of the two theories (in the sense of exactly marginal couplings).  The
type IIB compactification always has a universal marginal coupling,
the string coupling $\tau_S = {\chi\over 2\pi} + {2\pi i\over g}$,
where $\chi$ is the RR scalar.  In a
compactification on $AdS_5\times S^5$ this is the only modulus, and
the gauge coupling is simply given by the scaling argument as
$\tau_{++} =
\tau_S$, determined by the asymptotic value of the dilaton far from
the branes.  Extra moduli in this example must thus come from the twisted
sectors, and indeed there are three of these.  Each has the
structure (transverse to the fixed circle) $\BC^2/\BZ_2$ which will,
according to our general discussion, produce precisely one marginal
deformation, given by the periods of the two-form potentials about
the one two-cycle in the resolution of this singularity.   We label
these moduli $b_{\pm\pm}$ according to their sectors;
the global symmetry and the scaling argument then determine
\eqn\map{\eqalign{
\tau_{++} &= \tau_S\cr
e^{2\pi i\tau_{\pm\pm}} &= f(e^{2\pi ib_{\pm\pm}})\ ,\cr
}}
where in the second line the combination $++$ does not appear.
We note that the couplings $b$ here should be interpreted as the
deviation of the periods from their values at the $\BZ_2\times\BZ_2$
invariant orbifold point, i.e. 1/2.  One interesting feature of this
mapping is that the periods naturally parameterize a two-torus.  This
periodicity in principle predicts that the ${\cal N}=1$ theory that 
we are studying has an $S$-duality! 
Extracting a precise form of this prediction would require knowledge of
the function $f$ above, however.  In \lnv\ this predicted type of duality was
shown to agree with the known $S$-duality of $\CN{=}2$
orbifold models.  

As a final check of the conjecture we compare the spectrum of
conformal dimensions in the \CFT\ to the mass spectrum in the \AdS\
model.  In this paper, we restrict attention to relevant and marginal
scalars.  A scalar field of conformal dimension $\Delta$ on the
boundary couples \Whol\ to a scalar field of mass $m^2 =
\Delta(\Delta-4)$.  In the untwisted sector the matching spectra are
no surprise, since fundamentally both follow by projection from the
spectra in the covering theory, and the latter are known to agree.
The twisted-sector fields can be studied using the
methods of \refs{\gukov,\AFM}.  The twisted sector fields propagate in
six dimensions along $AdS_5\times S^1$.  Precisely this situation was
studied in \gukov. (That this was done there in the context of
$\CN{=}2$ models is irrelevant; in the twisted sector as formulated
here no further projection is required and low-lying twisted states
are sensitive only to the local properties of the space.)  The
spectrum of chiral states is found by Kaluza--Klein reduction on $S^1$
of the fields in the well-known six-dimensional theory associated to a
$\BC^2/\BZ_2$ singularity, modified by the presence of curvature
and five-form flux.  The resulting five-dimensional spectrum
corresponds to $\CN{=}2$ supersymmetry in the four-dimensional \CFT .
The states are thus labeled by their charges under the $SU(2)\times
U(1)$ $R$-symmetry, where the $SU(2)$ is the six-dimensional $R$-symmetry
and the $U(1)$ represents rotations of the fixed circle.  In 
our case, considering, say, the twisted sector associated to $y=z=0$
(which we previously denoted the $--$ sector) we will find that since
$Y,Z$ are massive the light states will all be invariant under two of
the $U(1)^3$ generators.  Thus there is only one $U(1)$ action in this
sector, which will be a linear combination of $J_{S^1}$ and $J_3$.  We
identify
\eqn\gukid{ J_R = {4\over 3} J_3 + {1\over 3} J_{S^1}\ .}
Thus the $U(1)^3$ charges of all the states in this sector will be of
the form $({3\over 2} R,0,0)$.  Note that all the twisted states predicted
above indeed have charges of this form.

Gukov finds the following light scalars in the twisted sector.
There are fields with $m^2 = -4$ in the ${\bf 3}_0$.  Using \gukid ,
we see that only one of these has the right $R$-charge, and thus
we find the field coupling to $x_{--}$. There is also a
triplet of $m^2 = -3$ transforming as ${\bf 3}_2$.  Here again only
one field has the correct $R$-charge and couples to $w_{--}$ of
\wcharges .  There is yet another set at $m^2=0$, transforming as
${\bf 3}_4$.  This too leads to a unique chiral primary coupling to
$q_{--}$ of \xfourth .  All of these come from the multiplet of blowup
modes.  From the two-form period multiplet, we find one marginal
field, a singlet under the global symmetry.  We conjecture that this
couples to the exactly marginal operator $f_{--}$, in accordance with
our discussion of the moduli above.

Finally, we consider the spectrum of wrapped D3-branes in this
model.  As discussed in {\it 5.1}, these will correspond to the
twisted baryonic operators.  The three-cycles about which we propose
to wrap D3-branes will be the $C_i$ mentioned above.  The
antibaryons will be obtained by wrapping $-\Sigma_i^*\times S^1_i$.
In each sector, according to the discussion above, the charges are
restricted to a one-dimensional subgroup of $U(1)^3$ and the total
charge can be determined from the mass as predicted by \bmass\ above.
The charges under the baryon number symmetry follow from the
intersection between the two- and three-cycles in $H^5$.  This yields
precise agreement with the computation in the \CFT .

\newsec{New examples in four dimensions}

Our choice of the $\BZ_2\times\BZ_2$ quotient construction in the previous
section was 
motivated by the fact that several interesting singularities can be
found by considering partial resolutions of this particular
quotient.  This was pointed
out recently in \refs{\brg,\mukray}, where the methods of \dgm\ were
used to show
that the moduli space of deformations included conifold transitions.
Note that \dgm\ studies the case of a single brane ($N=1$).  In this
case, according to our discussion in section 5,
the gauge symmetry will be completely broken. Nevertheless it is
possible to study the moduli space as a fibration of torus bundles over 
 moduli spaces of a family of auxiliary Abelian gauge theories (obtained
by gauging the baryon number symmetries), parameterized by the space of FI
terms.  These auxiliary moduli spaces have descriptions as toric varieties
\dgm.  This is a nontrivial result as it requires that the $F$-term equations
be expressed in terms of toric data.  We have seen that the fiber thus
described is in fact a neighborhood of the singularity in the internal
manifold.  In a sense, the baryonic degrees of freedom ignored in this
construction represent the moduli of the singularity, while the
meson-like $U(N)$ invariants parameterize the positions of the branes.
Thus this description leads directly to a description of the horizon
geometry.  Also, the K\"ahler class of the moduli space is explicitly
parameterized by the FI terms introduced.  This enables a
direct determination of loci in the moduli space at which the
low-energy theory is nontrivial, as well as the corresponding
singularities in the partial resolutions of the orbifold. 

The way this works in practice is the following.  By choosing values
for the parameters $\zeta_i$ of the previous section we select a point
in the moduli space of the field theory.  We can then study the
low-energy theory about this vacuum.  At generic points, the result as
mentioned above is the $\CN{=}4$ theory of branes at smooth points.
There will be special values of $\zeta$ which lead to different
low-energy limits.  In the \AdS\ picture we need to recall that
low-energy physics on the worldvolume is related to small-distance
physics in $AdS_5$ \refs{\mal,\SusskindWitten}.  Thus, the flow to the
low-energy limit corresponds to shrinking the horizon manifold.  We
thus find that the special points in moduli space are precisely those
for which a small neighborhood is non-spherical---those points
corresponding to locating the branes at singularities of the
partially-resolved space.  The latter were described in section 4
(and appendix B)
following \dgm , and in this section we will use this tool to 
investigate the conjectures for these singularities.

This makes possible a direct computation of the low-energy theory.
The horizon manifolds were described in section 4.  What checks can
we perform on the conjectures for these examples?  The first test is a
comparison of the global symmetries of the field theory to the
isometry groups found geometrically.  As we have seen above, in
general there will be other sources for global symmetries.  The
baryonic symmetries will correspond to the periods of the four-form.
In orbifold models there will also be discrete quantum symmetries.
Finally, our derivation will lead to models in which the two-form
periods are one-half their maximum value, thus preserving the $\BZ_2$
center of $SL(2,\BZ)$ which will appear in the field theory as a
charge-conjugation symmetry. 

As a second check we can try to compare the spectrum of chiral primary
operators in the \CFT\ to the spectrum of states in the \AdS\
compactification.  In general, the former is an easy exercise but the
latter is difficult for the horizon manifolds in question.  What we
will find relatively easy to describe is the spectrum of wrapped
D3-brane states corresponding to baryons.  Requiring only an
understanding of the topology of $H^5$ and some assumptions on the
form of the metric (and hence on the parameter spaces of minimal-volume
representatives of certain homology cycles which we have already determined
in section 4.3), this will in fact be a
tractable problem and we will make this comparison for all the models
we study.

Finally, we can compare marginal operators in the field
theory to moduli of the \AdS\ model.  Here we will find an extension
of the pattern we observed in the quotient models.   For small
resolutions (blowups in which each singular point is replaced by a curve)
there will be one marginal operator, corresponding to the two-form
periods, and the blowup parameter will appear as an FI parameter (more
precisely as the baryonic mode corresponding to this).
For a large blowup (in which a point is replaced by a surface) we
will find one marginal operator again, given by two-form periods, and
associated to two FI terms subject to a reality condition like \real. 

We turn now to the four singularities discussed in section 4.3; for
each example we compute the worldvolume theory at low energies and
compare to the \AdS\ predictions. 

\subsec{The $\BZ_2$ quotient singularity}

The toric analysis shows that setting, say, $\zeta_1 +\zeta_2=0$
corresponds to leaving one of the $\BZ_2$ singularities
unresolved. In the scaling limit we thus expect to find the
$\CN{=}2$ theory associated to the quotient $\BC^2/\BZ_2$.

The most symmetric vacuum in this level set of the moment map is
represented by
\eqn\vevo{\eqalign{
Z_{12} &= \zeta_1^{1/2}\cr
Z_{43} &= \zeta_4^{1/2}\ ,\cr}}
all others vanishing.  The gauge symmetry is broken to $SU(N)\times
SU(N)$.
The fields in \vevo\ are eaten by the Higgs mechanism, while \w\ shows
that $Y_{24}$ and $Y_{31}$, say, get $F$-term masses.  The equations of
motion follow from \w\
\eqn\masseso{\eqalign{
\zeta_1^{1/2} X_{23} &= \zeta_4^{1/2} X_{14}\cr
\zeta_1^{1/2} X_{41} &= \zeta_4^{1/2} X_{32}\ .\cr
}}
Inserting these and integrating out the massive fields, we find that
the light fields and their charges can be taken as\foot{Note that
$\bk$ is the fundamental 
representation and $\bkb$ its complex conjugate; $\bf N^2-1$ is the adjoint
representation.}
\eqn\lighto{
\eqalign{
x_1 &= X_{14} \quad (\bk,\bkb)\cr
\tilde x_1 &= Y_{42} \quad (\bkb,\bk)\cr}
\qquad
\eqalign{
x_2 &= Y_{13} \quad (\bk,\bkb)\cr
\tilde x_2 &= X_{32} \quad (\bkb,\bk)\cr}
\qquad
\eqalign{
\phi_1 &= Z_{21} \quad ({\bf N^2-1},{\bf 1})\cr
\phi_2 &= Z_{34} \quad ({\bf 1},{\bf N^2-1})\ ,\cr
}}
and plugging \masseso\ in we find the superpotential
\eqn\wo{ W = \zeta^{-1/2}\tr\left( (\zeta_1^{1/2}\phi_1-\zeta_4^{1/2}\phi_2)
(x_1\tilde x_1 - x_2\tilde x_2)\right) \ ,}
where $\zeta = \zeta_1 + \zeta_4$.  Note that the trace
of $Z$ is given a mass by the supersymmetric completion of
the couplings which break the Abelian gauge symmetry.

The theory is observed to be in fact precisely an $\CN{=}2$
theory with the gauge group above and two hypermultiplets in the
$(\bk,\bkb)$ representation as expected.  These cases have been
studied in detail in \refs{\ks,\lnv,\gukov}.

As in the previous section, we find a Higgs branch of the moduli
parameterizing motion of the branes in the transverse directions, a
Coulomb branch describing ``fractional'' branes constrained to move
along the singular curve, and the expected mixed branches.

The dual IIB compactification on $AdS_5\times (S^5/\BZ_2)$ has been
studied in detail in, e.g., \refs{\ks,\lnv,\ot,\gukov}.  These works
have provided evidence for the conjecture in this context, essentially
along the lines of our discussion of the $\BZ_2\times\BZ_2$ quotient.

We can add here only that we expect baryons in this theory in
accordance with our discussion in {\it 5.1}; the corresponding
three-cycles arise as discussed there. 

\subsec{The conifold}

As discussed in \refs{\brg,\mukray} there is a different
codimension-one locus in the space of FI parameters at which we obtain
a conifold singularity.  This occurs at, say, $\zeta_1 = 0$.
To find the worldvolume theory, we take $\zeta_2,\zeta_3\gg 0$,
maintaining $\zeta_1=0$.  The most symmetric vacuum solution in this
level set is represented by the expectation values
\eqn\vevs{
\eqalign{
Y_{24} &= \zeta_2^{1/2}\cr
Z_{34} &= \zeta_3^{1/2}\ ,\cr}}
all others zero.  This breaks the gauge group down to $SU(N)\times
SU(N)$.  The Higgs mechanism eats the components of the multiplets with
nonzero vevs.  Inserting these expectation values into the
superpotential leads to masses for some of the other fields, which are
then integrated out by imposing their equations of motion.  This
leaves the light fields and their representation content as
\eqn\lighti{
\eqalign{
X_1 &= X_{14}\quad (\bk,\bkb)\cr
X_2 &= Z_{12}\quad (\bk,\bkb)\cr}
\qquad
\eqalign{
Y_1 &= Y_{31}\quad (\bkb,\bk)\cr Y_2 &= Z_{21}\quad (\bkb,\bk)\ ,\cr
}}
where $Z_{21}$ represents the remaining massless degree of freedom
after this field mixes with $Y_{13}$.  Imposing the equations of
motion for the massive fields we are left with the superpotential
\eqn\wi{
W = \half \zeta^{-1/2}
\tr\left( \epsilon^{ij}\epsilon^{kl}X_i Y_k X_j Y_l\right)\ ,
}
with $\zeta = \zeta_2 + \zeta_3$.
Note that for $N=1$ we obtain precisely the charges and fields of
\conilbuoneaction .  This will not be true in general, since the
singularity is in general obtained at the origin of a moduli space
determined by the gauge symmetry as well as the superpotential.  It is
a coincidence in this case that for $N=1$ \wi\ in fact vanishes. 

This description agrees precisely with the one found independently in \kw\
by different methods.  Our approach yields an explicit derivation of this
field theory, including the superpotential proposed in \kw.

The model has a global symmetry group $U(2)\times U(2)$, 
under which the fields
transform as 
\eqn\reps{
(\bk,\bkb,{\bf 2},{\bf 1})\oplus (\bkb,\bk,{\bf 1},{\bf 2})\ .
} 
The diagonal $U(1)$ in this is in fact anomalous, but this can be
combined with the na\"\i ve $U(1)_R$ symmetry to yield a non-anomalous
$U(1)'_R$, under which all the lowest components of chiral superfields
have charge $1/2$.  The other  
$U(1)$ generator generates a non-$R$ baryonic symmetry, under which
$X$ and $Y$ have opposite charges $\pm 1$.  
There is an additional discrete $\BZ_2\times\BZ_2$
symmetry.  We can take the generators to act as follows.  One
exchanges the two factors of the gauge group and acts on the chiral
fields by exchanging $X$ and $Y$, and the other acts in the same way
on the chiral multiplets but acts on the vector multiplets by charge
conjugation.  (Their product thus acts by exchanging the factors in
the gauge group and charge conjugation, leaving the lowest components
of the chiral multiplets untouched).  Note that the first of these is
in fact an $R$-symmetry, since the superpotential changes sign.
The global symmetry group is thus
\eqn\globali{
G = U(2)^2\rtimes (\BZ_2\times\BZ_2)
\ .}
Comparing this to \susyisomconi\ we see that as expected the baryonic
symmetry as well as the discrete symmetry related to charge
conjugation are not realized as isometries.  The former corresponds to
the four-form periods about the generator of $H_3$ and the latter to
the center of $SL(2,\BZ)$ as in our general discussion.

The classical moduli space of vacua can be found here as above.  The
invariants are given by traces of the following four generators 
\eqn\invntsi{a_{ij} = X_i Y_j\ ,}
which inside traces all commute (using the equations of motion
following from \wi ) and are subject to the relation
\eqn\relni{
a_{12}a_{21} = a_{11} a_{22}\ .
}
To these we must add the baryonic invariants
\eqn\barsi{\eqalign{
B_s &\sim X_1^s X_2^{N-s}\cr
\tilde B_s &\sim Y_1^s Y_2^{N-s}\ ,\cr
}}
where both gauge indices are antisymmetrized, so that the flavor
indices are symmetrized.  These transform in the spin-$N/2$
representation of the two $SU(2)$ global symmetry groups.
They satisfy relations of the form 
\eqn\barlnsi{
B_s\tilde B_t \sim a_{11}^s a_{12}^{t-s} a_{22}^{N-t}
}
for $s\le t$. 

As in the quotient singularity of section five, 
the meson-like invariant traces \invntsi\ parameterize
the space of $S_N$ invariants in $N$ solutions of \relni .
Comparing to the third line of Table 4, we see that this is precisely
the $N^{\rm th}$ symmetric product of the
conifold singularity, as expected.  Including the baryonic invariants
\barsi\ will, it is expected, lead to a classical moduli space
fibered over the space of moduli of the singularity, with the fibers
describing the locations of $N$ branes in the resolved space. 
The gauge symmetry is generically
broken to $U(1)^{N-1}$.  There are no ``fractional'' branches in this
model.  We do not expect such branches since the singularity is
isolated.  

We can identify the baryons in the spectrum of the \CFT\ with
particles in the \AdS\ compactification.   As discussed above, these
correspond to D3-branes wrapped about three-cycles in $H^5$.  The
discussion of {\it 4.3\/} shows that this is topologically just
$S^2\times S^3$, so there is a natural identification of the
three-cycle.  In fact, there are two distinct ways to write $H^5$ as
such a product, corresponding to the two $\BC\BP^1$ components of the
base of the $S^1$ fibration.  These correspond to $B$ and $\tilde B$
of \barsi .  The space of three-cycles in either
class is simply $S^2$, as discussed in section 4.3.  The wavefunctions on
this space will be 
sections of a line bundle of degree $N$ as determined by the five-form
flux through $H^5$ \witbaryons .  
Under the $U(1)_B\times SU(2)\times SU(2)$
isometry group, this will lead to spins $(0,N/2)_N$ and
$(N/2,0)_{-N}$ in exact agreement with the predictions from the \CFT .

The gauge theory with no superpotential is known to flow in the
infrared to an interacting conformal fixed point (the nonabelian
Coulomb phase).  To analyze the situation in the presence of the
superpotential we can use once more the methods of \ls .  If we
preserve the full global symmetry mentioned above, then the scaling
coefficients for the two $SU(N)$ gauge couplings 
are identical; likewise, the anomalous dimensions for
all the fields are the same.  The superpotential is the unique one
preserving the symmetry.  The scaling coefficients are thus
\eqn\ais{\eqalign{
A_g &= -2N(1+2\gamma)\cr
A_h &= 1 + 2\gamma\ .\cr }}
Since they are
proportional, we expect a (complex) line of fixed points.  Note that
in this case $\gamma=0$ is {\it not\/} a solution.  The fixed points
are all interacting and the line of fixed points does not extend out
to weak coupling.  Thus we have one modulus even under the $\BZ_2$.
We can once more relax the symmetry requirement, demanding only that
$SU(2)\times SU(2)$ be unbroken.  We then have two independent
anomalous dimensions, and still only one superpotential coupling
$h$. The scaling factors are now
\eqn\ai{\eqalign{
A_{g_1} &= -2N(1 + \gamma_X + \gamma_Y)\cr
A_{g_2} &= -2N(1 + \gamma_X + \gamma_Y)\cr
A_h &= 1 + \gamma_X + \gamma_Y\ .\cr
}}
Since in this case we have one condition on three couplings, there is
an additional, $\BZ_2$-odd, marginal operator.  This is naturally
identified with the difference of the two gauge couplings.  More
precisely, as discussed in \kw , for this theory we should
form dimensionless combinations of $h$ and the dynamical scales
(complexified by $\theta$ angles as usual), $\lambda_i = h\Lambda_i$.
The even and odd couplings are then naturally
\eqn\lamdas{\lambda_\pm = \lambda_1\pm\lambda_2\ .}

Coordinates on the critical surface should correspond to moduli of the
background.  One of these, the string coupling and axion, is clearly
invariant under the parity symmetry implementing ${\bf Z}_2$, so maps
to the even operator.  In the case at hand, the derivation from the
quotient singularity of the previous subsection allows us to verify
this directly. If we set $\zeta_2 = \zeta_3$, and if further in the
quotient model we set all gauge couplings equal $\tau_i=\tau=\tau_S$,
then the model we obtain will be at the $\BZ_2$ invariant point.  Then
\ais\ shows that in fact the dimensionless coupling defined above
satisfies the one-loop matching condition $\lambda = \Lambda/\zeta_2 =
e^{2\pi i\tau_S}$.
The odd coupling should correspond to the periods of $B_{NS}$ and
$B_{RR}$ about the vanishing two-cycle.  This changes sign under the
${\bf Z}_2$ (acting on our invariants as $a\to -a$, it clearly induces
the flop).  

The type IIB compactification has as always one modulus, invariant
under all the symmetries, the string coupling $\tau_S$.  Above, we
showed that the worldvolume coupling $\lambda_+$ is determined by
the asymptotic value of the dilaton in the ambient space.  The scaling
argument relates this to the string coupling in the dual IIB model.
As is clear in the description as a topological product space,
$H^5$ has one nontrivial two-cycle. The
interpretation of the odd coupling, based upon previous experience, is
that it reflects the two-form periods about this $S^2$.  This is the
same cycle we identified in the previous discussion (on the
six-manifold) and as explained there is indeed $\BZ_2$-odd.  Thus we
are led to identify
\eqn\mapsii{\eqalign{
\lambda_+&=  e^{2\pi i\tau_S} \cr
\lambda_-&= f(b)\ ,\cr
}}
for some function $f$.  As before, the periodicity of $b$ predicts
$S$-dualities of the conformal field theory; in this case these are
the more interesting because we have an interacting fixed point.  On
the other hand, we do not know the precise form of $f$.  

There is a particularly interesting point in the space of deformations
of this theory, namely the point at which the two-form periods both
vanish.  At this point the conformal field theory describing the
closed-string sector is singular.  The singularity is signalled by the
fact that the tension of the wrapped D3-brane strings vanishes.  We
thus expect a singularity in the worldvolume theory.  This point
presumably corresponds to the limit in which the gauge couplings are
infinite; the $\BZ_2$ symmetry will be restored at this point.

\subsec{The suspended pinch point singularity}

The toric analysis of section 4 shows that in codimension two in the
parameter space, where the $\BZ_2$ locus intersects the conifold
locus, we find a suspended pinch point singularity.  We can locate
this at $\zeta_1 = \zeta_2 = 0$.
To find the worldvolume theory in this case we choose $\zeta_3\gg 0$.
The maximally symmetric ground state is then represented by
\eqn\vevsii{Z_{34} = \zeta_3^{1/2}\ ,}
breaking the gauge symmetry down to $SU(N)\times SU(N)\times SU(N)$.
Proceeding as
above we find that the light fields and their representations are
\eqna\lightii{
$$\eqalignno{
\eqalign{
X &= X_{14} \quad (\bk,{\bf 1},\bkb)\cr
\tilde X &= Y_{31} \quad (\bkb,{\bf 1},\bk)\cr}
\qquad&\eqalign{
Y &= Y_{24} \quad ({\bf 1},\bk,\bkb)\cr
\tilde Y&= X_{32} \quad (1,\bkb,\bk)\cr}
\qquad\eqalign{
Z &= Z_{12} \quad (\bk,\bkb,{\bf 1})\cr
\tilde Z&= Z_{21} \quad (\bkb,\bk,{\bf 1})\cr}&
\lightii{}\cr
\noalign{and}
&\phi = Z_{43} \quad ({\bf 1},{\bf 1},{\bf N^2-1})\ .\cr }$$}

\vskip-12pt

\noindent
Integrating the massive fields out by imposing their
equations of motion we find the superpotential
\eqn\wii{W = \tr\left(\phi(\tilde YY - \tilde XX ) +
\zeta_3^{-1/2}( Z\tilde ZX  \tilde X - \tilde Z Z \tilde Y Y)
\right)\ .}
Comparing with the discussion in section four we see that in this case
setting $N=1$ will not lead directly to an identical description.
Indeed, \wii\ does not vanish in this case. 

The model has a $U(1)^5$ global symmetry group preserving the
superpotential up to an overall phase.  Of this, one $U(1)$ is
anomalous and combines as usual with the na\"\i ve $U(1)_R$ to yield the
non-anomalous $U(1)'_R$ under which the bifundamental fields have
unit charge and $\phi$ charge 2.  Of the non-$R$ symmetries, a
$U(1)^2$ subgroup are baryonic.  There are
additional discrete symmetries of the model.  One of these exchanges
the first two factors of the gauge group, simultaneously exchanging
$X$ with $Y$ (and $\tilde X$ with $\tilde Y$) and $Z$ with $\tilde Z$.
This is an $R$-symmetry
because the superpotential changes sign and the previous footnotes
apply.  The other symmetry acts as complex conjugation on the vector
multiplets and on the components of $\phi$
while replacing $X^I$ with $\tilde X^I$.  This preserves
the superpotential.  The total global symmetry group is thus
\eqn\globalii{U(1)^5\rtimes (\BZ_2\times\BZ_2)\ .}

Comparing to \susyisomspp\ we see that as expected, the baryonic
$U(1)^2$ and charge conjugation are not realized as isometries.
The latter is the center of the duality group.  The former should
correspond to periods of the RR four-form about three-cycles in $H^5$.
Because the horizon in this case is not smooth we will need to resort
to our heuristic description of the cycles, justified
by its success in describing non-isolated quotients.  The construction
of section 4 suggests, along these lines, that there are two kinds of
three-cycles on $H^5$.  The first, which we denote $A$, arising from
the fiber structure as 
in the case of the conifold, is topologically $S^3$.  The second,
arising as in the non-isolated quotient case, is $C = \Sigma\times S^1$ where
the $S^1$ is the singular circle $x_0=y_0=0$ and $\Sigma$ is the
two-sphere resolving the $\BC^2/\BZ_2$ singularity in the transverse
space to the circle.  This description suggests that one of the
$\BZ_2$ generators in \globalii\ be interpreted as the quantum
symmetry associated to this singularity.  This is the first generator
mentioned above; the baryon current associated to $C$
cycle should be reversed by this, identifying this as
$J_1-J_2$, where $J_i$ couples to the $U(1)$ in the $i^{\rm th}$ factor of
the gauge group.  We complete our basis with the even current $J_3$.
The basis of three-cycles will be discussed below.

The moduli space of vacua is found once more by constructing the
holomorphic invariants in the fields.  Once more we find several
branches.  The invariants are traces of products of
\eqna\invntsii{
$$\eqalignno{
\eqalign{
a &= \tilde XX  = \tilde YY \cr
b &= \tilde ZZ \cr}
&\qquad\eqalign{
c &= X \tilde Y\tilde Z\cr
d &= Y\tilde X Z\cr}\cr
\noalign{and}
&\phi\ ,&\invntsii{}\cr
}$$}

\vskip-12pt

\noindent
where the second equality in the first line uses \wii .  Using the
rest of \wii\ we see that under traces these all commute.  They
satisfy the relation
\eqn\relii{a^2 b = cd\ .}
To these we must add baryonic invariants, in this case we have one
baryon for each bifundamental chiral multiplet, satisfying the
relations
\eqn\barlnsii{
\eqalign{
B_X B_{\tilde X} &\sim B_Y B_{\tilde Y} \sim a^N\cr
B_Z B_{\tilde Z} &\sim b^N\ .\cr
}}

The traces of products
of \invntsii{} parameterize a symmetric product of $N$ copies of the
space determined by \relii .  Comparing to the second line of Table 4,
we see that this is precisely the suspended pinch
point.  Thus these moduli describe the motion of $N$
branes in the vicinity of the singularity.  At generic points in the
space, $\phi$ is massive.
The gauge symmetry is broken down to $U(1)^{N-1}$, and all massless
matter is neutral, leading to the expected accidental $\CN{=}4$ SUSY.
Incorporating the baryonic fields we expect to find as above that the
moduli of the singularity itself, as well as the positions of the
branes on the resolved geometry, comprise the moduli space. 

There is another branch of the moduli space along which $\phi$ is
nonvanishing.  Then \wii\ implies that of the invariants above only
$b$ can be nonzero, while $X$ and $Y$ are massive.   This branch thus
has the form $\left( 
\BC^N/S_N\right)^2$ and describes the splitting of branes into pairs of
fractional branes constrained to move along the $z$ axis, where there
is a vanishing cycle.  The gauge symmetry is
generically $U(1)^{2N-1}$ and there is no charged massless matter.  
Along this branch, only $B_Z$ and $B_{\tilde Z}$ are massless.
Along this branch the moment map satisfies $\zeta_3=0$.
As usual there are mixed branches in which some of the branes have
split and $\phi$ has smaller rank.

We can identify the baryons in the \CFT\ with particles in the \AdS\
compactification representing D3-branes wrapping three-cycles in the
horizon.  The discussion above allows us to identify these.  The fact
that $B_Z$ and $B_{\tilde Z}$ remain massless for vacua corresponding
to generic $\zeta_1$
provided $\zeta_3=0$ shows they are related to the three-cycle $C$
from our discussion above.  They are thus interpreted as
D3-branes wrapped about $\pm C$.  Note that these two
are indeed exchanged by the quantum symmetry, as one would expect.
The cycle $A$, as in the conifold case, has two natural
representatives. Recall from section 4 that the horizon $H^5$ is a $U(1)$
bundle 
over $ {\Bbb W}\BC{\Bbb P}^{1,2}\times {\Bbb W}\BC{\Bbb P}^{1,2}$.  Over a point in
the first factor in the base we find a three-cycle we can denote $A$.
Over a point in the second factor we find a cycle which one can show
is homologically $\tilde A=-A$.  We keep these two distinct since they
transform differently under the global symmetry group (again in
analogy with the conifold).
The quantum symmetry acts on the homology as 
$C\to -C$ and $A\to A-C$, and the charges under the baryonic
symmetries show that the baryonic states correspond to D3-branes
wrapping three-cycles as follows
\eqn\barcycles{
\eqalign{
B_X &\to A\cr
B_{\tilde X} &\to \tilde A\cr}
\qquad
\eqalign{
B_Y &\to A-C\cr
B_{\tilde Y} &\to \tilde A+C\cr}
\qquad
\eqalign{
B_Z &\to C\cr
B_{\tilde Z} &\to -C\ .\cr}
}

To look for candidate truly marginal operators we once more appeal to
the methods of \ls .  The most general gauge-invariant superpotential
preserving the full global symmetry has two couplings multiplying the
two kinds of terms 
in \wii .  The anomalous dimensions of $X$ and $Y$ are constrained to
be the same, while $Z$ can have a different dimension.  The scaling
coefficients are
\eqn\aiis{
\eqalign{
A_{g_{1,2}} &= -2N(1 + \gamma_{X,Y} + \gamma_Z)\cr
A_{g_3} &= -2N(\gamma_0 + 2\gamma_{X,Y})\cr}
\qquad\eqalign{
A_\lambda &= \gamma_0/2 + \gamma_{X,Y}\cr
A_h &= 1 + \gamma_Z + \gamma_{X,Y}\ ,\cr
}}
where $\gamma_0$ is the dimension of $\phi$.  These satisfy two
linear relations, so we predict two marginal
couplings neutral under all of \globalii .  Relaxing the symmetry
conditions and considering also terms breaking the quantum symmetry
mentioned above, the four terms in $W$
now have independent couplings, and the scaling coefficients are
\eqn\aii{
\eqalign{
A_{g_1} &= -2N(1 +\gamma_{Z}+\gamma_{X})\cr
A_{g_2} &= -2N(1 +\gamma_{Z}+\gamma_{Y})\cr
A_{g_3} &= -2N(\gamma_0+\gamma_{X}+\gamma_{Y})\cr}
\qquad\eqalign{
A_{\lambda_1} &= \gamma_0/2 + \gamma_{Y}\cr
A_{\lambda_2} &= \gamma_0/2 + \gamma_{X}\cr
A_{h_1} &= 1 + \gamma_{Z}+ \gamma_{X}\cr
A_{h_2} &= 1 + \gamma_{Z} + \gamma_{Y}\ .\cr
}}
The seven functions satisfy three linear relations so that we predict
one additional, $\BZ_2$-odd, marginal coupling.

Interpreting these along the lines of the previous subsection, the
one-loop $\beta$-function for $g_3$ vanishes, so we can take this as
one of the even couplings.  The $\beta$-functions for the other two
factors do not vanish and as usual the couplings are transmuted into
dynamical scales $\Lambda_{1,2}$.  Together with the superpotential
couplings we can form the dimensionless quantities
$\lambda_i=h_i\Lambda_i$.  These are permuted by $\BZ_2$, so lead to
one even and one odd coupling.  As above, we can directly identify the
even coupling by recalling the derivation from the quotient model.
One-loop matching yields (working at the $\BZ_2$ invariant point)
\eqn\matchii{g_3 = \Lambda/\zeta = e^{2\pi i\tau_S}\ ,}
so that all three couplings $g_i$ are equal to the string coupling.
The other two couplings will arise from two-form periods.
The odd coupling is clearly associated to the two-cycle resolving the
$\BZ_2$ singularity; the even coupling can be associated to either of
the two factors in the base. 

\subsec{The complex cone over ${\Bbb F}_0$}

We can apply the methods here to study one more singularity.  This is
described as a $\BZ_2$ quotient of the conifold singularity, the
action on the coordinates of \relni\ being $(a,b,c,d)\to
(-a,-b,-c,-d)$.
We can realize this model with our techniques because it can be
described as a $\BZ_2$ quotient of the conifold as mentioned earlier.  We
thus apply 
once more the methods of \dgm\ for dealing with quotients.  To study
$N$ branes at this point we need to study $2N$ branes at the
conifold, a theory we have discussed above.
We take the $\BZ_2$ action on the fields of \lighti\
to be $X\to -X$.  Note that on the
invariants of \invntsi\ this acts by reversing all signs, preserving
\relni\ and fixing only the origin.

The gauge fields surviving the projection correspond to the subgroup
$SU(N)^4$.  The surviving matter multiplets and their charges are
\eqn\lightiii{
\eqalign{
{X_i}_{12}&\quad (\bk,{\bf 1},{\bf 1},\bkb)\cr
{X_i}_{21}&\quad ({\bf 1},\bk,\bkb,{\bf 1})\cr}
\qquad\eqalign{
{Y_i}_{11}&\quad (\bkb,{\bf 1},\bk,{\bf 1})\cr
{Y_i}_{22}&\quad ({\bf 1},\bkb,{\bf 1},\bk)\ ,\cr}
}
and the superpotential takes the form
\eqn\wiii{
W = \tr(\epsilon^{ij}\epsilon^{kl} {X_i}_{12} {Y_k}_{22} {X_j}_{21}
{Y_l}_{11})\ .}

This leaves unbroken a $U(2)\times U(2)\times U(1)\times U(1)$
global symmetry.  The $U(2)^2$ acts as in \reps.  
The diagonal $U(1)$ subgroup of this is
anomalous, and as usual combines with the na\"\i ve $R$-symmetry to
form the non-anomalous $U(1)'_R$ under which all lowest components of
the chiral fields have charge $1/2$.  The remaining $U(1)^3$ are the
baryon number symmetries.  Note that in this chiral theory we find
that in fact two of these are in fact broken by gauge anomalies, so
that only one baryon number symmetry survives.  In addition,
there is a discrete $\BZ_4$ symmetry which permutes the factors in the
gauge group according to the cycle (1423) and permutes the four types
of chiral multiplets cyclically so that the trace in \wiii\ is
unchanged.  Including action on the vector multiplets by charge
conjugation, together with a permutation, say $(12)$ so that the
representation content is preserved, and an action on the matter
multiplets by the permutation determined by the representations (in
the case at hand this is $X_{12}\leftrightarrow Y_{22}$ and
$X_{21}\leftrightarrow Y_{11}$) extends this to
an entire $D_4$ group.  Two of the elements of order two are associated
to charge conjugation.  The third, the square of the order four
element, is in fact the quantum symmetry associated to the $\BZ_2$
quotient we used to construct the model from the conifold.  Note that
two of the order four elements (the ones involving charge conjugation)
do not preserve the superpotential and must be accompanied by
appropriate $U(1)_R$ rotations.
The correct global symmetry group is thus 
\eqn\globiii{
U(2)\times U(2)\rtimes D_4\ .
}
Comparing to \susyisomconefzero\ we find the expected differences.
The non-anomalous baryon number symmetry corresponds to the gauge
symmetry carried by the period of the RR four-form about the unique
(non-torsion) three-cycle on the horizon.  The $\BZ_2$ isometry found
in section 4 squares in fact to the quantum symmetry, generating the
$\BZ_4$ discrete symmetry.  The additional $\BZ_2$ action 
associated to charge conjugation corresponds to the center of $SL(2,\BZ)$.

The classical moduli space of vacua for this theory is parameterized
by the invariant traces.  In this case the generators are in fact
quartic expressions, of the form 
\eqn\invntsiii{ 
c_{ijkl} = {X_i}_{12} {Y_j}_{22} {X_k}_{21} {Y_l}_{11}\ .
}
The equations of motion following from \wiii\ show that these satisfy 
\eqn\reliii{\eqalign{
c_{ijkl} &= c_{kjil} = c_{ilkj}\cr
c_{ijkl}c_{mnrs} &= c_{ijrs} c_{mnkl}\ ,\cr
}}
so that they can be written as quadratic expressions
$c_{ijkl} = a_{ij} a_{kl}$ with the $a_{ij}$ satisfying \relni .  This
corresponds to our representation of the transverse space as a $\BZ_2$
quotient of the conifold.  To these we must add the baryonic
invariants
\eqn\barsiii{
\eqalign{
{B_s}_{12} &\sim {X_1}_{12}^s {X_2}_{12}^{N-s}\cr
{B_s}_{21} &\sim {X_1}_{21}^s {X_2}_{21}^{N-s}\cr
}\qquad
\eqalign{
{B_s}_{11} &\sim {Y_1}_{11}^s {Y_2}_{11}^{N-s}\cr
{B_s}_{22} &\sim {Y_1}_{22}^s {Y_2}_{22}^{N-s}\ .\cr
}}
These transform in two copies of $({\bf N+1},{\bf 1})_N\oplus 
({\bf 1},{\bf N+1})_{-N}$ under the continuous global symmetry (we
label the representation by $SU(2)^2\times U(1)_B$ content; all these
state have $U(1)'_R$ charge $N/2$).  Their transformation properties
under the discrete group follow from those of the chiral multiplets
\lightiii .  They 
satisfy relations of which a representative is 
\eqn\barelnsiii{
{B_s}_{12}{B_r}_{22}{B_t}_{21}{B_u}_{11}\sim c_{1111}^u
c_{1112}^{t-u}c_{1122}^{r-t}c_{1222}^{s-r}c_{2222}^{N-s}\ ,
}
valid when $s\ge r\ge t\ge u$ and $s-r\le r-t$. 

As in the previous cases, the meson-like degrees of freedom
\invntsiii\ parameterize the $N^{\rm th}$ symmetric product
of the singularity in question.  Incorporating the baryons, we expect
to find the moduli of the singularity as well. 
At generic points on this branch the
gauge symmetry is $U(1)^{N-1}$; the $\CN{=}4$ supersymmetry is
restored, as expected.

However, there is a
subtlety in this case that was absent in the previous ones. There are
additional branches in the moduli space which are purely baryonic,
hence absent from the description we have given.  To see these, note
that the symmetries of $c_{ijkl}$ show that \barelnsiii\ depends on
the $U(2)$ indices only through the combinations $s+t$ and $u+r$.
In fact, \barelnsiii\ together with \reliii\ are generically
sufficient to yield
\eqn\barsame{
{B_s}_{12} = {B_s}_{21}\qquad {\rm and}\qquad {B_s}_{11} = {B_s}_{22}\ .
}
If we now consider regions of moduli space in which all the $Y$
variables vanish, however, we find that $c_{ijkl}=B_{11}=B_{22}=0$,
and the remaining baryons are subject to no such restriction.
Thus, in this region, new baryonic degrees of freedom are light and
can parameterize new branches of the moduli space. 
To understand the low-energy physics along these branches we return to
the description in terms of the charged fields, and assign expectation
values to ${X_1}_{12}$ and ${X_1}_{21}$.  We then see
that the gauge group is broken to
$SU(N)\times SU(N)$, and the light matter surviving the Higgs
mechanism is precisely that of an $\CN{=}2$ theory with this gauge
group and two hypermultiplets in the $(\bk,\bkb)$ representation;
inserting the expectation values into \wiii\ leaves the expected cubic
superpotential.  The baryonic moduli in the full theory map out the
Coulomb branch of the low-energy theory.  The Higgs branch of the
latter is a subspace of the large branch discussed above.  The
couplings in the $\CN{=}2$ model are determined by the remaining
baryonic moduli. 
There is of course a similar story when the $X$ variables vanish.
Since the new branches are baryonic we cannot form mixed branches.

This description of new branches is consistent with the geometry described
in section 4.  Recall that for each of the partial resolutions of this
singularity there is a curve of $\BZ_2$
quotient singularities.  In this situation we would expect
``fractional'' branches in which wrapped D5-branes are confined to
the singular curve, in agreement with the description we have given.

We now turn to the counting of marginal operators in the theory.  This
is simplified in this case by the fact that \wiii\ is the unique
superpotential preserving the continuous global symmetry.  Thus we have
\eqn\aiii{\eqalign{
A_{g_1} &= -2N(1 + \gamma_{12}+\gamma_{11})\cr
A_{g_2} &= -2N(1 + \gamma_{21}+\gamma_{22})\cr
A_{g_3} &= -2N(1 + \gamma_{21}+\gamma_{11})\cr
A_{g_4} &= -2N(1 + \gamma_{12}+\gamma_{22})\cr
A_h &= 1 + \half(\gamma_{12}+\gamma_{22}+\gamma_{21}+\gamma_{11})\ .
}}
These are easily seen to obey two linear relations, leading to a
two-dimensional space of fixed points.  As usual, one of these,
preserving the full discrete symmetry, corresponds to the string
coupling, while the other, transforming by a sign under all order four
elements of the discrete group, should be associated to two-form
periods.
The marginal operators in the \AdS\ description will be
the IIB string coupling as well as the two-form periods.  The relevant
cohomology group in this case is $H^2_{\rm DR}(H^5) = \BZ$ so we
indeed expect precisely one such coupling. 

We can also compare the baryon spectrum to the homology cycles on
$H^5$ and the natural parameter spaces for their representatives.
As discussed in section 4, the relevant homology group $H_3(H^5)$ is
$\BZ\oplus\BZ_2$.  The natural representative cycles are two families
of three-spheres, such that their sum is the torsion element.  The parameter
space for each of these is a two-sphere.  This is in agreement with
the discussion above.   In the ``large'' branch of the moduli space,
there are indeed two flavors of baryons, since $B_{12}\sim B_{21}$ and
$B_{11}\sim B_{22}$.   These have opposite charge under the baryonic
$U(1)$ symmetry, so that a state with two baryons, one from each
family, is neutral.  They also transform in the spin-$N/2$
representation of (distinct) $SU(2)$ global symmetry groups,
corresponding to the quantization of collective coordinates on the
two-spheres.  

\newsec{Discussion}

Extending the \AdS /\CFT\ correspondence to describe branes at
singularities is, as we have stressed, required for a complete
description of the duality.  Since configurations with branes at
singularities occur at a finite distance in moduli space, the
modifications to the extreme low-energy theory at these points should
be captured by an appropriate modification of the \AdS\
compactification.  We have argued that the modification in question is
to replace the spherical horizon with a (non-spherical) horizon manifold
determined by 
the local geometry at the singular point.  We expect this statement to
hold quite generally for the types of singularities encountered at
finite distance. 
In this paper, we have offered some evidence to support this extension
of the conjecture.  In complete generality, we find that the $R$-symmetry
group of the \CFT\ is realized in the dual model by isometries of 
the horizon.  This correspondence is predicted by the conjectures and
the fact that it obtains is very encouraging.  Further 
evidence is found by considering individual models in detail.

We have presented such detailed studies of a small set of examples in the
D3-brane case.  (It will be interesting to extend this work to the
case of M5-branes \nonspherII\ as well as to the more challenging
case of M2-branes.)  Our approach to these D3-brane models allows us to
derive the worldvolume Lagrangian (including interactions) from an 
explicit construction.  In each case we have studied, comparisons with the
predictions of the dual 
\AdS\ model give additional evidence for the extended conjecture.  The
agreement in the baryon spectrum, in particular, is noteworthy.  These
states are chiral primaries, but their conformal weight grows at
large $N$ like $N$; most tests of the conjectures thus far were
restricted to operators whose weight remains finite in this limit.  As
in \refs{\witbaryons,\gubserklebanov} we interpret this agreement as
evidence that finite $N$ effects are captured as stringy corrections
to the SUGRA theory on \AdS .
Our discussion of these states has necessitated a careful study of the fate
of Abelian factors in the gauge group.  Extending our observations on
orbifold models, we have formulated a conjecture according to which these
behave in one of two ways, depending upon the nature of the blowup to which
they correspond (``small'' or ``big'').  An explicit computation supporting
this, giving evidence from additional examples, is expected to clarify this
issue soon.
Another feature of our approach is that we directly find relations
between the various models by embedding them all in the original
orbifold theory.  Thus, we can describe deformations of one into the
other quite explicitly.  

Our discussion of these examples, while encouraging, has not been
complete.  Most dramatically, we have not discussed the most
interesting singularities---the points in the moduli space
at which the two-form periods vanish.  At these points, the conformal
field theory describing the closed-string perturbation theory is
singular, and we expect new light degrees of freedom, described by
wrapped branes, to propagate in the brane worldvolume.  It is not
clear if these will decouple from the low-energy theory on the
worldvolume but this appears unlikely.  A more explicit description of
these theories than the one we have obtained---as the
infinite-coupling limit of one of our gauge theories---would
presumably lead to an understanding of this issue.   

Related to this is the question of the quantum corrections to the
classical moduli spaces we have constructed.   These are important
near the interacting conformal theories and understanding the
corrections in the dual \AdS\ model would be useful.  More
speculatively, it should be possible to make more explicit statements
about the UV/IR connection that has played such an important role in
our construction.  In particular, some properties of RG flow should
find corresponding properties in the radial dependence of the
background fields in the associated SUGRA solutions.

\bigbreak\bigskip\bigskip\centerline{{\bf Acknowledgements}}\nobreak
We would like to thank Eva Silverstein for collaboration in the early
stages of this project.
We would also like to thank Ofer Aharony, Paul Aspinwall,
Robert Bryant, Brian Greene, Jeff Harvey,
Juan Maldacena, Greg Moore, Bill Pardon,
and Nick Warner for discussions.
We gratefully acknowledge the hospitality and support of 
 the Institute for Theoretical Physics, Santa
Barbara (D.R.M.) and the Aspen Center for Physics (D.R.M.\ 
and M.R.P.).
The work of D.R.M.\ was supported in part
by NSF grant DMS-9401447, and the work of M.R.P.\ was supported in part by
the Israel Academy of Sciences and Humanities Centres of Excellence
Programme.

\appendix{A}{Sasaki and $G_2$ structures}

We collect here the definitions of the various geometric structures which
occur on horizon manifolds, taken from \refs{\FKone,\FKtwo,\baer}.
All of these are implied by existence of
certain numbers of Killing spinors, as we explained in section 2.

A {\it metric contact structure}\/ on a Riemannian manifold $H^{2n-1}$ (with
metric $\langle\,,\,\rangle$)
consists of a vector field $X$,
a one-form $\eta$, and a $(1,1)$ tensor field $\phi$ such that
\item{(i)} $\eta \wedge (d\eta)^{n-1}\ne0$
\item{(ii)} $\eta(X)=1, \phi(X)=0$
\item{(iii)} $\phi^2=-\boldone+\eta\otimes X$
\item{(iv)}$\langle \phi(V),\phi(W)\rangle = \langle V,W \rangle -
\eta(V)\eta(W)$
\item{(v)} $d\eta(V,W)=2\langle V,\phi(W)\rangle$ where
$d\eta(V,W)=V(\eta(W))-W(\eta(V))-\eta([V,W])$.

A {\it Sasaki structure}\/ is a metric contact structure such that
\item{(vi)} $X$ is a Killing vector field, or equivalently,
$\nabla_VX=-\phi(V)$
\item{(vii)} $(\nabla_V\phi)(W)=\langle V,W\rangle X-\eta(W)V$.

\noindent
A K\"ahler structure on the cone $C(H)$ determines a Sasaki structure on $H$ as
follows.
Let $J$ be the parallel
complex structure on $C(H)$ determined by the K\"ahler structure. Then
\eqn\sasex{
X=J(\partial_r),\quad \eta(V)=\langle X,V\rangle,\quad \phi=\nabla X
}
defines the Sasaki structure, where $\partial_r$ is the radial vector field.

A {\it $3$-Sasaki structure}\foot{These structures have recently made an
appearance in the study of rigid $N=2$ superconformal hypermultiplets
\GibbonsRych.}  consists of a triple of Sasaki structures
$(\phi_i,X_i,\eta_i)$ such that
\item{(i)}$X_1$, $X_2$, $X_3$ are orthonormal
\item{(ii)} $[X_1,X_2]=2X_3$, $[X_2,X_3]=2X_1$, $[X_3,X_1]=2X_2$
\item{(iii)} $\phi_3\phi_2=-\phi_1+\eta_2\otimes X_3$,\quad
$\phi_2\phi_3=\phi_1+\eta_3\otimes X_2$,
\item{\ } $\phi_1\phi_3=-\phi_2+\eta_3\otimes X_1$,\quad
$\phi_3\phi_1=\phi_2+\eta_1\otimes X_3$,
\item{\ } $\phi_2\phi_1=-\phi_3+\eta_1\otimes X_2$,\quad
$\phi_1\phi_2=\phi_3+\eta_2\otimes X_1$.

\noindent
A hyper-K\"ahler structure on the cone $C(H)$ determines a $3$-Sasaki structure
on
$H$ as follows.  Let $I$, $J$, $K$ be  parallel complex
structures on $C(H)$ such that $IJ=-JI=K$. Then the three Sasaki structures are
given
by \eqn\threeex{
X_1=I(\partial_r),\quad X_2=J(\partial_r),\quad X_3=-K(\partial_r)
}
with $\eta_i$ and $\phi_i$ determined from $X_i$ as in \sasex.

To describe  nearly-parallel $G_2$-structures, we must first recall that
the action of 
$SO(7)$
on $\Lambda^3(\Bbb{R}^7)^*$ has two open orbits, the {\it definite}\/
three-forms and
the {\it indefinite}\/ three-forms \bryant; the stabilizer of any definite
three-form is
(conjugate to) the compact group $G_2$ whereas the stabilizer of an indefinite
three-form is the non-compact
form of $G_2$.  A three-form on a seven-manifold is called {\it definite}\/ if
it is definite
in each tangent space.

A {\it nearly-parallel $G_2$-manifold}\/\foot{We thank K. Galicki for a
discussion on terminology, and for pointing us to his expository paper with
C.~P. Boyer (hep-th/9810250) which contains an extensive bibliography on
this subject.}
is a seven-manifold with a definite
three-form $\phi$ such that $\nabla\phi=\star\phi$.  The Riemannian metric is
automatically Einstein.
There is a one-to-one correspondence between nearly-parallel
$G_2$-structures on $H$ 
and $\Spin(7)$-structures
on $C(H)$.

\appendix{B}{$D$-term equations and Moduli Spaces}

The standard construction of the moduli space for the
$\BZ_2\times\BZ_2$ quotient proceeds by constructing all the
holomorphic $SU(N)^4$ invariants in the chiral fields, modulo the
$F$-term relations.  In addition to \invnts{} we then have the baryons
$B_{ij} \sim X_{ij}^N$, subject to relations of the form (suppressing
numerical factors)
\eqn\brelns{
\eqalign{
B_{14} B_{41} &= x_1^N\cr
B_{23} B_{32} &= x_2^N\cr
}
\qquad\eqalign{
B_{13} B_{31} &= y_1^N\cr
B_{24} B_{42} &= y_2^N\cr
}
\qquad\eqalign{
B_{12} B_{21} &= z_1^N\cr
B_{34} B_{43} &= z_2^N\ .\cr
}}
Together with \invnts{} these give coordinates on the full moduli
space. 

The $D$-term equations for this case are
\eqn\d{\eqalign{
X_{14} X^\dagger_{14} + Y_{13} Y^\dagger_{13} + Z_{12} Z^\dagger_{12}
- X^\dagger_{41}X_{41} - Y^\dagger_{31}Y_{31} - Z^\dagger_{21}Z_{21}
&= \zeta_1 \cr
X_{23} X^\dagger_{23} + Y_{24} Y^\dagger_{24} + Z_{21} Z^\dagger_{21}
- X^\dagger_{32}X_{32} - Y^\dagger_{42}Y_{42} - Z^\dagger_{12}Z_{12}
&= \zeta_2 \cr
X_{32} X^\dagger_{32} + Y_{31} Y^\dagger_{31} + Z_{34} Z^\dagger_{34}
- X^\dagger_{23}X_{23} - Y^\dagger_{13}Y_{13} - Z^\dagger_{43}Z_{43}
&= \zeta_3 \cr
X_{41} X^\dagger_{41} + Y_{42} Y^\dagger_{42} + Z_{43} Z^\dagger_{43}
- X^\dagger_{14}X_{14} - Y^\dagger_{24}Y_{24} - Z^\dagger_{34}Z_{34}
&= \zeta_4\ ,\cr
}}
where the $\zeta_i$ are free to vary subject to $\sum_i\zeta_i = 0$.
As in section 5, it is useful to study the various ``slices'' of this
moduli space in which the $\zeta_i$'s are held constant.

To make contact with the description of this space in terms of toric
geometry, and in particular to correctly identify the points in moduli
space corresponding to interesting singular geometries,
we review from \refs{\dgm,\brg,\mukray} another description
of this space in the case $N=1$.\foot{This description does not
appear to directly generalize past $N=1$.}
In that case, the $F$-flatness conditions arising from \w\ themselves
describe a toric variety, as they are all of the form ``one monomial equals
another monomial.''  Some combinatorial manipulations (\refs{\brg,\mukray},
following
the method of \refs{\sardoinfirrione})
show that we can describe this toric variety as a quotient of $\Bbb{C}^9$
with (homogeneous)
coordinates $p_1$, \dots, $p_9$ by $U(1)^3$, acting with charge matrix:\foot{We
have used different bases than either \brg\ or \mukray, in order to clarify
certain points below.}
\eqn\greenecharge{
\pmatrix{p_1&p_2&p_3&p_4&p_5&p_6&p_7&p_8&p_9\cr
0&-1&-1&1&0&0&1&0&0\cr-1&0&-1&0&1&0&0&1&0\cr-1&-1&0&0&0&1&0&0&1}.
}
The invariant coordinates are:
\eqn\greeneinvs{
\matrix{
X_{14}=p_1p_8p_9 & X_{23}=p_1p_5p_9 & X_{32}=p_1p_6p_8 & X_{41}=p_1p_5p_6 \cr
Y_{13}=p_2p_4p_9 & Y_{24}=p_2p_7p_9 & Y_{31}=p_2p_6p_7 & Y_{42}=p_2p_4p_6 \cr
Z_{12}=p_3p_4p_8 & Z_{21}=p_3p_5p_7 & Z_{34}=p_3p_7p_8 & Z_{43}=p_3p_4p_5}.
}
These satisfy all of the $F$-term equations, viz.,
\eqn\greeneF{\matrix{\matrix{
X_{14}Y_{42}=p_1p_2p_4p_6p_8p_9=Y_{13}X_{32}, &
X_{14}Z_{43}=p_1p_3p_4p_5p_8p_9=Z_{12}X_{23}, \cr
X_{23}Y_{31}=p_1p_2p_5p_6p_7p_9=Y_{24}X_{41}, &
X_{23}Z_{34}=p_1p_3p_5p_7p_8p_9=Z_{21}X_{14}, \cr
X_{32}Y_{24}=p_1p_2p_6p_7p_8p_9=Y_{31}X_{14}, &
X_{32}Z_{21}=p_1p_3p_5p_6p_7p_8=Z_{34}X_{41}, \cr
X_{41}Y_{13}=p_1p_2p_4p_5p_6p_9=Y_{42}X_{23}, &
X_{41}Z_{12}=p_1p_3p_4p_5p_6p_8=Z_{43}X_{32}.
}\cr\phantom{M}\cr
\matrix{
Y_{13}Z_{34}=p_2p_3p_4p_7p_8p_9=Z_{12}Y_{24}, \cr
Y_{24}Z_{43}=p_2p_3p_4p_5p_7p_9=Z_{21}Y_{13}, \cr
Y_{31}Z_{12}=p_2p_3p_4p_6p_7p_8=Z_{34}Y_{42}, \cr
Y_{42}Z_{21}=p_2p_3p_4p_5p_6p_7=Z_{43}Y_{31}.
}}}

The $D$-term equations from our theory are associated to a $U(1)^4$ action
on the fields
$X_{ij}$, $Y_{ij}$, $Z_{ij}$, and this can be lifted to a second $U(1)^4$
action on the fields $p_\alpha$,
with charge matrix:
\eqn\greenechrgbis{
\pmatrix{p_1&p_2&p_3&p_4&p_5&p_6&p_7&p_8&p_9\cr
1&0&0&1&-1&-1&0&0&0\cr 0&1&0&-1&1&-1&0&0&0\cr 0&0&1&-1&-1&1&0&0&0\cr
-1&-1&-1&1&1&1&0&0&0}.
}
(Of course, the diagonal $U(1)$ acts trivially as expected.)
To verify that we have lifted the charge assignments correctly, we calculate
the induced charges on the invariant coordinates, from \greeneinvs\ and
\greenechrgbis:
\eqn\greenechrgverify{\pmatrix{
X_{14}&X_{23}&X_{32}&X_{41}& Y_{13}&Y_{24}&Y_{31}&Y_{42}&
Z_{12}&Z_{21}&Z_{34}&Z_{43}\cr
1&0&0&-1&1&0&-1&0&1&-1&0&0\cr
0&1&-1&0&0&1&0&-1&-1&1&0&0\cr
0&-1&1&0&-1&0&1&0&0&0&1&-1\cr
-1&0&0&1&0&-1&0&1&0&0&-1&1}
}
These are exactly the expected charge assignments for our theory.

The four $U(1)$'s in \greenechrgbis\
have associated FI terms with coefficients $\zeta_1$,
$\zeta_2$, $\zeta_3$, $\zeta_4$.
If we collect the charges from all seven $U(1)$'s into a single matrix, and
include the FI coefficients
as an extra column, we obtain
\eqn\greeneFI{
\pmatrix{1&0&0&1&-1&-1&0&0&0&\zeta_1\cr 0&1&0&-1&1&-1&0&0&0&\zeta_2\cr
0&0&1&-1&-1&1&0&0&0&\zeta_3\cr -1&-1&-1&1&1&1&0&0&0&\zeta_4\cr
0&-1&-1&1&0&0&1&0&0&0\cr-1&0&-1&0&1&0&0&1&0&0\cr-1&-1&0&0&0&1&0&0&1&0
}}
Doing a few row operations, this becomes:
\eqn\greeneperm{
\pmatrix{1&0&0&1&-1&-1&0&0&0&\zeta_1\cr 0&1&0&-1&1&-1&0&0&0&\zeta_2\cr
0&0&1&-1&-1&1&0&0&0&\zeta_3\cr
0&0&0&0&0&0&0&0&0&\sum\zeta_i\cr
0&0&0&-1&0&0&1&0&0&\zeta_2+\zeta_3\cr
0&0&0&0&-1&0&0&1&0&\zeta_1+\zeta_3\cr
0&0&0&0&0&-1&0&0&1&\zeta_1+\zeta_2
}.}
(Notice that as expected we must have $\sum\zeta_i=0$.)  In an appropriate
region of
$\zeta$ space, we can use the last three rows of \greeneperm\ to
eliminate $p_7$, $p_8$, $p_9$, and we are left with precisely the toric
description
of a $\Bbb{Z}_2\times \Bbb{Z}_2$ quotient singularity given in section 4.2,
with
the $D$-terms precisely identified.  (The notational coincidence of using
$\zeta_i$ for the $D$-terms \d\ 
is now seen to be a precise correspondence with the moment map of section
4.2.) 

\listrefs

\bye